\documentclass[sigconf,nonacm]{acmart}
\usepackage{amsmath,amsfonts}
\usepackage{textcomp}
\usepackage{graphicx}
\usepackage{booktabs}
\usepackage{xspace}
\usepackage{algorithm}
\usepackage{algpseudocode}
\usepackage{enumitem}

\makeatletter
\if@ACM@newfonts\else
  \RequirePackage[T1]{fontenc}
  \RequirePackage{lmodern}
\fi
\makeatother

\newcommand{\tokenstack}{\textsc{TokenStack}\xspace}

\title{\tokenstack: A Heterogeneous HBM-PIM Architecture and Runtime for Efficient LLM Inference}

\author{Zhuoran Li}
\affiliation{
  \institution{Peking University}
  \city{Beijing}
  \country{China}}
\email{zhuoranli@stu.pku.edu.cn}

\author{Zhuohang Bian}
\affiliation{
  \institution{Peking University}
  \city{Beijing}
  \country{China}}
\email{bianzhuohang26@stu.pku.edu.cn}

\author{Zihao Huang}
\affiliation{
  \institution{University of Electronic Science and Technology of China}
  \city{Chengdu}
  \country{China}}
\email{zhhuang@std.uestc.edu.cn}

\author{Yibo Zhao}
\affiliation{
  \institution{Peking University}
  \city{Beijing}
  \country{China}}
\email{yibozhao@stu.pku.edu.cn}

\author{Xueqi Li}
\affiliation{
  \institution{Institute of Computing Technology, Chinese Academy of Sciences}
  \city{Beijing}
  \country{China}}
\email{lixueqi@ict.ac.cn}

\author{Guangyu Sun}
\affiliation{
  \institution{Peking University}
  \city{Beijing}
  \country{China}}
\email{gsun@pku.edu.cn}

\author{Youwei Zhuo}
\affiliation{
  \institution{Peking University}
  \city{Beijing}
  \country{China}}
\email{youwei@pku.edu.cn}

\renewcommand{\shortauthors}{Zhuoran Li, Zhuohang Bian, Zihao Huang, Yibo Zhao, Xueqi Li, Guangyu Sun, and Youwei Zhuo}

\begin{document}

\begin{abstract}
Large language model (LLM) serving is now limited by the key--value (KV) cache. During decode, each new token rereads prior KV state, so attention becomes a bandwidth- and capacity-heavy memory task. HBM-PIM helps by moving attention closer to memory, but current stack organizations still waste resources. In practice, only hot KV blocks benefit from near-memory compute. Weights, activations, and cold KV mainly need dense storage and GPU-visible bandwidth. A uniform HBM-PIM stack makes all layers pay for PIM logic, while a dedicated-PIM design such as AttAcc recovers capacity but shrinks the HBM bandwidth left for GPU-side work.

We propose \textbf{\tokenstack}, a vertically heterogeneous HBM-PIM architecture for KV-centric LLM serving that leverages HBM4's logic-die substrate. \tokenstack separates each stack into dense \emph{capacity layers} and PIM-enabled \emph{compute layers}, then uses the logic base die as a stack-local control point that manages cross-layer movement without host-side overhead. The base-die controller handles cross-layer DMA, layered address translation, attention-side gather/broadcast coordination, and inline quantization during migration. On top of this hardware, \tokenstack uses topology-aware KV placement, workload-aware eviction, and bounded replication to keep hot KV near PIM compute while moving colder state to dense layers. Using production-derived traces across four models, completed multi-QPS runs show that \tokenstack increases geometric-mean token throughput by $1.62\times$ and SLO-compliant serving capacity by $1.70\times$ over AttAcc, and reduces per-token energy by 30-47\%.
\end{abstract}

\ccsdesc[500]{Computer systems organization~Heterogeneous (hybrid) systems}
\ccsdesc[300]{Computer systems organization~Processors and memory architectures}
\ccsdesc[300]{Hardware~Memory and dense storage}

\keywords{processing-in-memory, high-bandwidth memory, large language model inference, KV cache, memory hierarchy}

\maketitle

\section{Introduction}
\label{sec:introduction}

Large language model (LLM) serving has become a memory-system problem.
During autoregressive decode, each newly generated token must reread the
entire accumulated key--value (KV) context; the arithmetic intensity of
this step drops to $O(1)$, making decode-phase attention fundamentally
bandwidth-limited.  As sequence
lengths and request concurrency grow, the KV Cache dominates both the
capacity and bandwidth demands of the serving
pipeline~\cite{Park2024ASPLOSAttAcc, Yun2024Duplex, Gu2025ASPLOSCENT}.

Processing-in-memory (PIM) offers a natural response: by placing
lightweight compute units adjacent to the memory arrays that store KV
data, PIM reduces the data-movement cost of attention.
AttAcc offloads attention to bank-level HBM-PIM engines while the GPU
executes compute-heavy projection and feed-forward
layers~\cite{Park2024ASPLOSAttAcc}, and subsequent designs extend the
approach to mixture-of-experts and grouped-query
attention~\cite{Yun2024Duplex}.
These designs accelerate the attention path but leave a structural
question unanswered: \emph{given that only a fraction of the data in the
memory stack benefits from PIM, how should the stack itself be
organized?}

\begin{figure*}[t]
\centering
\includegraphics[width=\textwidth]{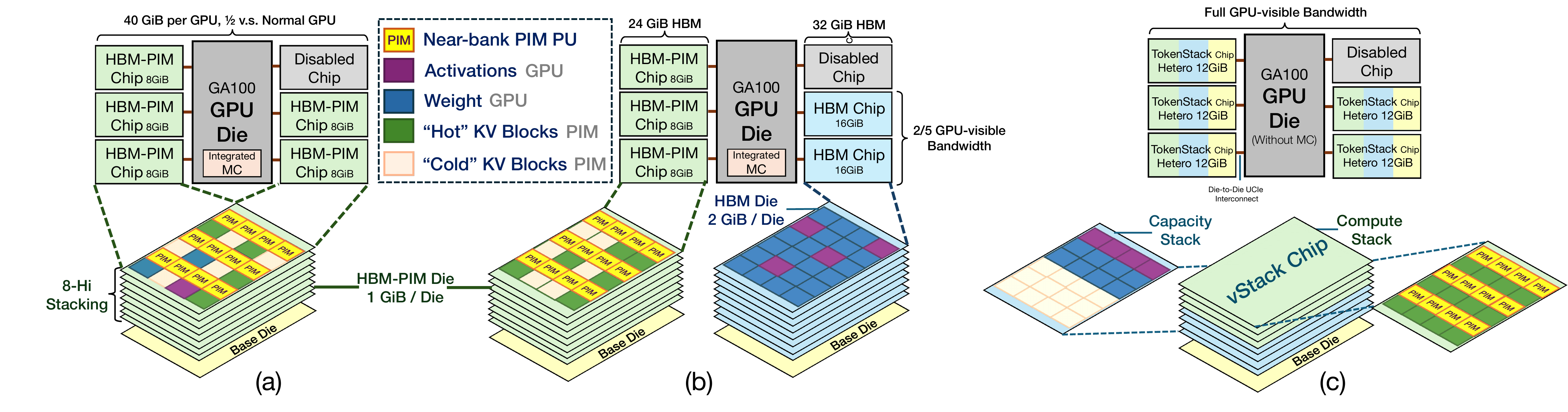}
\caption{HBM-PIM organizations. \textbf{(a)} Uniform PIM sacrifices
capacity. \textbf{(b)} Dedicated PIM exposes only part of memory BW
to the GPU. \textbf{(c)} \tokenstack combines dense capacity and PIM-enabled compute
layers within each stack, preserving full GPU-visible BW and reserving
PIM for hot KV.}
\Description{Uniform HBM-PIM loses capacity, dedicated PIM loses GPU-visible
bandwidth, and TokenStack combines capacity and compute layers within each stack to
preserve bandwidth while applying PIM only to hot KV blocks.}
\label{fig:intro_tokenstack}
\end{figure*}

The data mix of KV-centric serving is inherently heterogeneous.  Model
weights, activations, and runtime metadata require dense storage and
high GPU-visible bandwidth but never execute on PIM units.  Among KV
blocks themselves, reuse is highly skewed: production traces report that
roughly 10\% of KV blocks account for 77\% of reuse
events~\cite{Wang2025KVCacheWild}.
The reuse distribution also varies by workload category.
Short-form API traffic exhibits high single-turn reuse, while long-form
reasoning shows almost no cross-request sharing~\cite{Wang2025KVCacheWild}.
A serving architecture should therefore differentiate between \emph{hot}
KV that benefits from near-memory compute and the remaining data that
primarily needs capacity.

Recent HBM-PIM organizations fail to provide this differentiation because
they operate within homogeneous memory regions
(Fig.~\ref{fig:intro_tokenstack}(a)--(b)).  A
\emph{uniform} HBM-PIM stack equips every layer with PIM,
forcing cold KV, weights, and activations to pay the PIM area cost even
though those objects never execute on PIM units; the resulting 40\,GiB
configuration provides only half the capacity of a conventional GPU.
At the other extreme, a \emph{dedicated-PIM} organization assigns a
fixed subset of dies to PIM and retains the rest as standard
HBM~\cite{Park2024ASPLOSAttAcc}.  This preserves 32\,GiB of dense storage
beside 24\,GiB of PIM memory, but exposes only two-fifths of the HBM
bandwidth to the GPU.  Neither organization matches the heterogeneous
data mix described above.

This organizational mismatch cannot be resolved through software tiering
alone because its root cause is physical: when all layers share the same
substrate, every byte pays the same area and density cost regardless of
its access pattern.  The fundamental solution is a stack in which different
layers serve different roles---dense capacity in some and PIM-enabled
compute in others---and a local controller migrates data between them
without routing routine traffic through the host.  Historically, the host
GPU had to manage such migration because stack-local control logic was
unavailable, adding latency and bus contention to the critical path.  HBM4
makes heterogeneous stacking practical by replacing the traditionally
passive base die with a CMOS logic die capable of hosting active control
logic and die-to-die communication with the host
~\cite{JEDEC2025HBM4, DasSharma2025UCIeMemory}, thereby enabling stack-local
data management.

We propose \textbf{\tokenstack{}}, a vertically heterogeneous HBM-PIM
architecture for KV-centric LLM serving that leverages HBM4's logic-die
substrate to enable efficient stack-local management (Fig.~\ref{fig:intro_tokenstack}(c)).  At the hardware level,
\tokenstack{} divides each memory stack into two layer types.
\emph{Capacity layers} are dense HBM layers that store weights, activations,
and cold KV. \emph{Compute layers} are PIM-enabled layers that hold the
hot KV working set adjacent to bank-level attention engines.  The logic
base die serves as a stack-local control point, managing cross-layer DMA,
layered address translation, attention-side gather/broadcast coordination,
and inline K8V4 quantization during KV migration.  By confining PIM logic
to only the compute layers, \tokenstack{} matches the hardware cost of each
layer to the access patterns it serves: capacity density where needed,
compute proximity where beneficial.

The heterogeneity is only effective if software policies keep hot KV
aligned with the compute layers and cold KV in capacity layers.
\tokenstack{}'s runtime exploits the stack topology to maximize
this alignment.  Topology-aware home assignment places each request's KV
blocks near the stack location where its prefix already resides,
reducing initial migration cost.  Workload-aware eviction uses per-block
metadata (request category, reuse recency, and prompt
position) to estimate short-horizon reuse probability, demoting
low-value blocks to capacity layers via stack-internal DMA.  Bounded
replication creates a second copy of a frequently accessed block on a
remote card only when the predicted callback savings exceed the fanout
cost.  Continuous batching exposes reuse opportunities incrementally,
allowing newly arrived
requests to share prefix blocks that are already resident in compute
layers.

We evaluate \tokenstack{} using a cycle-accurate simulator extended for
heterogeneous stacks and trace-driven serving, across four
production-derived traces and four models ranging from Qwen3-4B to
GPT-175B (\S\ref{sec:methodology}).  Across all 16 model--trace
combinations, \tokenstack{} delivers a geometric-mean token-throughput
improvement of $1.62\times$ over AttAcc, with per-pair gains ranging
from $1.03\times$ to $2.32\times$.

Under a $2\times$ latency SLO,
\tokenstack{} serves $1.70\times$ more requests than AttAcc (geometric
mean), with the capacity advantage widest on the largest models where
KV pressure is most severe.

This paper makes four contributions:
\begin{enumerate}
  \item \textbf{Problem characterization.}
    We analyze the data-placement requirements of KV-centric LLM serving
    and show that homogeneous stack designs cannot simultaneously provide
    the capacity and compute density needed; a heterogeneous stack is
    required~(\S\ref{sec:background}).

  \item \textbf{Heterogeneous stack architecture.}
    Leveraging HBM4's logic-die control substrate, we design a vertically
    specialized HBM stack that combines dense capacity layers, PIM-enabled
    compute layers, and stack-local data management, including asymmetric
    Key/Value layouts matched to the attention dataflow and an inline
    quantization path~(\S\ref{sec:design}, \S\ref{sec:layout}).

  \item \textbf{Topology- and workload-aware runtime.}
    We develop runtime policies (topology-aware placement,
    category-aware eviction, and bounded replication) that keep the highest-value blocks in the
    compute-visible domain under dynamic request
    arrivals~(\S\ref{sec:runtime}).

  \item \textbf{Comprehensive evaluation.}
    We evaluate \tokenstack{} across 16 model--trace pairs and show
    consistent latency and throughput improvements over uniform-PIM,
    dedicated-PIM, and GPU, with the largest gains on the
    models and workloads where KV pressure is severe~(\S\ref{sec:methodology}, \S\ref{sec:evaluation}).
\end{enumerate}

\section{Background and Motivation}
\label{sec:background}

This section establishes the two observations that motivate
\tokenstack{}'s design.  First, autoregressive decode creates coupled
bandwidth and capacity pressure on the KV Cache
(\S\ref{ssec:llm-serving}).  Second, existing HBM-PIM stack
organizations address one dimension of this pressure at the expense of
the other (\S\ref{ssec:limitations}).

\subsection{Decode-Time KV Cache Pressure}
\label{ssec:llm-serving}

Autoregressive LLM inference proceeds in two
phases~\cite{Park2024ASPLOSAttAcc,Yun2024Duplex,Liu2026PAM}.
\emph{Prefill} processes the entire prompt in parallel; its attention
operates on a dense $L{\times}L$ score matrix that amortizes memory
accesses over $O(L^{2})$ arithmetic.  \emph{Decode} generates one token
per step, rereading the full accumulated Key and Value state for each
new query.  Its arithmetic intensity drops to $O(1)$ per cached token,
making decode attention memory-bandwidth-limited.

This bandwidth demand is compounded by capacity demand that grows along three axes: the per-token KV footprint scales with hidden dimension and layer count, the per-request state grows linearly with context length, and concurrent requests multiply the aggregate live footprint.
KV blocks still being actively attended by in-flight requests constitute the \emph{hot} set;
all other cached blocks awaiting potential future reuse are \emph{cold}.
The two pressures are tightly coupled: when cold KV, weights, and activations compete for the same memory as hot KV, effective attention bandwidth diminishes even with a fast interface.

Production trace studies confirm that the resulting management problem
cannot be solved with a simple recency-based eviction
policy~\cite{Wang2025KVCacheWild}.  Three characteristics of real
KV reuse make this clear.

\noindent\textbf{(1) Skew.}  A small fraction of KV blocks accounts for a
    disproportionate share of reuse events; the majority of blocks are
    accessed only by the request that created them.
\noindent\textbf{(2) Category dependence.}  Short-form API traffic exhibits
    high single-turn prefix reuse concentrated in the first few hundred
    tokens; code-editing workloads show moderate cross-request sharing
    over file-context prefixes; long-form reasoning workloads generate
    large per-request KV states with almost no inter-request reuse.
\noindent\textbf{(3) Transience.}  Even among blocks that are reused, the
    reuse window is often short: a block that is not accessed within a
    category-specific time horizon is unlikely to be accessed again.

Together, these characteristics define a clear requirement for the
memory substrate: the system must differentiate between a small, dynamic
hot set of KV blocks that benefits from in-situ computation and a much
larger body of data---cold KV, weights, and activations---that primarily
needs dense, GPU-visible storage.
\subsection{Limitations of Current HBM-PIM Organizations}
\label{ssec:limitations}
PIM has been applied to graph processing~\cite{Zhuo2018GraphP,Zhuo2019GraphQ,Dai2019GraphH,AhnScalable,Zhuo2018GraphR,Zhuo2021DistGraphPIM,Shin2025Piccolo,Yenimol2026CoGraf,Chen2025Anchor,Kim2025EOD,Chen2025HEAT},
linear algebra~\cite{Giannoula2022SparseP,Devic2022PIMGeneralPurposeDDR,Kwon2019TensorDIMM},
neural networks~\cite{liu2025l3dimmpim,zhou2022gnnearacceleratingfullbatchtraining,Kim2023SamsungPIMPNMTransformerAI,Imani2019FloatPIM,Zhou2022TransPIM,Kim2016NeuroCube,Li2024PIMDL},
recommendation systems~\cite{Liu2021ENMCNearMemoryClassification,Ke2020RecNMPNearMemoryProcessing,Liu2023ISCAAccel,Ke2022AxDIMM,Chen2024UpDLRM,Liu2024NearCXL},

retrieval and ANN search~\cite{Liu2025SeIM,Li2025ANSMET,Quinn2025DReX,Zou2026NasZip,Liu2025HeterRAG,Liu2026MERIDIAN},
database analytics~\cite{Zhao2025PUSHtap,Devic2026AnalyticBank,Ma2024Moctopus},
security and privacy~\cite{Xiong2022SecNDP,Zhang2025FENIX,Lin2025Ironman,Kim2025Anaheim,Park2025FHENDI,Nam2025AffinityPIM},
and other domains including genomics~\cite{Kim2025NMPPaK}, ray tracing~\cite{Saed2025RayN}, circuit simulation~\cite{lee2025pimutation}, and on-device inference~\cite{jeun2025lpddr5foldpim,park2024lpddrcxl}.

Among these efforts, HBM-PIM\cite{KwonHBM2PIM,Samsung2021ISSCCHBM2PIM,Lee2021HBM2PIMSamsung} for LLM serving stands out as a
particularly compelling use case: decode-phase attention is
bandwidth-bound yet touches only a small fraction of total memory,
making it a natural fit for near-memory execution.
Indeed, AttAcc~\cite{Park2024ASPLOSAttAcc} has demonstrated that
placing lightweight compute units adjacent to HBM arrays allows
attention to run at bank-level bandwidth without traversing the
memory bus to the
GPU~\cite{Samsung2021ISSCCHBM2PIM,Park2024ASPLOSAttAcc}, and
Duplex~\cite{Yun2024Duplex} extends this idea to MoE
and GQA.  The open question, therefore, is not
\emph{whether} near-memory attention is beneficial, but \emph{how}
PIM capability should be distributed across the layers of the memory
stack and which data should reside in PIM-enabled regions.

\begin{figure}[t]
\centering
\includegraphics[width=\columnwidth]{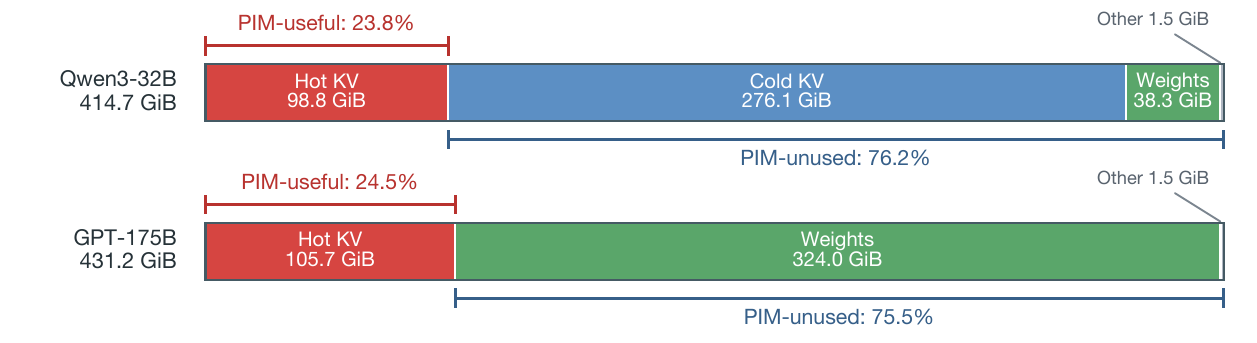}
\caption{Resident-data mix under near-full Uniform HBM-PIM pressure while
replaying traceA for 1800\,s. Only active/hot KV executes on PIM; all other
resident data consumes PIM-enabled capacity without using PIM compute.}
\Description{Stacked bars show the resident data mix for Uniform HBM-PIM. Active hot KV occupies roughly one quarter of memory, while cold KV, weights, activations, and runtime overhead occupy the remaining three quarters.}
\label{fig:pim_useful_data_mix}
\end{figure}

\noindent\textbf{Uniform HBM-PIM.}
In a uniform organization, every memory layer carries PIM logic.  This
provides a single homogeneous substrate and maximizes aggregate
near-memory bandwidth.
The uniform organization in Fig.~\ref{fig:intro_tokenstack}(a)
shows the physical cost: PIM circuitry reduces each die from 2\,GiB to
1\,GiB in the illustrated eight-high stack, leaving only 40\,GiB per GPU.
Our data-mix breakdown in Fig.~\ref{fig:pim_useful_data_mix} shows
that only 23.8--24.5\% of occupied HBM is PIM-useful active KV.
The remaining 75.5--76.2\% is cold KV, weights, activations, and runtime overhead that never executes on PIM units.
Uniform HBM-PIM thus pays the area and density cost of PIM circuitry on roughly three-quarters of its resident bytes.
Furthermore, uniform PIM introduces an
operational constraint: banks must alternate between all-bank (AB) mode
for PIM attention and single-bank (SB) mode for regular data access,
creating synchronization stalls on every decode step.

\noindent\textbf{Dedicated-PIM} assigns a fixed subset of dies to PIM and retains the remaining
dies as standard HBM for GPU-side data, as in AttAcc~\cite{Park2024ASPLOSAttAcc}.
The illustrated organization provides 24\,GiB of PIM memory and 32\,GiB
of dense HBM.  This separation avoids storing non-KV data in
compute-enabled area, but only the conventional HBM dies serve GPU
accesses, reducing GPU-visible bandwidth to two-fifths of the full stack.
As model size or batch concurrency increases, this bandwidth loss becomes
the dominant bottleneck.

\noindent\textbf{The common limitation.}
Both share a structural deficiency: the granularity of differentiation is the \emph{die}. A software-managed caching hierarchy can prioritize placement but cannot change the underlying cost---in a uniform stack every byte of cold KV occupies compute-enabled area; in a dedicated-PIM stack every added PIM die reduces GPU-visible bandwidth.

\section{\tokenstack{} Design Overview}
\label{sec:overview}

\begin{figure*}[!t]
\centering
\includegraphics[width=0.85\textwidth]{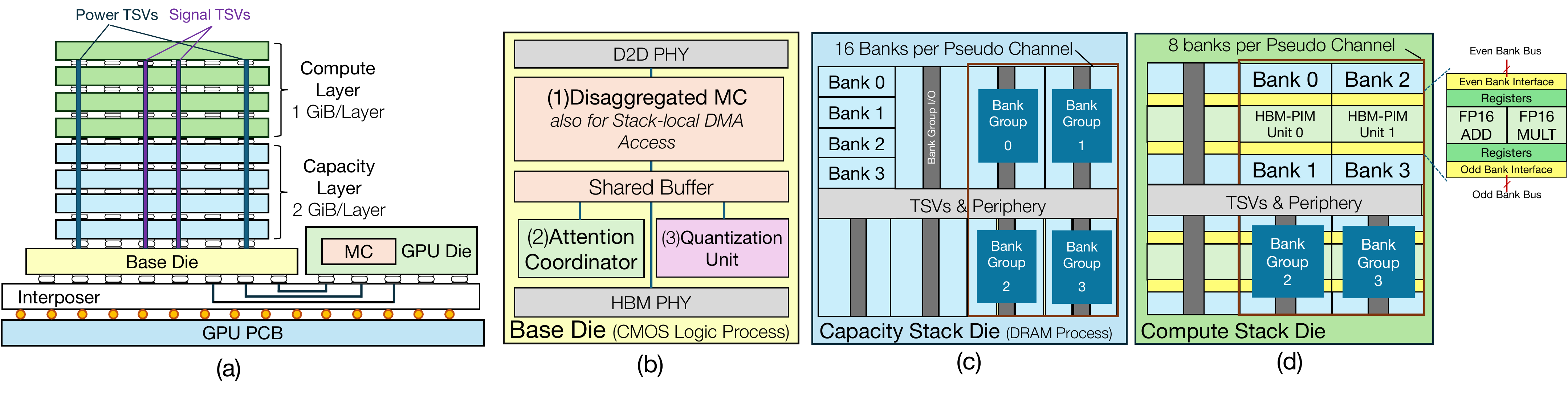}
\caption{\tokenstack{}'s heterogeneous capacity and compute layers.}
\Description{A GPU connected to TokenStack HBM chips.  Each chip combines dense capacity layers for weights, activations, and cold KV blocks with PIM-enabled compute layers for hot KV blocks.}
\label{fig:system_arch}
\end{figure*}

\tokenstack{} separates mechanism from policy across 4 layers.

\noindent\textbf{Heterogeneous stacking.}
Each HBM stack is partitioned into dense capacity layers
and PIM-enabled compute layers (\S\ref{ssec:stack-org}).
Capacity layers store weights, activations, and cold KV;
compute layers hold the hot KV working set beside bank-level
attention engines.
This avoids both the density penalty of uniform PIM and the bandwidth
loss of a dedicated-PIM.

\noindent\textbf{Base-die control.}
The HBM4 logic base die provides control:
cross-layer DMA, layered address translation,
attention-side coordination, and inline K8V4 quantization,
all managed within the stack (\S\ref{ssec:base-die}).

\noindent\textbf{Static data layout} assigns each data class to a domain
and fixes the block arrangement for the entire run (\S\ref{sec:layout}).
In the compute layers, Keys take a token-major layout and Values
a feature-interleaved layout, both matched to the attention dataflow.
In the capacity layers, weights and activations are chunk-interleaved
for bandwidth.
Cold KV is paged across dies for aggregate TSV throughput.

\noindent\textbf{Dynamic runtime} manages individual blocks continuously
via topology-aware assignment (\S\ref{ssec:serving-sched}),
category-aware eviction (\S\ref{ssec:eviction}), and bounded replication
(\S\ref{ssec:selective-replica}).  As requests arrive and contexts grow,
these policies revise which blocks occupy the compute domain,
keeping the highest-value KV resident near PIM engines.

\section{\tokenstack Hardware Design}
\label{sec:design}

\subsection{Heterogeneous Stack Organization}
\label{ssec:stack-org}

Each \tokenstack stack contains three zones: a logic base die at the bottom,
capacity layers above it, and compute layers at the top.
The two DRAM zones match the data mix of LLM serving: capacity layers provide
density for weights, activations, and cold KV, while compute layers provide
PIM proximity for hot KV.

\noindent\textbf{Capacity layers.}
Standard high-density HBM layers hold weights, activations, metadata, and
cold KV---data requiring density over compute.
They sit nearest the base die to minimize TSV hops for GPU-originated reads.

\noindent\textbf{Compute layers.}
Embedding FP16 MAC units, register files, and control logic in each Bank
Group consumes ${\sim}$50\% of the silicon area, halving per-bank storage
capacity.
This trade-off enables in-situ attention execution for hot KV, eliminating
the GPU round-trip.
Placing these dies at the stack top also provides the shortest path to the
heat sink.

\noindent\textbf{Capacity accounting.}
A stack with $C$ capacity and $P$ compute layers exposes
$C \!\times\! B_{\mathrm{full}} + P \!\times\! B_{\mathrm{half}}$ addressable bytes.
The host GPU sees a standard HBM device with an auxiliary PIM command interface.

\subsection{Base-Die Control Substrate}
\label{ssec:base-die}

Migration between layers must stay off the critical path and require no host
involvement.
Historically, HBM base dies were passive interfaces; data migration had to be
managed by the GPU, adding latency and bus contention.
The HBM4 standard introduces a base die fabricated on an advanced CMOS logic
process
node~\cite{JEDEC2025HBM4,samsung2026hbm,rambus2026hbm4e,marvell2024customhbm}.
Combined with a die-to-die interconnect such as
UCIe~\cite{DasSharma2025UCIeMemory,ucie2026specifications}, this base die
turns the heterogeneous layers into an actively managed memory system and
serves as the sole interface between the DRAM stack and the rest of the
system.  The base die contains three functional blocks.

\noindent(1)~\textbf{Disaggregated memory controller (MC).}
Relocated from the GPU die to the base die, the MC sits between the
HBM TSV PHY connecting the DRAM layers and the UCIe die-to-die link
connecting the host GPU.
It arbitrates GPU-originated and PIM-originated requests through separate
queues, preventing either path from starving the other.
The MC implements
\emph{layered address translation}, mapping each logical KV block to its
physical capacity- or compute-layer location so that promotion and
demotion remain transparent to the host.  By directly bridging both PHY
interfaces, it also performs stack-local DMA through a streaming,
page-at-a-time pipeline.  Each transfer processes one KV page
($T_{\mathrm{page}}$ tokens) through the quantization path and requires
only a small on-die SRAM page buffer per head group.  This granularity
matches the paged-interleaved capacity-layer layout
(\S\ref{ssec:hist-kv-layout}), supporting arbitrarily large logical
blocks without proportional SRAM growth.
(2)~\textbf{Attention coordinator.}  The coordinator manages the
gather, broadcast, and partial-reduce steps of distributed attention
across banks and channels.  It sequences PIM commands to the compute
layers and collects partial results before returning them to the host
or forwarding them to another stack.
(3)~\textbf{Quantization / dequantization unit.}
This unit applies inline K8V4 compression when demoting a KV block from a
compute layer to a capacity layer, and reverses the process during promotion.
This expands effective capacity-layer KV space without adding a separate
compression kernel to the GPU timeline.

\noindent\textbf{External interface.}
The base die connects to the GPU through a UCIe die-to-die
(D2D) link rather than the HBM PHY.  UCIe D2D replaces the wide, low-speed HBM signal bus with a narrower,
serialized point-to-point interface that runs at up to 32\,GT/s per lane,
delivering comparable aggregate bandwidth at higher bandwidth density per
millimeter of package shoreline.
Moving the MC onto the base die frees the corresponding GPU-die
silicon---the 2,048-bit HBM PHY and its controller logic occupy
${\approx}$15\,mm$^2$ per
stack~(\S\ref{sec:methodology})---which the GPU can instead dedicate to
compute logic such as CUDA cores, tensor cores, or on-chip SRAM.

\subsection{Data-Transfer Paths}
\label{ssec:transfer-path}

\tokenstack supports two classes of data movement: \emph{intra-stack}
transfers between capacity and compute layers, and \emph{off-chip}
transfers that leave the package.

\noindent\textbf{Intra-stack path.}
A capacity-to-compute promotion (or the reverse demotion) follows a
two-hop route: source layer $\rightarrow$ base die
$\rightarrow$ target layer, using the through-silicon vias
(TSVs) that connect every die in the stack.  The base-die MC
redirects the data to the target layer via a second TSV write,
keeping the entire migration inside the package.

Promotions are tagged \emph{latency-critical} because the requester stalls
until the block arrives; demotions and replica fanout are tagged
\emph{background} and fill idle TSV slots, so migration throughput can grow
without inflating tail latency for active attention requests.

\noindent\textbf{Off-chip paths.}
Data that must leave the stack exits through the base-die UCIe D2D
link to the host GPU, which provides two onward paths.
(1)~\emph{Intra-package:} the GPU routes the data over its on-die crossbar
to another stack on the same interposer, enabling cross-stack KV migration or
remote attention gather without leaving the package.
(2)~\emph{Inter-node:} the GPU forwards the data through its
NVLink ports to a remote GPU on another card, supporting cross-node KV
replication and distributed attention when a request's KV blocks span
multiple cards.
Because both off-chip paths reuse the GPU's existing interconnect fabric,
\tokenstack does not require any additional inter-die links beyond the UCIe D2D
interface on the base die.

\section{Data Layout Policy}
\label{sec:layout}
\tokenstack's data layout maps each class of live data to the memory-layer type
that matches its consumer and arranges objects within that layer to keep the
dominant reduction bank-local.
The mapping follows from who consumes the data:
\begin{itemize}[leftmargin=1.2em,itemsep=2pt,topsep=2pt,parsep=0pt]
  \item \textbf{GPU-streamed data} (weights, activations) requires high
    GPU-visible bandwidth but no PIM compute, so chunk-interleaving across
    capacity layers is sufficient.
  \item \textbf{PIM-resident hot KV} demands bank-local reductions: splitting a
    single dot product across banks would route partial sums through the base die
    on every decode step, erasing the PIM bandwidth advantage.
    The compute layers therefore use asymmetric Key/Value layouts matched to the
    two attention phases~(\S\ref{sec:bank-layout}).
  \item \textbf{Historical KV} resides in capacity layers and is promoted to
    compute on demand.
    To maximize throughput during promotion, pages are round-robin interleaved
    across capacity-layer dies, exposing aggregate TSV
    bandwidth~(\S\ref{ssec:hist-kv-layout}).
\end{itemize}

This mapping differs from generic caching or software-managed tiering:
placement follows from a data class's role in the attention dataflow, not from
recency.
The quality metric is the compute-layer hit rate for PIM-eligible accesses,
which the runtime maximizes online~(\S\ref{sec:runtime}).
Table~\ref{tab:data_layout} and Fig.~\ref{fig:intro_tokenstack}(c) summarize the
resulting placement.

\noindent\textbf{Placement objective.}
The placement decision can be stated as a capacity-constrained optimization
over a single stack.
Let $\mathcal{D}=\{d_1,\dots,d_N\}$ be the set of live data objects in one
stack.
Each object~$d_i$ carries a normalized access frequency
$\alpha_i\!\in\![0,1]$ and a PIM affinity flag $\beta_i\!\in\!\{0,1\}$
($\beta_i\!=\!1$ iff $d_i$ is a KV block consumed by PIM-side attention).
\tokenstack chooses the placement function
$\pi:\mathcal{D}\!\to\!\{\textsc{Compute},\,\textsc{Capacity}\}$ to maximize
the total access-weighted PIM affinity served from compute layers, subject to
their aggregate capacity:
\begingroup\small
\begin{equation}
\label{eq:placement}
  \max_{\pi}\;\sum_{d_i:\,\pi(d_i)=\textsc{Compute}}
    \!\alpha_i\,\beta_i\,,
  \quad\text{s.t.}
  \sum_{d_i:\,\pi(d_i)=\textsc{Compute}}
    \!\lvert d_i\rvert
  \;\le\; C_{\mathrm{comp}},
\end{equation}
\endgroup
where $|d_i|$ is the size of object~$d_i$ in bytes and
$C_{\mathrm{comp}}=P\!\times\!B_{\mathrm{half}}$ is the aggregate compute-layer
capacity ($P$~compute-layer dies, each with per-die capacity
$B_{\mathrm{half}}$; see~\S\ref{sec:design}).
This section and \S\ref{sec:runtime} divide Eq.~\eqref{eq:placement} by factor.
The affinity flag $\beta_i$ is a static class property fixed here: objects with
$\beta_i\!=\!0$ (weights, activations, metadata) contribute zero to the
objective regardless of placement, so an optimal $\pi$ excludes them whenever the
capacity constraint binds.
The access frequency $\alpha_i$ is a per-block property observable only while
serving; the runtime estimates it from reuse history and re-solves the
remaining knapsack online~(\S\ref{sec:runtime}).

\begin{table*}[tb]
\centering
\caption{\tokenstack data-class placement.}
\label{tab:data_layout}
\scriptsize
\renewcommand{\arraystretch}{1.05}
\setlength{\tabcolsep}{3pt}
\begin{tabular}{@{}llll@{}}
\toprule
\textbf{Data Class} & \textbf{Layer} & \textbf{Layout}
  & \textbf{Rationale} \\
\midrule
Weights ($W_Q$, \\ $W_K,W_V,W_O$, FFN)
  & Capacity & Interleaved (chunk)      & Max.\ GPU BW \\
Key cache
  & Compute  & Token-major (bank-local) & Score $s{=}qK^\top$ \\
Value cache
  & Compute  & Feature-interleaved      & Context $o{=}aV$ \\
Activations
  & Capacity & Interleaved (chunk)      & GPU element-wise \\
Historical KV
  & Capacity & Paged-interleaved        & Max.\ TSV BW \\
\bottomrule
\end{tabular}
\end{table*}

\subsection{Bank Layout of the KV Cache}
\label{sec:bank-layout}

\noindent\textbf{Why K and V need different layouts.}
For a query $q \in \mathbb{R}^{d}$ and a KV block of $L$ tokens,
attention performs two reductions whose contraction axes differ, requiring
different bank layouts.
Score computation contracts the \emph{feature} dimension,
\begingroup\small
\begin{equation}
  \mathit{score}_n \;=\; \sum_{d'} q_{d'}\, k_{n,d'},
  \qquad n = 0,\dots,L-1,
  \label{eq:score}
\end{equation}
\endgroup
producing one scalar per token by summing over $d$. Output computation,
in contrast, contracts the \emph{token} dimension,
\begingroup\small
\begin{equation}
  \mathit{out}_{d'} \;=\; \sum_{n} a_n\, v_{n,d'},
  \qquad d' = 0,\dots,d-1,
  \label{eq:av}
\end{equation}
\endgroup
producing one value per feature channel by summing over the $L$ tokens,
where $a_n$ are the (softmax-normalized) attention weights. Because the summation index differs ($d$ in Eq.~\eqref{eq:score} versus $n$
in Eq.~\eqref{eq:av}), the bank layout that keeps each reduction local must
also differ.

\begin{figure}[tb]
\centering
\includegraphics[width=\columnwidth]{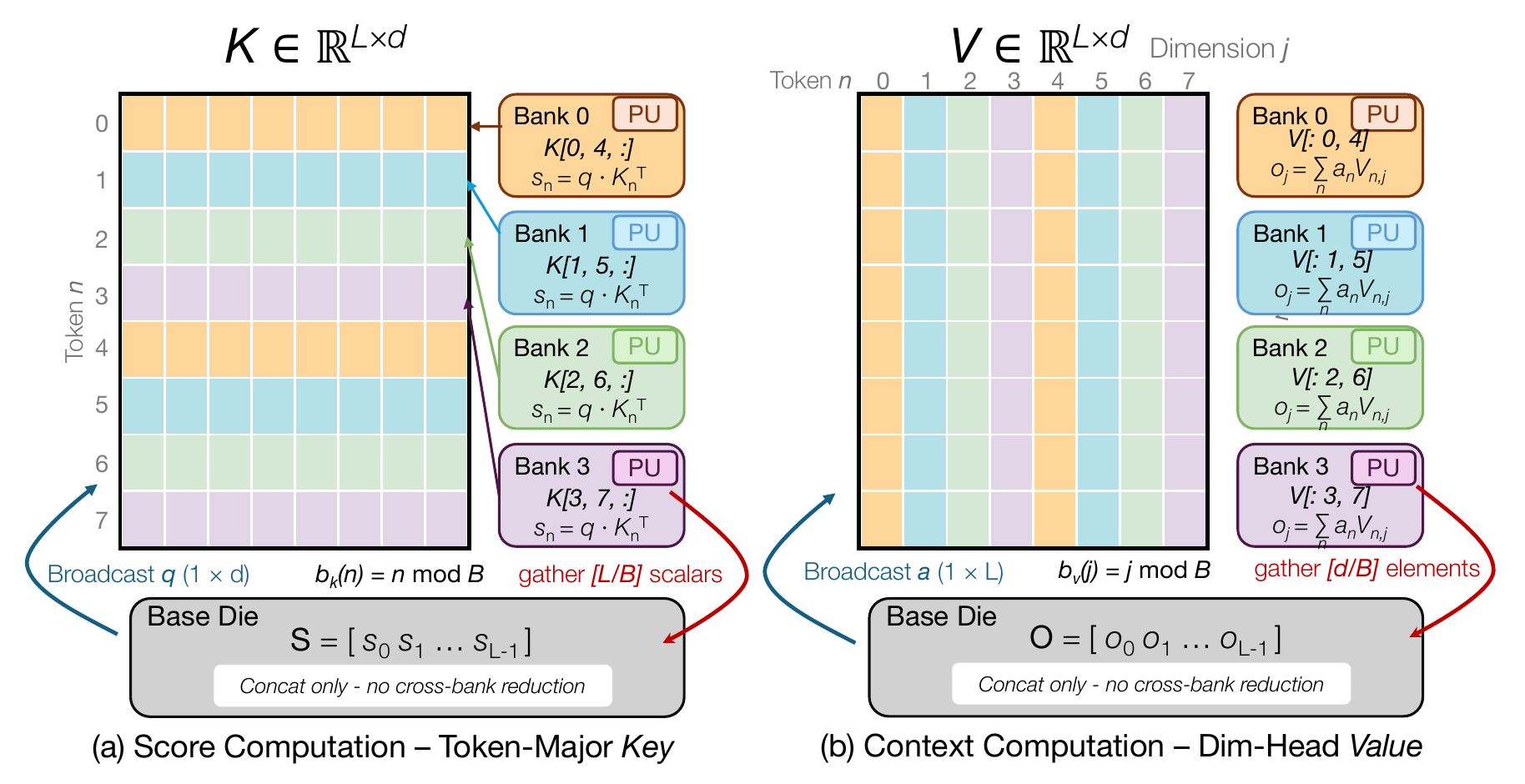}
\caption{Bank layouts for the K and V, with the $B$
  banks distinguished by color.
  \textbf{(a) K is token-major:} each token's full $d$-dimensional key
  vector resides in a single bank (one color per column), so the
  reduction over the feature dimension $d$ in $QK^{\top}$ is bank-local;
  \textbf{(b) V is feature-interleaved:} the hidden dimension $d$ is
  striped across all banks (all colors per column), so the reduction
  over the token dimension $n$ in $AV$ is bank-local.}
\Description{Two diagrams compare KV bank layouts. In the Key layout, each token vector occupies one bank so feature reduction is local. In the Value layout, feature dimensions are striped across banks so token reduction is local.}
\label{fig:kv-layout}
\end{figure}

\noindent\textbf{Token-major \textit{Key} layout.}
To evaluate Eq.~\eqref{eq:score} without crossing bank boundaries, all
$d$ feature channels of a given token must be co-located in one bank;
otherwise a single dot product would have to gather partial sums from
multiple banks. We therefore store K \emph{token-major}: the entire
$1\times d$ key vector of token $n$ is placed in bank
$n \bmod B$ (Fig.~\ref{fig:kv-layout}(a)).
Each bank then computes complete dot products for its $\frac{L}{B}$ resident
tokens independently, with no cross-bank reduction.

\noindent\textbf{Feature-interleaved \textit{Value} layout.}
Eq.~\eqref{eq:av} reduces over tokens instead, so the bank-local
requirement is the opposite: for each output channel $d'$, all $L$
token contributions $v_{n,d'}$ must reside in the same bank. We
therefore store V \emph{feature-interleaved}, partitioning the hidden
dimension across banks so that bank $b$ holds a $\frac{d}{B}\times L$ slice
spanning every token but only $\frac{d}{B}$ of the channels
(Fig.~\ref{fig:kv-layout}(b)).
A single token's value vector is therefore \emph{striped across all $B$ banks}.
The scalar weights $a_n$ from Eq.~\eqref{eq:av} are broadcast to all
banks, and each bank independently accumulates its $\frac{d}{B}$-wide slice of
the output, again with no cross-bank reduction.

\noindent\textbf{The two layouts are transposes.}
K partitions by token and V partitions by feature, making the two layouts
transposes of each other (Fig.~\ref{fig:kv-layout}).
This aligns each tensor's bank-partitioning axis with the axis
that is \emph{not} reduced, allowing both $QK^{\top}$ and $AV$ to contract
along a dimension fully resident within one bank.
Consequently, neither attention phase requires a cross-bank reduction; the only
inter-bank traffic is the broadcast of $q$ in Eq.~\eqref{eq:score} and the
scalar weights $a_n$ in Eq.~\eqref{eq:av}.

\subsection{Paged-Interleaved Layout for Historical KV}
\label{ssec:hist-kv-layout}

When a KV block is demoted from compute to capacity~(\S\ref{sec:runtime}), it
must be arranged so that a later promotion can read it back at full TSV
bandwidth.
Following vLLM-style paging~\cite{Kwon2023vLLM}, each logical block is sliced
into pages of $T_{\mathrm{page}}$ tokens, and pages are round-robin interleaved
across capacity layers.
Concretely, page~$p$ of a logical block is placed on capacity layer bank

\begingroup\small
\begin{equation}
\label{eq:cap-interleave}
  \mathrm{bank}_{\mathrm{cap}}(p) \;=\; p \bmod B_{\mathrm{cap}},
\end{equation}
\endgroup

where $B_{\mathrm{cap}}$ is the number of capacity-layer banks in the stack.
Interleaving maximizes parallel TSV bandwidth (up to $C\!\times$ single-die
throughput, where $C$ is the number of capacity-layer dies;
see~\S\ref{ssec:stack-org}) and balances load across those dies.

\noindent\textbf{Streaming promotion and demotion.}
The runtime triggers promotions and demotions~(\S\ref{sec:runtime}); the layout
determines what each transfer must do, because compute and capacity layers
differ in both data ordering and numeric format (FP16 vs.\ K8V4).
Key data can be scattered directly: each token's Key resides in one bank, so
promotion and demotion have no cross-bank dependency.
Value data, however, must be transposed between token-first ordering (capacity
pages) and dimension-first ordering (PIM banks), using an on-die page buffer.
On promotion, the MC dequantizes K8V4$\to$FP16, scatters Keys, and transposes
Values; consecutive-page reads overlap with the transpose of the prior page.
Demotion reverses the path, quantizing FP16$\to$K8V4 before writing compressed
pages.
The page buffer holds $T_{\mathrm{page}}\!\times\!d\!\times\!2\!\times\!2$\,bytes
per head group (one page of K and one of V in FP16; a few KB), so the pipeline
scales to large blocks without proportional SRAM growth.
The host GPU sees only logical block identifiers; page-level placement is managed entirely on the base die.

\section{Runtime Optimizations}
\label{sec:runtime}

\begin{figure}[tb]
\centering
\includegraphics[width=\columnwidth]{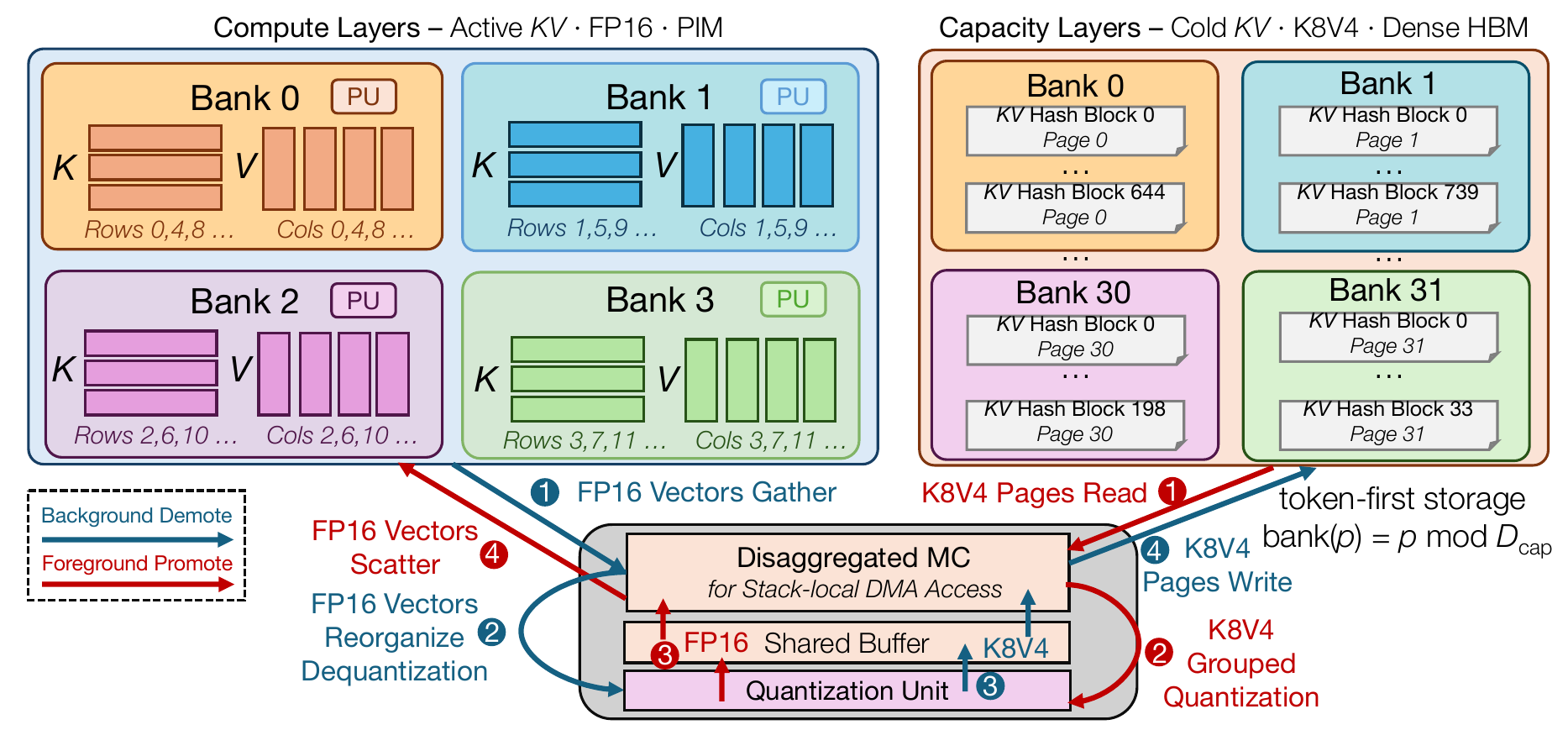}
\caption{KV block lifecycle.}
\Description{A state diagram follows a KV block from allocation in a compute layer through access, replication, demotion to a capacity layer, and later promotion or eviction.}

\label{fig:kv_lifecycle}
\end{figure}

The static layout determines where each data class belongs, but not which KV
blocks should occupy the scarce compute layers at runtime.
Because the hot KV working set shifts with every arrival and context
extension, the runtime estimates each block's access frequency $\alpha_i$ from
observed reuse and re-solves Eq.~\eqref{eq:placement} online.
A request's home is fixed at admission beside resident prefixes.
Three policies then govern what enters, leaves, and is replicated in compute
layers, all executed on the base-die MC~(\S\ref{ssec:base-die}) and kept off
the critical path:
\begin{itemize}[leftmargin=1.2em,itemsep=2pt,topsep=2pt,parsep=0pt]
  \item \textbf{Admission.} Compute capacity is deliberately scarce, so
    category-specific reuse windows determine whether a block enters compute or
    stays in dense capacity.
  \item \textbf{Eviction.} Demotion is cheap---a block moves into adjacent
    compressed capacity and is restored in ${\sim}$1--2\,\textmu s rather than
    recomputed---so scarce compute can turn over aggressively.
  \item \textbf{Replication.} Repeated cross-stack callbacks to the same block
    cause foreground stalls, so a small replica reserve converts the most
    frequent remote accesses into local ones.
\end{itemize}
Fig.~\ref{fig:kv_lifecycle} illustrates the lifecycle of a KV block through
the three physical domains.

\subsection{Serving Model and Request Scheduling}
\label{ssec:serving-sched}

\noindent\textbf{Continuous batching.}
Requests are admitted in arrival-time order, processed through chunked
prefill, and decoded with immediate backfill.
This exposes prefix-sharing incrementally: a newly arrived request can reuse
blocks already resident in compute layers from an active request without
waiting for a batch boundary.
All evaluated configurations use the same scheduler so that throughput
differences isolate the stack organization and KV policies.

\noindent\textbf{Topology-aware home assignment (admission-time).}
A request's home $(card, die)$ determines whether a shared prefix is served
by local PIM or requires a cross-stack or cross-card callback.
The scheduler uses \texttt{hash\_id} to place requests beside resident
prefixes, reducing initial migration cost.
This co-location matters because each \tokenstack home couples a PIM domain with
an adjacent capacity tier, so locality translates directly into fewer
promotions.

\noindent\textbf{Category-aware admission.}
The scheduler tags each request with its workload category (API, text, code,
thinking, etc.)\ and applies category-specific admission windows.  Short-form
API traffic exhibits high single-turn reuse concentrated in the first few
hundred tokens~\cite{Wang2025KVCacheWild}, so prefix blocks are admitted aggressively with
short lifespan windows.  Long-form reasoning has almost no inter-request
sharing; the runtime therefore keeps the \emph{current} request's KV resident
for the duration of its decode rather than speculatively caching for future
reuse.  Code traffic shows moderate cross-request sharing over file-context prefixes;
these prefixes receive longer lifespan windows and are eligible for bounded
replication~(\S\ref{ssec:selective-replica}).  Per-category parameters are
derived from exponential fits of the reuse-time distribution, updated
periodically from recent history (\S\ref{ssec:eviction}).

\subsection{Architecture-Aware Eviction}
\label{ssec:eviction}

\noindent\textbf{Per-block metadata.}
The base-die MC maintains a compact record (${\leq}$32\,B) per compute-layer
block: category~$w$, last-access timestamp $t_{\text{last}}$, prompt-position
offset, remote-hit count $n_{\text{remote}}$, and distinct-card count
$n_{\text{cards}}$---all updated inline with zero GPU overhead.

\noindent\textbf{Reuse probability.}
The runtime fits a category-specific CDF $F_w(t)$ from recent serving
history~\cite{Wang2025KVCacheWild}.
For a block of category~$w$ last accessed $\Delta t$ ago, the probability of
reuse within lifespan $\ell_w$ is
\begingroup\scriptsize
\begin{equation}
\label{eq:reuse-prob}
\textsc{ReuseProb}_w(\Delta t,\;\ell_w)
  \;=\; F_w(\Delta t + \ell_w) - F_w(\Delta t).
\end{equation}
\endgroup

\noindent\textbf{Demotion score.}
\tokenstack ranks blocks for demotion using a lexicographic score that combines the
reuse estimate with two topology-aware signals:

\begingroup\scriptsize
\begin{equation}
\label{eq:demotion-score}
\begin{aligned}
\textsc{Score}(b) = \bigl(&
  -\textsc{ReuseProb}_{b.w}(\Delta t_b,\ell_{b.w}),
  &+b.\mathit{offset},\;-b.n_{\text{remote}}\bigr).
\end{aligned}
\end{equation}
\endgroup

Lexicographic ordering: (1) lowest reuse probability first, (2) highest
prompt-position offset first~\cite{Wang2025KVCacheWild}, (3) fewest remote hits
first.
Algorithm~\ref{alg:eviction} executes whenever occupancy exceeds a high-water
mark; per-category priority queues reduce the per-victim selection cost to
$O(|W|)$, where $|W|$ is the number of active categories.
Demotions are background DMAs overlapped with the decode.

\begin{algorithm}[tb]
\caption{Compute-layer demotion on the base-die MC.}
\label{alg:eviction}
\scriptsize
\begin{algorithmic}[1]
\Require $\mathcal{Q}_w$: per-category queue of compute-layer blocks
         ordered by $t_{\text{last}}$;
         $F_w(\cdot)$, $\ell_w$: fitted CDF and lifespan for category $w$;
         $\theta_{\text{hi}}, \theta_{\text{lo}}$: high/low water marks
\Ensure  Demote blocks until occupancy $\leq \theta_{\text{lo}}$
\While{\textsc{Occupancy}() $> \theta_{\text{lo}}$}
    \State $b^* \gets \textsc{null}$;\;
           $s^* \gets (+\infty,\;-\infty,\;+\infty)$
    \For{each active category $w$}
        \State $b \gets \mathcal{Q}_w.\text{front}()$
        \State $\Delta t \gets t_{\text{now}} - b.t_{\text{last}}$
        \State $r \gets F_w(\Delta t + \ell_w) - F_w(\Delta t)$
        \State $s \gets (-r,\; +b.\mathit{offset},\; -b.n_{\text{remote}})$
        \If{$s <_{\text{lex}} s^*$}
            $b^* \gets b$;\; $s^* \gets s$
        \EndIf
    \EndFor
    \State \textsc{Pop}($\mathcal{Q}_{b^*.w}$)
    \State \textsc{EnqueueBgDMA}($b^*$, compute$\to$capacity, K8V4)
           \Comment{non-blocking}
\EndWhile
\end{algorithmic}
\end{algorithm}

\noindent\textbf{On-the-fly KV quantization.}
Cold KV needs density rather than PIM arithmetic, making \tokenstack's
compute-to-capacity transition the natural compression point.
The base-die engine compresses each block inline with the stack-local DMA
that moves it; homogeneous organizations lack a distinct transition point, and
host-side schemes consume GPU cycles and external-link bandwidth.
On demotion, the engine applies asymmetric K8V4---converting Keys
FP16$\!\to\!$INT8 ($2\times$) and Values FP16$\!\to\!$INT4 ($4\times$)---for an
overall $2.667\times$ capacity-layer expansion.  On promotion, it dequantizes
the block back to full FP16 \emph{before} it reaches any PIM compute unit,
making quantization invisible to the attention path.  K8V4 achieves
near-lossless accuracy relative to the FP16 baseline, with an average
degradation of only 0.3\% across Llama\,3 8B/70B and Qwen2.5 7B/32B on
GSM8K and HumanEval+~\cite{Zhang2025OSDIDiffKV,Liu2024ICMLKIVI}.
The engine is provisioned to match the internal TSV transfer bandwidth, so
the DMA channel, not the engine, is the throughput bottleneck.
Quantization therefore adds no critical-path latency beyond the DMA transfer
itself, unlike GPU-side
schemes~\cite{Liu2024ICMLKIVI,Hooper2024NeurIPSKVQuant,Zhang2025OSDIDiffKV}
that launch explicit compress/decompress kernels competing for GPU cycles.

\subsection{Selective Replication}
\label{ssec:selective-replica}

Since 10\% of blocks account for 77\% of reuse~\cite{Wang2025KVCacheWild}, a
small replica reserve converts repeated remote callbacks into local accesses
on \tokenstack's stack-local path (896\,GB/s intra-stack vs.\ 512\,GB/s
inter-stack, at ${\sim}$40\% of the callback latency).

\noindent\textbf{Identifying replica-worthy blocks.}
The base-die MC applies a three-gate test using metadata already maintained
for demotion.  \emph{Gate~1 (position):} the block's token-position offset
must fall within the system-prompt region ($\mathit{offset} \leq
\tau_{\text{off}}$); blocks beyond this threshold are user-specific content
with negligible cross-request reuse.  \emph{Gate~2 (fan-out):} the
distinct-card count $n_{\text{cards}}$ must exceed $\tau_{\text{cards}}$,
ensuring that the callback cost is distributed across enough cards to justify
a replica.  \emph{Gate~3 (frequency):} the remote-hit count
$n_{\text{remote}}$ must exceed $\tau_{\text{hits}}$, ensuring that the
amortized DMA cost of creating the replica is below the cumulative callback
savings.  A block that passes all three gates is replicated into the
requesting card's compute-layer \emph{replica reserve} via a background DMA.

Replicas whose callback-elimination rate drops below a threshold are revoked,
returning capacity when access patterns shift.

\subsection{Other Runtime Optimizations}
\label{ssec:movement}

\tokenstack's base die owns DMA resources that operate independently of GPU and
PIM, so latency-critical promotions and callbacks run in the
foreground while demotions, replica fanout fill idle TSV slots.
Measured across all runs, contention between foreground runtime
actions and GPU-originated or PIM accesses is only 0.04\% at p50 and 0.50\%
at p95.

\label{sec:methodology}

\begin{figure*}[!t]
\centering
\includegraphics[width=\textwidth]{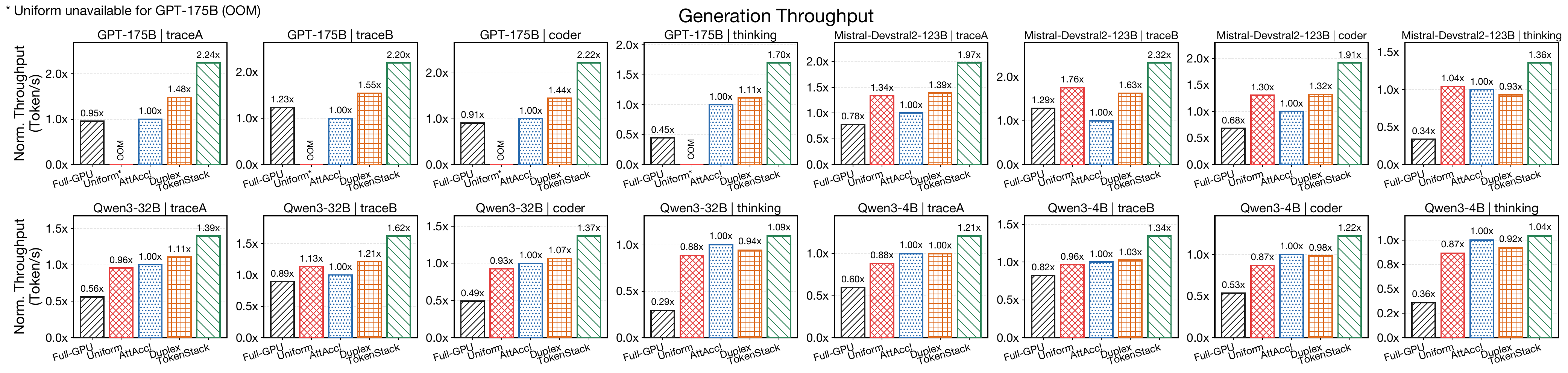}
\caption{Token throughput normalized to AttAcc. Uniform unavailable for GPT-175B (OOM).}
\Description{A 4-by-4 grid of bar charts. Each subplot shows four bars (Full-GPU, AttAcc, Uniform, and \tokenstack) with throughput normalized to AttAcc. \tokenstack bars are consistently the tallest, with speedup labels annotated above each bar.}
\label{fig:throughput}
\end{figure*}
\begin{table}[t]
\centering
\caption{Experimental setup. Comp./Cap.: per-card domain capacities before weight reservation.
$\bar{P}$/$\bar{G}$: mean prompt/generation length (tokens).}
\label{tab:setup}
\scriptsize
\renewcommand{\arraystretch}{0.95}
\setlength{\tabcolsep}{3pt}
\begin{tabular}{@{}llcrr@{}}
\toprule
\multicolumn{5}{@{}l}{\textbf{Platform:}
  DGX-A100, 8$\times$GPU, HBM3, bank-level PIM, W16A16} \\
\multicolumn{5}{@{}l}{\quad
  GPU: 312\,TF/s \quad
  UCIe-A BW: 512\,GB/s per stack $\times$ 5 stacks \quad
  } \\
\multicolumn{5}{@{}l}{\quad
  TSV DMA BW: 896\,GB/s per Stack \quad
  NVLink3: 600\,GB/s \quad
  Host: 512\,GB DDR5} \\
\midrule
\textbf{Mode}
  & \textbf{Die org.}
  & \textbf{Comp./Cap.}
  & \multicolumn{2}{l}{\textbf{KV policy}} \\
AttAcc
  & Ded.\ PIM+HBM
  & 32\,/\,16\,GB
  & \multicolumn{2}{l}{LRU, unique} \\
Full-GPU
  & No PIM
  & 80\,/\,0\,GB
  & \multicolumn{2}{l}{LRU, unique} \\
Uniform
  & All-PIM
  & 40\,/\,0\,GB
  & \multicolumn{2}{l}{LRU, unique} \\
  Duplex
    & All-PIM + GPU/PIM paths
    & 40\,/\,0\,GB
    & \multicolumn{2}{l}{LRU, unique} \\
\tokenstack
  & Hybrid stack
  & 20\,/\,40\,GB
  & \multicolumn{2}{l}{Aware, sel.\ rep., K8V4} \\
\midrule
\textbf{Trace}
  & \textbf{Type}
  & \textbf{Reqs}
  & {$\bar{P}$}
  & {$\bar{G}$} \\
\texttt{traceB}
  & API/text   & 172\,800 &     914 &     91 \\
\texttt{traceA}
  & Mixed      &  43\,058 &  2\,322 &    430 \\
\texttt{coder}
  & Code       &  27\,572 &  5\,832 &    802 \\
\texttt{thinking}
  & Reasoning  &  10\,812 &  4\,766 &  3\,644 \\
\midrule
\textbf{Model}
  & \textbf{Layers}
  & \textbf{Hidden}
  & \textbf{Heads}
  & \textbf{Params} \\
Qwen3-4B       & 36  & 2\,560  & 32  & 4\,B \\
Qwen3-32B      & 64  & 5\,120  & 64  & 32\,B \\
Mistral-\\Devstral2-123B  & 80  & 12\,288 & 96  & 123\,B \\
GPT-175B       & 96  & 12\,288 & 96  & 175\,B \\
\bottomrule
\end{tabular}
\end{table}

\section{Experimental Methodology}
\noindent\textbf{Simulation Framework.}
We evaluate \tokenstack with a cycle-accurate simulator built on
AttAcc's open-source frontend\footnote{https://github.com/scale-snu/attacc\_simulator} and Ramulator2\footnote{https://github.com/CMU-SAFARI/ramulator2}, both widely used in prior PIM-serving work~\cite{Park2024ASPLOSAttAcc,Luo2024Ramulator2}.
Ramulator2 provides cycle-level timing for bank-level PIM attention;
the Python frontend models tensor-parallel GPU execution, heterogeneous attention offload, request scheduling, KV placement, and end-to-end trace-driven serving on DGX-class nodes.
We validate the unmodified simulator by reproducing AttAcc's published throughput within 3\% across its reported configurations.

The key extension over prior PIM-serving simulators is support for
\emph{vertically heterogeneous} HBM stacks.
The extended model tracks stack-local DMA between capacity and compute layers, layered address mapping, workload-aware KV placement, bounded replication, and overlap-aware movement accounting across DMA, HBM, NVLink, and PCIe paths.
The base-die controller is evaluated through its system-level effects (latency, bandwidth utilization, and data-movement cost) rather than through a separate RTL model; this is appropriate because the controller's impact is fully captured by the resulting data-movement latencies and bandwidth sharing, which the simulator models cycle-accurately.
Inline K8V4 quantization is modeled as a pipelined transfer path with effective bandwidth and per-byte quantization/dequantization energy derived from published MANT unit characterizations~\cite{Hu2025MANT}.

\noindent\textbf{Configurations and Workloads.}
Table~\ref{tab:setup} lists the full setup on a single DGX-A100 node (8 GPUs).
To ensure a fair comparison, all configurations share the same continuous-batching scheduler (arrival admission, chunked prefill, immediate decode-slot backfill); only the stack organization and KV management policy differ.
The compared configurations bracket the HBM-PIM design space along two axes (capacity allocation and PIM granularity):
\textbf{Full-GPU} removes PIM entirely, serving as the GPU-only reference that isolates PIM's contribution;
\textbf{Uniform} makes every HBM layer PIM-enabled, representing the all-PIM extreme;
\textbf{Duplex}~\cite{Yun2024Duplex} extends Uniform with separate GPU/PIM execution paths and MoE/GQA-specific scheduling;
\textbf{AttAcc}\footnote{The A100-80GB package exposes 5 HBM stack positions (1 chip disabled), and AttAcc allocates each whole stack to either PIM or standard HBM. Among all legal whole-stack splits, 4 PIM stacks (8 GB each) plus 1 standard HBM stack (16 GB) yields the best geomean, giving the 32 GB compute / 16 GB capacity in Table~\ref{tab:setup}.}~\cite{Park2024ASPLOSAttAcc} dedicates a fixed subset of dies to PIM while retaining the rest as standard HBM, representing the dedicated-PIM split;
and \textbf{\tokenstack} adds the heterogeneous stack (\S\ref{sec:design}), KV-aware layout (\S\ref{sec:layout}), and the full runtime policy suite (\S\ref{sec:runtime}).
We configure each baseline according to its original publication and optimize AttAcc's stack split to its best-performing allocation.

We drive all experiments with four production traces from the Aliyun Qwen-Bailian deployment~\cite{Wang2025KVCacheWild} (Table~\ref{tab:setup}), selected to span the two dimensions that most affect PIM utilization: output length (91 to 3{,}644 mean tokens) and KV-reuse structure (zero to high multi-turn sharing).
Each trace is the full 2-hour productive window at native arrival rate, preserving request population, ordering, and KV-reuse structure; load intensity is varied separately by rescaling arrival timestamps.
Each request carries an arrival timestamp, prompt/generation lengths, request type, turn metadata, and hashed KV blocks at 16-token granularity, enabling faithful replay of real prefix-sharing patterns.
We pair these traces with four models from 4\,B to 175\,B parameters, yielding 16 model-trace combinations that cover the range where KV pressure grows from moderate (working set fits in a flat PIM tier) to dominant (KV state exceeds any single-tier allocation by $>$2$\times$).

\noindent\textbf{Load Sweep and Metrics.}
For each of the 16 model-trace pairs we sweep QPS from light load to saturation, reporting the standard metrics for latency-sensitive LLM serving~\cite{Park2024ASPLOSAttAcc, Yun2024Duplex}:
normalized p50 and p95 TTFT, TBT, and end-to-end request latency;
request and token throughput; and total energy per generated token.
We decompose energy into five components (off-chip DRAM, on-package cache, register file, ALU, inter-die communication) to isolate data-movement savings from compute savings.
We focus on p50 and p95 rather than p99 because the traces' native arrival variability already stresses the tail; p95 captures scheduling-induced latency inflation while remaining stable across runs.

\noindent\textbf{Hardware overhead.}
The base-die additions are area-negative overall: replacing the HBM PHY and memory controller with UCIe D2D frees ${\approx}$15\,mm$^2$ per stack on the GPU die~\cite{DasSharma2025UCIeMemory}, while the new logic occupies only 1.53\,mm$^2$ on the 121\,mm$^2$ HBM base die~\cite{Lee2022ISSCCHBM3}.
We estimate area with CACTI 7.0~\cite{HP2017CACTI} at 7\,nm.
The memory controller is relocated from the GPU die to the base die (as in HBM4's UCIe-attached design), not added as new silicon.
Of the three new blocks, the attention coordinator and shared SRAM buffer are the largest at ${\approx}$1.44\,mm$^2$, dominated by SRAM sized consistently with AttAcc's bank-group buffers~\cite{Park2024ASPLOSAttAcc}.
The K8V4 quantization/dequantization engine uses parallel group-wise units~\cite{Hu2025MANT} provisioned to saturate the 896\,GB/s per-stack TSV bandwidth~\cite{Lee2022ISSCCHBM3}, occupying ${\approx}$0.09\,mm$^2$ with latency fully pipelined behind the TSV transfer.
In total, the new blocks consume 1.3\% of the base-die area, while the freed GPU-die area returns net silicon to compute.

\section{Evaluation}
\label{sec:evaluation}
\begin{figure*}[t]
\centering
\includegraphics[width=\textwidth]{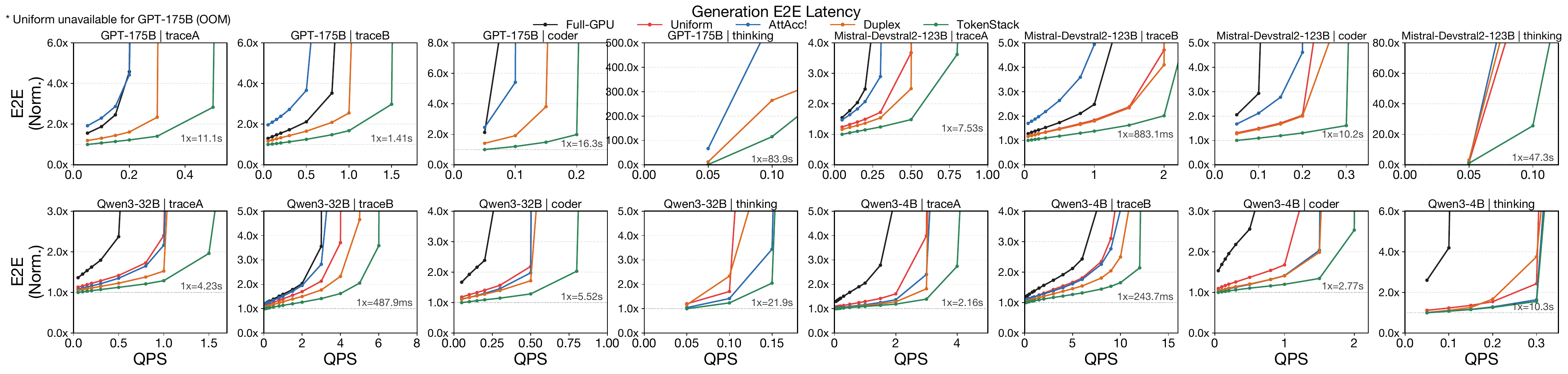}
\caption{Normalized p50 end-to-end latency vs.\ QPS (normalized to \tokenstack's minimum). Uniform unavailable for GPT-175B (OOM).}
\Description{A 4-by-4 grid of line charts showing normalized end-to-end latency versus QPS. Each subplot corresponds to one model--trace pair. The \tokenstack curve is consistently the lowest and flattest, while AttAcc rises steeply as QPS increases.}
\label{fig:e2e_latency}
\end{figure*}

We organize the evaluation around four questions.
First, does \tokenstack improve delivered throughput without sacrificing large-model support?
Second, does it reduce end-to-end request latency under realistic, trace-driven arrivals?
Third, does it reduce per-token serving energy?
Fourth, which components of the design contribute most to the observed gains?

\subsection{Throughput}

Fig.~\ref{fig:throughput} reports token throughput normalized to AttAcc.
Across all 16 model--trace pairs, \tokenstack delivers a geometric-mean improvement of $1.62\times$, ranging from $1.03\times$ (Qwen3-4B on \texttt{thinking}) to $2.32\times$ (Mistral-Devstral2-123B on \texttt{traceB}).
The gain scales monotonically with model size: per-model geometric means are $1.20\times$ (Qwen3-4B), $1.38\times$ (Qwen3-32B), $1.94\times$ (Devstral2-123B), and $2.15\times$ (GPT-175B).
This scaling reflects KV capacity overflow.
At 4\,B parameters, AttAcc's flat PIM tier holds the full KV working set even under load (L1~hit~rate~$\geq\,0.96$ at peak throughput), so \tokenstack's capacity layers absorb little additional traffic.
At 175\,B, the KV state far exceeds any single-tier allocation; \tokenstack's capacity layers absorb the overflow via stack-internal DMA at 896\,GB/s, keeping compute layers exclusively for the hottest blocks.

Per-trace geometric means vary from $1.36\times$ (\texttt{thinking}) to $1.83\times$ (\texttt{traceB}), revealing a second sensitivity axis: reuse structure.
\texttt{traceB} (short-form API traffic, 172\,K~requests, mean output 91~tokens) exhibits high temporal locality in system-prompt prefixes, producing a compact hot set that fits entirely in compute layers.
The \texttt{thinking} trace (mean output 3\,644~tokens) generates fresh KV per request with minimal cross-request sharing, weakening the hot--cold distinction that drives \tokenstack's advantage.
The $1.03\times$ floor case (Qwen3-4B/\texttt{thinking}) sits at the intersection of both minima: the smallest model and the least structured reuse.
Neither AttAcc nor \tokenstack overflows its compute tier at any load point, so vertical heterogeneity adds no value---this delineates the design's scope rather than indicating a failure.

Beyond peak throughput, \tokenstack sustains effective serving over a wider load range.
On GPT-175B and Devstral2-123B, the QPS at which throughput reaches 90\% of its maximum is $2.0$--$3.0\times$ higher for \tokenstack than for AttAcc (e.g., GPT-175B/\texttt{traceB}: QPS\,=\,3.0 vs.\ 1.0).
The wider range follows from reduced per-request service time: because \tokenstack resolves each attention step locally, decode slots free earlier, allowing the scheduler to admit the next request sooner and deferring the saturation knee.
Among baselines, Uniform achieves a modest $1.07\times$ geometric-mean speedup over AttAcc on the 12~pairs where it can run but falls below AttAcc on most small-model configurations where the halved capacity outweighs its bandwidth gain; it cannot serve GPT-175B (OOM).
Full-GPU retains the full 80\,GB GPU-visible pool and occasionally outperforms AttAcc on \texttt{traceB} for the two largest models, but without PIM it cannot approach \tokenstack at any load.

Improvement is smallest when the working set lacks clear hot--cold structure or when the model is small enough that a flat PIM organization holds the full set (Qwen3-4B on \texttt{thinking}, $1.03\times$).

\subsection{Latency}

The end-to-end latency divergence visible in Fig.~\ref{fig:e2e_latency} is rooted in queue delay.
For Mistral-Devstral2-123B on \texttt{traceA}, \tokenstack reduces average queue delay by 86\% at QPS\,=\,0.2, by 95\% at QPS\,=\,1.0 (126\,s vs.\ 2{,}572\,s), and still by 55\% at QPS\,=\,8.0.
As in the throughput analysis, shorter per-step attention latency frees decode slots earlier, cutting queue residence.
The advantage compounds under load: queueing delay grows superlinearly with utilization, so each service-time reduction yields a progressively larger absolute saving.

We define the SLO-compliant arrival rate as the maximum QPS at which p50
end-to-end latency remains within $2\times$ of its minimum.  Under this SLO,
\tokenstack serves $1.70\times$ more requests than AttAcc (geomean across all
16~pairs), with individual ratios from $1.0\times$ (Qwen3-4B/\texttt{thinking}) to
$3.8\times$ (Mistral-Devstral2-123B/\texttt{traceB}).  The capacity advantage scales
with model size---per-model geometric means are $1.35\times$ (4\,B),
$1.54\times$ (32\,B), $2.13\times$ (123\,B), and $1.95\times$ (175\,B). Beyond the SLO boundary the gap escalates
rapidly: at each subplot's cutoff QPS, AttAcc's p50 latency exceeds
\tokenstack's by a geomean of $25.2\times$, up to $939\times$
(GPT-175B/\texttt{traceB}).  This extreme is mechanistic. At the cutoff
QPS of 1.4, AttAcc has already plateaued (80.7\,tok/s, reached near
QPS\,$\approx$\,1.0) and sits in exponential queueing blowup
(p50$ = 2{,}029$\,s), while \tokenstack stays in its linear regime
(108.5\,tok/s, plateau of 178.8\,tok/s not reached until QPS\,$\approx$\,3.0)
below its knee (p50$ = 2.2$\,s).

\begin{figure}[bt]
    \centering
    \includegraphics[width=\linewidth]{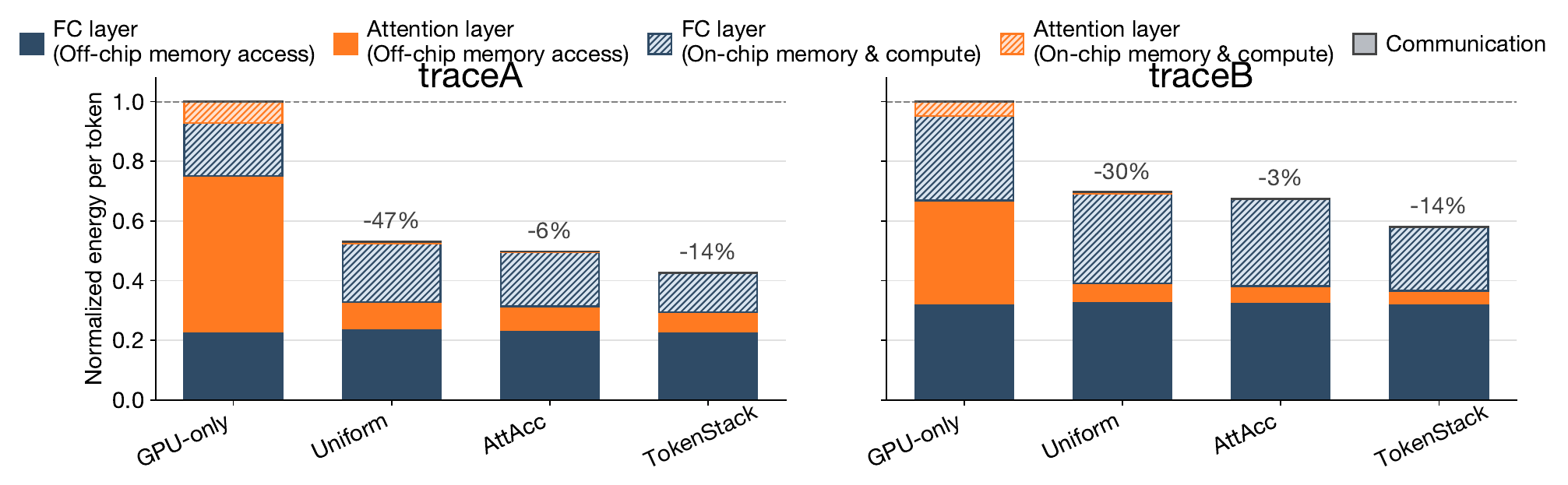}
    \caption{Normalized energy breakdown per token for Mistral-Devstral2-123B at QPS\,=\,32 on \texttt{traceA} and \texttt{traceB}.}
    \Description{Two groups of four stacked bars (one group per trace). Each bar decomposes energy per token into five components: FC off-chip, attention off-chip, FC on-chip, attention on-chip, and communication. \tokenstack's bar is visibly shorter than AttAcc's, with the attention off-chip component most reduced.}
    \label{fig:energy}
\end{figure}

\begin{figure*}[t]
\centering
\includegraphics[width=\textwidth]{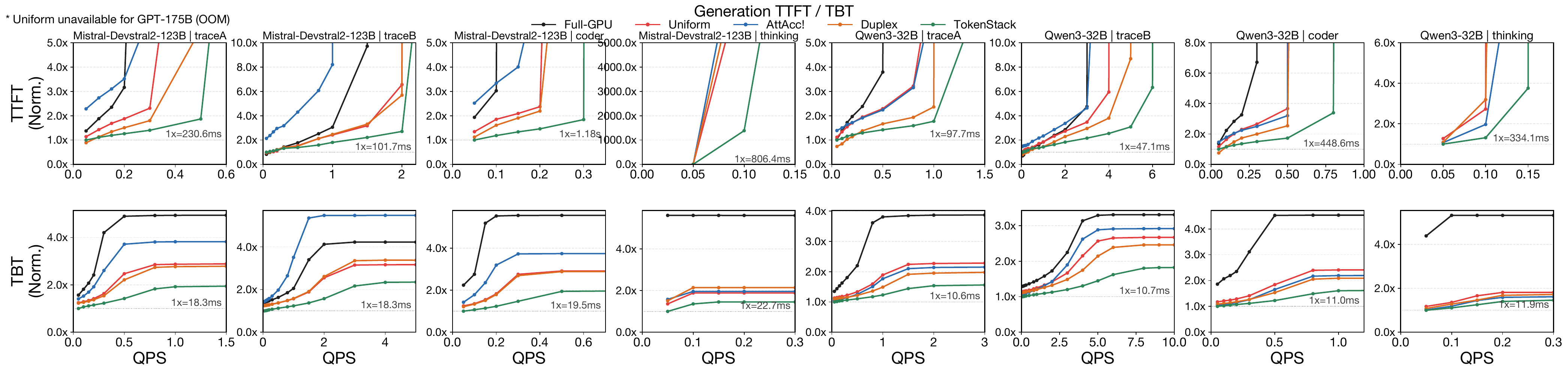}
\caption{Normalized p50 TTFT (top) and TBT (bottom) vs.\ QPS for Devstral2-123B and Qwen3-32B. TTFT dominates the gap.}
\Description{Line charts compare normalized median time to first token and time between tokens across QPS values for Devstral2-123B and Qwen3-32B. TokenStack shows substantially lower TTFT and moderately lower TBT than the baselines.}
\label{fig:ttft_tbt}
\end{figure*}

\noindent\textbf{TTFT and TBT.}
Fig.~\ref{fig:ttft_tbt} decomposes latency for Devstral2-123B and Qwen3-32B.
TTFT dominates the gap: AttAcc's p50~TTFT exceeds \tokenstack's by a geomean of $127\times$, peaking above $4{,}500\times$ on Devstral2-123B/\texttt{traceB}.
The TTFT improvement and the $1.62\times$ throughput improvement are manifestations of the same mechanism (capacity--bandwidth relief) measured at different granularities: TTFT captures the per-request latency reduction at the point of maximum divergence, while throughput averages it across operating points.
TBT differences are smaller (geomean $1.8\times$) because decode is inherently sequential, touching only one new token's KV per step; the gain arises because shorter per-step attention latency lets the scheduler pack more decode slots per scheduling interval.
The TBT advantage is widest on short-output, high-QPS traces (e.g., \texttt{traceB}) where large decode batches amplify the per-step cost.

\subsection{Energy Efficiency}
Fig.~\ref{fig:energy} reports normalized energy per generated token for Devstral2-123B at QPS\,=\,32, decomposed into five components: off-chip and on-chip energy for FC~layers and attention~layers, plus inter-die communication.
\tokenstack reduces energy per token by 47\% on \texttt{traceA} and 30\% on \texttt{traceB} relative to AttAcc.
The savings concentrate in the attention layer's off-chip memory-access component: by keeping hot KV adjacent to PIM engines in compute layers, \tokenstack replaces costly off-chip DRAM transfers with stack-internal reads, directly targeting the dominant energy term in bandwidth-bound attention.
FC-layer energy is comparable across all configurations because weights and activations follow the same GPU-visible bandwidth path regardless of PIM organization.
Within each model--trace pair, per-token energy tracks inversely with throughput ($r > 0.98$ across QPS points), confirming that both improvements stem from the same root cause---reduced off-chip data movement during attention.
GPU-only saves 14\% by avoiding AttAcc's KV-forwarding overhead, but its attention on-chip energy is higher because all attention arithmetic remains on the GPU without near-memory execution.

\subsection{Ablation and Discussion}

\noindent\textbf{Duplex Baseline.} Duplex improves over AttAcc and
  Uniform by optimizing GQA and continuous batching through xPU and
  Logic-PIM execution paths. However, its memory organization remains a variant
  of Uniform, every HBM bank is PIM-enabled within the same reduced-capacity
  organization. Consequently, weights, activations, metadata, cold KV, and other
  PIM-unused data still occupy expensive PIM banks, while PIM area overhead
  reduces the total capacity available for the KV cache. Duplex also relies on
  static bank-bundle partitioning, introducing additional cross-bank reduction and data movement. Finally, its flat PIM pool provides no dense capacity tier for cold KV, forcing hot and cold blocks to compete for the same limited PIM-enabled capacity.

\begin{figure}[t]
    \centering
    \includegraphics[width=\linewidth]{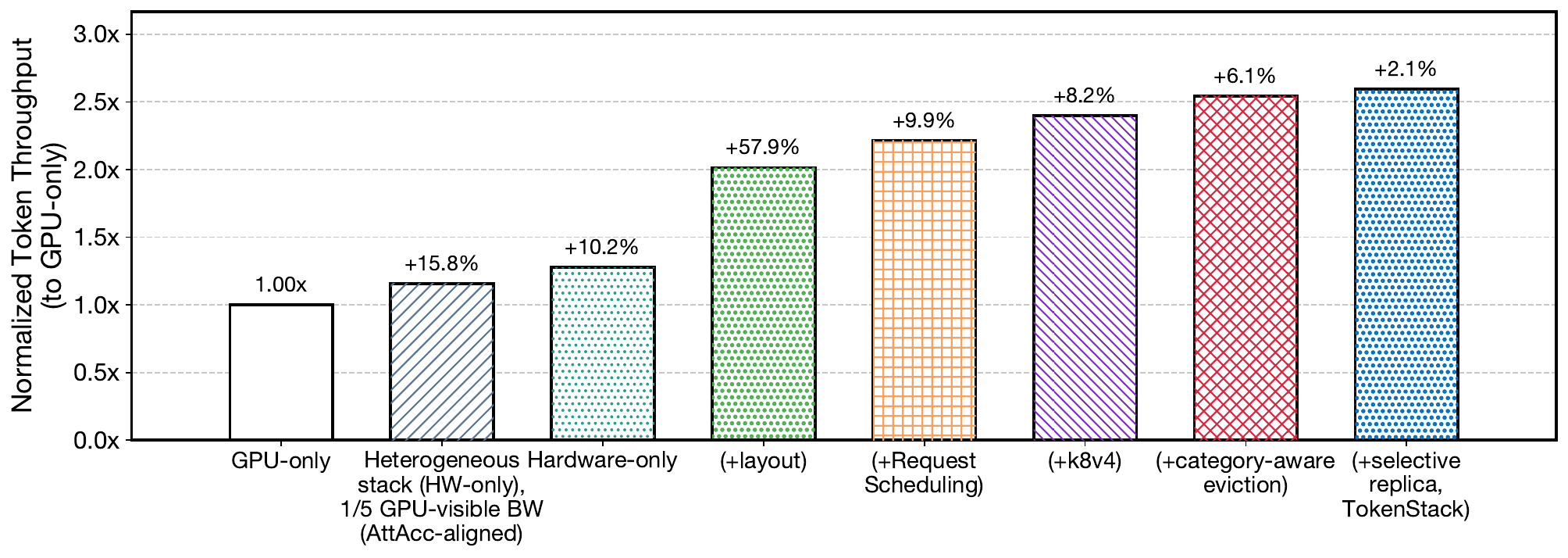}
    \caption{Cumulative throughput contribution of each \tokenstack component (Devstral2-123B, QPS\,=\,32, \texttt{traceA}), normalized to GPU-only.}
    \Description{Eight bars increase cumulatively from GPU-only at 1.0 times through heterogeneous hardware, KV-aware layout, request scheduling, K8V4 quantization, category-aware eviction, and selective replication. Each added component is labeled with its relative percentage gain.}
    \label{fig:ablation}
\end{figure}

\noindent\textbf{Component contributions.}
Fig.~\ref{fig:ablation} isolates each design component's throughput contribution for Devstral2-123B at QPS\,=\,32 on \texttt{traceA}, adding features cumulatively atop the bare heterogeneous hardware.
The KV-aware data layout~(\S\ref{sec:layout}) delivers the single largest gain.
Without asymmetric Key/Value placement---token-major Keys for reduction-free score gather, dim-head Values for reduction-free output concatenation---PIM banks cannot fully exploit their local bandwidth because KV data is misaligned with the attention dataflow.
The remaining three components---inline K8V4 quantization~(\S\ref{ssec:base-die}), category-aware eviction~(\S\ref{ssec:eviction}), and selective replication~(\S\ref{ssec:selective-replica})---are refinements that squeeze additional throughput from the same physical resources.
Together, the five features raise throughput to $\sim2\times$ the hardware-only baseline; for reference, GPU-only achieves only $0.78\times$ of that.

\noindent\textbf{Sensitivity Study.}
\begin{figure}[t]
    \centering
    \includegraphics[width=\linewidth]{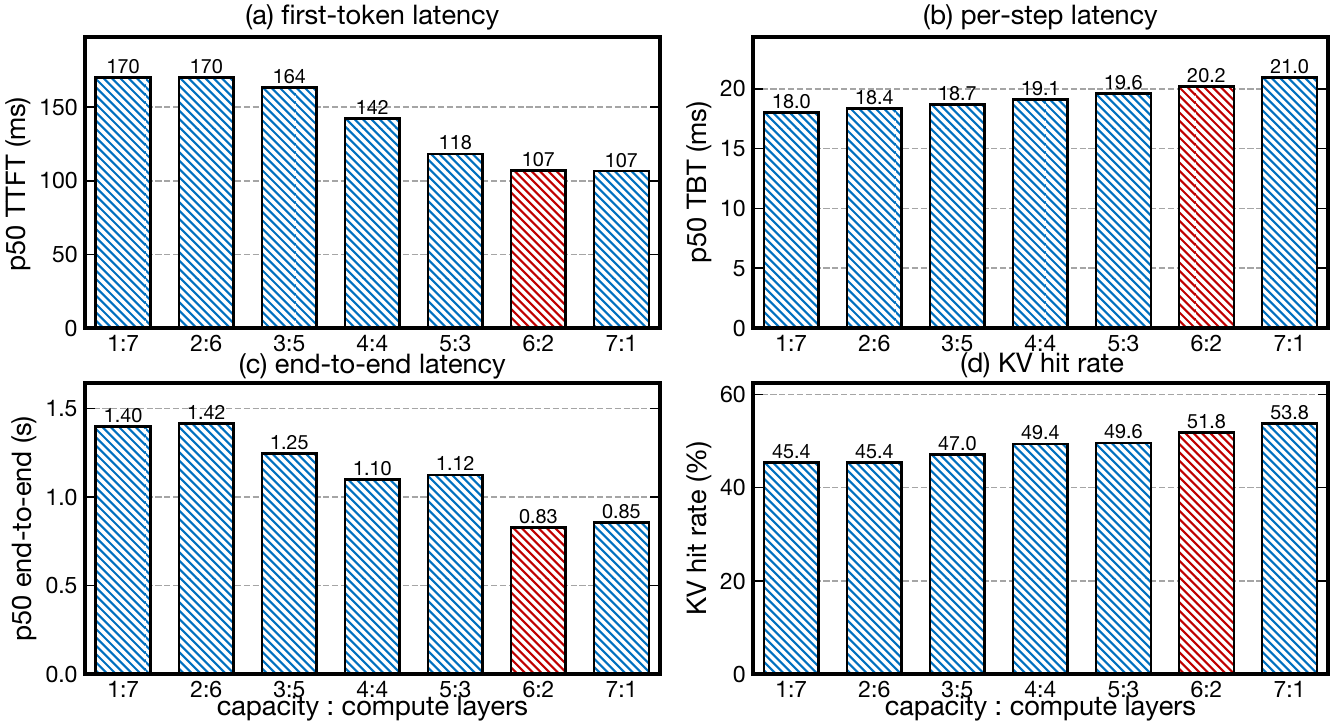}
    \caption{Sensitivity for Devstral2-123B on \texttt{traceB} at QPS\,=\,0.1.}
    \Description{Four bar charts compare TokenStack capacity-to-compute layer splits for Devstral2-123B on traceB at QPS 0.1, showing median first-token, per-step, and end-to-end latency plus KV hit rate.}
    \label{fig:sensitivity}
\end{figure}
We sweep seven per-stack topologies from \texttt{cap1-comp7} to \texttt{cap7-comp1} for \tokenstack{} at each trace's SLO-constrained goodput. Since each compute die spends ${\sim}$50\% of its area on PIM logic (\S\ref{ssec:stack-org}), \texttt{4:4} is the 20/40\,GB point of Table~\ref{tab:setup}. The split is governed by \emph{decode recomputation}: when capacity is under-provisioned, evicted KV blocks must be recomputed during decode, and the resulting stalls dominate the TTFT and TBT \emph{tails}. Adding capacity layers first eliminates this recomputation---collapsing the tails and lifting sustainable throughput even as PIM compute shrinks---after which further capacity only strips decode-attention parallelism and slowly raises the median step. The optimum is therefore the \emph{knee}: the smallest capacity ratio that clears recomputation for the target workload. This knee tracks memory pressure. Qwen3-32B/\texttt{traceB}, whose compact hot set barely pressures the tier, plateaus by \texttt{2:6} (p95 TBT stays 11--13\,ms across the sweep); Qwen3-32B/\texttt{coder} and \texttt{traceA} reach it at \texttt{4:4}, where p95 TBT collapses from 559 to 16\,ms (\texttt{coder}) as recomputation falls to zero; and Devstral2-123B/\texttt{traceB}, the heaviest, does not clear recomputation until \texttt{6:2}, its \texttt{1:7} and \texttt{2:6} splits leave \emph{no} usable capacity tier and stall at ${\sim}$1\,s p95 TBT. Provisioning \emph{past} the knee is not free, p50 TBT rises toward \texttt{7:1} as decode parallelism shrinks (Qwen3-32B/\texttt{traceA}, 11.2\,ms at \texttt{4:4} to 12.2\,ms at \texttt{7:1}).

\noindent\textbf{Config selection.} The split trades first-token latency against per-step latency. Capacity layers absorb prompt-KV writes and eliminate decode recomputation (cutting prefill $2.6$--$8\times$ on the long-prompt \texttt{coder} and Mistral workloads), while compute layers lower TBT. The knee is therefore where added capacity stops helping prefill/recomputation; its cost past that point is $\Delta$TBT $\times$ decode length. This yields a two-line policy: (i) size capacity to the prompt-KV footprint---\texttt{4:4} suffices for 32B on all traces, the 123B model needs \texttt{5:3} (\texttt{traceA}) to \texttt{6:2} (\texttt{traceB}); (ii) never exceed the knee, and exceed it least for long-decode workloads (\texttt{coder}'s \texttt{7:1} costs $\sim$3.8\,s of extra decode) since \texttt{7:1} wins no median metric and can re-trigger recomputation.

\section{Related Work}
\label{sec:related}

\subsection{PIM for LLM Inference}

  Prior systems accelerate LLM inference by offloading memory-intensive
  operators to near-memory compute while retaining GPUs or xPUs for compute-heavy
  layers; they also explore
  compiler optimization, dynamic scheduling, augmented retrieval,
  and CXL-based or GPU-free execution paths~\cite{Zhou2022TransPIM,Zhou2021MAT,Choi2024CALAttAcc,
  Kim2023SamsungPIMPNMTransformerAI,
  liu2025l3dimmpim,kwon2026pimphony,
  Kim2025PIMQuant,chen2025p3llmintegratednpupimaccelerator,
  malekar2025pimllmhighthroughputhybridpim,
  Han2025NearMemoryLLM,Liu2025McPAL,Hu2025PLAIN,
  Kim2025Pimba,Quinn2025LongSight,He2025PAPI,
  Hong2026REPA,Seo2025FACIL,Liu2025NDPDIMM,
  Wang2026AdaptiveDraft,Matsushima2026AQPIM,
  Jang2025PIMPAL,Hong2026LoCaLUT,
  Huang2025HDMoE,Pan2025Stratum,Bai2025AccelStack,
  Li2025H2LLM,Sun2025LincolnR5,
  jeun2025lpddr5foldpim,park2024lpddrcxl,
  Liu2025HeterRAG,Liu2026MERIDIAN,
  Liu2026CHIME,Huang2026HybridSpec,Chen2026P3LLM,
  Li2026HybridPIM,Nair2026Raptor,he2025lpspecleveraginglpddrpim,
cong2024specpim,heo2024neupim,IanusNPUPIM2024,Jang2025PIMRAG,
kim2024mondemixtureneardataexperts,Gu2025ASPLOSCENT}. AttAcc~\cite{Park2024ASPLOSAttAcc} applies bank-level HBM-PIM to
  attention, while Duplex~\cite{Yun2024Duplex} extends this execution model to
  MoE, GQA, and continuous batching.
  PAISE~\cite{Lee2025PAISE} improves GPU--PIM kernel scheduling on commercial
  HBM-PIM, while STARC~\cite{Fan2026STARC} clusters and remaps KV pairs to reduce
  the number of banks accessed by sparse attention. These systems optimize execution or data placement on a fixed, homogeneous PIM substrate; \tokenstack instead changes how capacity and compute are distributed within each stack and manages KV movement between layer types. PAISE-style scheduling and STARC-style clustering are thus complementary, while AttAcc and Duplex remain the closest same-substrate baselines.

\subsection{KV Management and Serving}

Recent work has shown that KV-cache behavior depends strongly on real workloads~\cite{Kimi2025MoonCake,Bin2024CachedAttention,Gim2024promptcache}. \emph{KVCache Cache in the Wild} reports that reuse is skewed, category-dependent, and often predictable within a request class~\cite{Wang2025KVCacheWild}. PAM argues that LLM serving must manage bandwidth and capacity together across a memory hierarchy~\cite{Liu2026PAM}. \tokenstack builds on these observations, but it does not stop at software policy. It uses the trace-driven insights to decide which blocks deserve compute-layer space inside a physically heterogeneous HBM stack.

\subsection{HBM Evolution and Near-Data Control}

Commercial HBM-PIM work has already shown that bank-level processing
is practical~\cite{Samsung2021ISSCCHBM2PIM,Kwon2022GDDR6AiM}.  What
changes with HBM4 is the replacement of the traditionally passive
base die with a CMOS logic die capable of hosting active control
circuitry~\cite{JEDEC2025HBM4,samsung2026hbm,rambus2026hbm4e,marvell2024customhbm}.  Together with die-to-die links such
as UCIe~\cite{DasSharma2025UCIeMemory,ucie2026specifications}, this logic base die creates a
realistic control point for stack-local DMA, layered address
translation, attention-side coordination, and inline data
transformation---functions that would otherwise require host-side
round-trips on every KV migration.  \tokenstack{} adds the next
step for KV-centric serving: it gives different parts of the same
stack different roles and uses the base die to coordinate them.

\section{Conclusion}
\label{sec:conclusion}

KV-centric LLM serving needs more than a fast attention engine: it needs
a memory stack that gives hot KV, cold KV, weights, and activations different
physical homes. Uniform HBM-PIM wastes compute-enabled space on data that
does not benefit from near-memory execution, while dedicated-PIM splits
reduce the HBM resources left for GPU-visible work. \tokenstack addresses this
mismatch with vertically heterogeneous stacks, a logic base die that manages
stack-local movement, and runtime policies that keep only the highest-value
KV in the compute-visible domain.

\bibliographystyle{ACM-Reference-Format}
\bibliography{reference}


\begin{thebibliography}{110}


\ifx \showCODEN    \undefined \def \showCODEN     #1{\unskip}     \fi
\ifx \showISBNx    \undefined \def \showISBNx     #1{\unskip}     \fi
\ifx \showISBNxiii \undefined \def \showISBNxiii  #1{\unskip}     \fi
\ifx \showISSN     \undefined \def \showISSN      #1{\unskip}     \fi
\ifx \showLCCN     \undefined \def \showLCCN      #1{\unskip}     \fi
\ifx \shownote     \undefined \def \shownote      #1{#1}          \fi
\ifx \showarticletitle \undefined \def \showarticletitle #1{#1}   \fi
\ifx \showURL      \undefined \def \showURL       {\relax}        \fi
\providecommand\bibfield[2]{#2}
\providecommand\bibinfo[2]{#2}
\providecommand\natexlab[1]{#1}
\providecommand\showeprint[2][]{arXiv:#2}

\bibitem[Ahn et~al\mbox{.}(2015)]%
        {AhnScalable}
\bibfield{author}{\bibinfo{person}{Junwhan Ahn}, \bibinfo{person}{Sungpack
  Hong}, \bibinfo{person}{Sungjoo Yoo}, \bibinfo{person}{Onur Mutlu}, {and}
  \bibinfo{person}{Kiyoung Choi}.} \bibinfo{year}{2015}\natexlab{}.
\newblock \showarticletitle{A scalable processing-in-memory accelerator for
  parallel graph processing}. In \bibinfo{booktitle}{\emph{2015 ACM/IEEE 42nd
  Annual International Symposium on Computer Architecture (ISCA)}} (2015).
  \bibinfo{pages}{105--117}.
\newblock
\href{https://doi.org/10.1145/2749469.2750386}{doi:\nolinkurl{10.1145/2749469.2750386}}


\bibitem[Bai et~al\mbox{.}(2025)]%
        {Bai2025AccelStack}
\bibfield{author}{\bibinfo{person}{Chen Bai}, \bibinfo{person}{Xin Fan},
  \bibinfo{person}{Zhenhua Zhu}, \bibinfo{person}{Wei Zhang}, {and}
  \bibinfo{person}{Yuan Xie}.} \bibinfo{year}{2025}\natexlab{}.
\newblock \showarticletitle{AccelStack: A Cost-Driven Analysis of 3D-Stacked
  LLM Accelerators}. In \bibinfo{booktitle}{\emph{2025 IEEE/ACM International
  Conference On Computer Aided Design (ICCAD)}} (2025). \bibinfo{pages}{1--9}.
\newblock
\href{https://doi.org/10.1109/ICCAD66269.2025.11240867}{doi:\nolinkurl{10.1109/ICCAD66269.2025.11240867}}


\bibitem[Balasubramonian et~al\mbox{.}(2017)]%
        {HP2017CACTI}
\bibfield{author}{\bibinfo{person}{Rajeev Balasubramonian},
  \bibinfo{person}{Andrew~B. Kahng}, \bibinfo{person}{Naveen Muralimanohar},
  \bibinfo{person}{Ali Shafiee}, {and} \bibinfo{person}{Vaishnav Srinivas}.}
  \bibinfo{year}{2017}\natexlab{}.
\newblock \showarticletitle{CACTI 7: New Tools for Interconnect Exploration in
  Innovative Off-Chip Memories}.
\newblock  \bibinfo{volume}{14}, \bibinfo{number}{2} (\bibinfo{year}{2017}).
\newblock
\showISSN{1544-3566}
\href{https://doi.org/10.1145/3085572}{doi:\nolinkurl{10.1145/3085572}}


\bibitem[Chen et~al\mbox{.}(2025c)]%
        {Chen2025Anchor}
\bibfield{author}{\bibinfo{person}{Jiaxian Chen}, \bibinfo{person}{Yuxuan Qi},
  \bibinfo{person}{Yongbiao Zhu}, \bibinfo{person}{Jianan Yuan},
  \bibinfo{person}{Kaoyi Sun}, \bibinfo{person}{Tianyu Wang},
  \bibinfo{person}{Chenlin Ma}, {and} \bibinfo{person}{Yi Wang}.}
  \bibinfo{year}{2025}\natexlab{c}.
\newblock \showarticletitle{Anchor First, Accelerate Next: Revolutionizing GNNs
  with PIM by Harnessing Stationary Data}. In \bibinfo{booktitle}{\emph{2025
  62nd ACM/IEEE Design Automation Conference (DAC)}} (2025).
  \bibinfo{publisher}{IEEE}, \bibinfo{pages}{1--7}.
\newblock
\href{https://doi.org/10.1109/DAC63849.2025.11132411}{doi:\nolinkurl{10.1109/DAC63849.2025.11132411}}


\bibitem[Chen et~al\mbox{.}(2025d)]%
        {Chen2025HEAT}
\bibfield{author}{\bibinfo{person}{Ruiyang Chen}, \bibinfo{person}{Zhuoran
  Song}, \bibinfo{person}{Yicheng Zheng}, \bibinfo{person}{Zeyu Zhu},
  \bibinfo{person}{Gang Li}, \bibinfo{person}{Naifeng Jing},
  \bibinfo{person}{Xiaoyao Liang}, {and} \bibinfo{person}{Haibing Guan}.}
  \bibinfo{year}{2025}\natexlab{d}.
\newblock \showarticletitle{HEAT: NPU-NDP HEterogeneous Architecture for
  Transformer-Empowered Graph Neural Networks}. In
  \bibinfo{booktitle}{\emph{Proceedings of the 58th IEEE/ACM International
  Symposium on Microarchitecture}} (Seoul, Republic of Korea, 2025)
  \emph{(\bibinfo{series}{MICRO '25})}. \bibinfo{publisher}{Association for
  Computing Machinery}, \bibinfo{pages}{263--276}.
\newblock
\showISBNx{9798400715730}
\href{https://doi.org/10.1145/3725843.3756117}{doi:\nolinkurl{10.1145/3725843.3756117}}


\bibitem[Chen et~al\mbox{.}(2024a)]%
        {Chen2024UpDLRM}
\bibfield{author}{\bibinfo{person}{Sitian Chen}, \bibinfo{person}{Haobin Tan},
  \bibinfo{person}{Amelie~Chi Zhou}, \bibinfo{person}{Yusen Li}, {and}
  \bibinfo{person}{Pavan Balaji}.} \bibinfo{year}{2024}\natexlab{a}.
\newblock \showarticletitle{UpDLRM: Accelerating Personalized Recommendation
  using Real-World PIM Architecture}. In \bibinfo{booktitle}{\emph{Proceedings
  of the 61st ACM/IEEE Design Automation Conference}} (San Francisco, CA, USA,
  2024) \emph{(\bibinfo{series}{DAC '24})}. \bibinfo{publisher}{Association for
  Computing Machinery}.
\newblock
\href{https://doi.org/10.1145/3649329.3658266}{doi:\nolinkurl{10.1145/3649329.3658266}}


\bibitem[Chen et~al\mbox{.}(2026e)]%
        {Chen2026P3LLM}
\bibfield{author}{\bibinfo{person}{Yuzong Chen}, \bibinfo{person}{Chao Fang},
  \bibinfo{person}{Xilai Dai}, \bibinfo{person}{Yuheng Wu},
  \bibinfo{person}{Thierry Tambe}, \bibinfo{person}{Marian Verhelst}, {and}
  \bibinfo{person}{Mohamed~S. Abdelfattah}.} \bibinfo{year}{2026}\natexlab{e}.
\newblock \showarticletitle{P3-LLM: An Integrated NPU-PIM Accelerator for Edge
  LLM Inference Using Hybrid Numerical Formats}. In
  \bibinfo{booktitle}{\emph{53rd Annual International Symposium on Computer
  Architecture (ISCA '26)}} (2026).
\newblock
\urldef\tempurl%
\url{https://www.iscaconf.org/isca2026/program/}
\showURL{%
\tempurl}
\newblock
\shownote{ISCA 2026 technical program}.


\bibitem[Chen et~al\mbox{.}(2025b)]%
        {chen2025p3llmintegratednpupimaccelerator}
\bibfield{author}{\bibinfo{person}{Yuzong Chen}, \bibinfo{person}{Chao Fang},
  \bibinfo{person}{Xilai Dai}, \bibinfo{person}{Yuheng Wu},
  \bibinfo{person}{Thierry Tambe}, \bibinfo{person}{Marian Verhelst}, {and}
  \bibinfo{person}{Mohamed~S. Abdelfattah}.} \bibinfo{year}{2025}\natexlab{b}.
\newblock \bibinfo{title}{P3-LLM: An Integrated NPU-PIM Accelerator for LLM
  Inference Using Hybrid Numerical Formats}.
\newblock
\showeprint[arXiv]{2511.06838}~[cs.AR]
\urldef\tempurl%
\url{https://arxiv.org/abs/2511.06838}
\showURL{%
\tempurl}


\bibitem[Choi et~al\mbox{.}(2023)]%
        {Choi2024CALAttAcc}
\bibfield{author}{\bibinfo{person}{Jaewan Choi}, \bibinfo{person}{Jaehyun
  Park}, \bibinfo{person}{Kwanhee Kyung}, \bibinfo{person}{Nam~Sung Kim}, {and}
  \bibinfo{person}{Jung~Ho Ahn}.} \bibinfo{year}{2023}\natexlab{}.
\newblock \showarticletitle{Unleashing the Potential of PIM: Accelerating Large
  Batched Inference of Transformer-Based Generative Models}.
\newblock  \bibinfo{volume}{22}, \bibinfo{number}{2} (\bibinfo{year}{2023}),
  \bibinfo{pages}{113--116}.
\newblock
\href{https://doi.org/10.1109/LCA.2023.3305386}{doi:\nolinkurl{10.1109/LCA.2023.3305386}}


\bibitem[Dai et~al\mbox{.}(2019)]%
        {Dai2019GraphH}
\bibfield{author}{\bibinfo{person}{Guohao Dai}, \bibinfo{person}{Tianhao
  Huang}, \bibinfo{person}{Yuze Chi}, \bibinfo{person}{Jishen Zhao},
  \bibinfo{person}{Guangyu Sun}, \bibinfo{person}{Yongpan Liu},
  \bibinfo{person}{Yu Wang}, \bibinfo{person}{Yuan Xie}, {and}
  \bibinfo{person}{Huazhong Yang}.} \bibinfo{year}{2019}\natexlab{}.
\newblock \showarticletitle{GraphH: A Processing-in-Memory Architecture for
  Large-Scale Graph Processing}.
\newblock  \bibinfo{volume}{38}, \bibinfo{number}{4} (\bibinfo{year}{2019}),
  \bibinfo{pages}{640--653}.
\newblock
\href{https://doi.org/10.1109/TCAD.2018.2821565}{doi:\nolinkurl{10.1109/TCAD.2018.2821565}}


\bibitem[Das~Sharma et~al\mbox{.}(2025)]%
        {DasSharma2025UCIeMemory}
\bibfield{author}{\bibinfo{person}{Debendra Das~Sharma},
  \bibinfo{person}{Swadesh Choudhary}, \bibinfo{person}{Peter Onufryk}, {and}
  \bibinfo{person}{Rob Pelt}.} \bibinfo{year}{2025}\natexlab{}.
\newblock \showarticletitle{On-Package Memory with Universal Chiplet
  Interconnect Express ({UCIe}): A Low Power, High Bandwidth, Low Latency and
  Low Cost Approach}. In \bibinfo{booktitle}{\emph{2025 Hot Interconnects}}
  (2025).
\newblock
\newblock
\shownote{arXiv:2510.06513}.


\bibitem[Devic et~al\mbox{.}(2026b)]%
        {Devic2026AnalyticBank}
\bibfield{author}{\bibinfo{person}{Alexandar Devic}, \bibinfo{person}{Martin
  Prammer}, \bibinfo{person}{Kevin Gaffney},
  \bibinfo{person}{Siddhartha~Balakrishna Rai}, \bibinfo{person}{Anand
  Sivasubramaniam}, \bibinfo{person}{Jignesh Patel}, {and}
  \bibinfo{person}{Ameen Akel}.} \bibinfo{year}{2026}\natexlab{b}.
\newblock \showarticletitle{Taking Analytic Databases to the Bank}. In
  \bibinfo{booktitle}{\emph{53rd Annual International Symposium on Computer
  Architecture (ISCA '26)}} (2026).
\newblock
\urldef\tempurl%
\url{https://www.iscaconf.org/isca2026/program/}
\showURL{%
\tempurl}
\newblock
\shownote{ISCA 2026 technical program}.


\bibitem[Devic et~al\mbox{.}(2022a)]%
        {Devic2022PIMGeneralPurposeDDR}
\bibfield{author}{\bibinfo{person}{Alexandar Devic},
  \bibinfo{person}{Siddhartha~Balakrishna Rai}, \bibinfo{person}{Anand
  Sivasubramaniam}, \bibinfo{person}{Ameen Akel}, \bibinfo{person}{Sean
  Eilert}, {and} \bibinfo{person}{Justin Eno}.}
  \bibinfo{year}{2022}\natexlab{a}.
\newblock \showarticletitle{To PIM or not for emerging general purpose
  processing in DDR memory systems}. In \bibinfo{booktitle}{\emph{Proceedings
  of the 49th Annual International Symposium on Computer Architecture}} (New
  York, New York, 2022) \emph{(\bibinfo{series}{ISCA '22})}.
  \bibinfo{publisher}{Association for Computing Machinery},
  \bibinfo{pages}{231--244}.
\newblock
\showISBNx{9781450386104}
\href{https://doi.org/10.1145/3470496.3527431}{doi:\nolinkurl{10.1145/3470496.3527431}}


\bibitem[Fan et~al\mbox{.}(2026)]%
        {Fan2026STARC}
\bibfield{author}{\bibinfo{person}{Zehao Fan}, \bibinfo{person}{Yunzhen Liu},
  \bibinfo{person}{Garrett Gagnon}, \bibinfo{person}{Zhenyu Liu},
  \bibinfo{person}{Yayue Hou}, \bibinfo{person}{Hadjer Benmeziane},
  \bibinfo{person}{Kaoutar~El Maghraoui}, {and} \bibinfo{person}{Liu Liu}.}
  \bibinfo{year}{2026}\natexlab{}.
\newblock \showarticletitle{STARC: Selective Token Access with Remapping and
  Clustering for Efficient LLM Decoding on PIM Systems}. In
  \bibinfo{booktitle}{\emph{Proceedings of the 31st ACM International
  Conference on Architectural Support for Programming Languages and Operating
  Systems, Volume 2}} (USA, 2026) \emph{(\bibinfo{series}{ASPLOS '26})}.
  \bibinfo{publisher}{Association for Computing Machinery},
  \bibinfo{pages}{1863--1879}.
\newblock
\showISBNx{9798400723599}
\href{https://doi.org/10.1145/3779212.3790226}{doi:\nolinkurl{10.1145/3779212.3790226}}


\bibitem[Gao et~al\mbox{.}(2024)]%
        {Bin2024CachedAttention}
\bibfield{author}{\bibinfo{person}{Bin Gao}, \bibinfo{person}{Zhuomin He},
  \bibinfo{person}{Puru Sharma}, \bibinfo{person}{Qingxuan Kang},
  \bibinfo{person}{Djordje Jevdjic}, \bibinfo{person}{Junbo Deng},
  \bibinfo{person}{Xingkun Yang}, \bibinfo{person}{Zhou Yu}, {and}
  \bibinfo{person}{Pengfei Zuo}.} \bibinfo{year}{2024}\natexlab{}.
\newblock \showarticletitle{{Cost-Efficient} Large Language Model Serving for
  Multi-turn Conversations with {CachedAttention}}. In
  \bibinfo{booktitle}{\emph{2024 USENIX Annual Technical Conference (USENIX ATC
  24)}} (Santa Clara, CA, 2024-07). \bibinfo{publisher}{USENIX Association},
  \bibinfo{pages}{111--126}.
\newblock
\showISBNx{978-1-939133-41-0}
\urldef\tempurl%
\url{https://www.usenix.org/conference/atc24/presentation/gao-bin-cost}
\showURL{%
\tempurl}


\bibitem[Giannoula et~al\mbox{.}(2022)]%
        {Giannoula2022SparseP}
\bibfield{author}{\bibinfo{person}{Christina Giannoula}, \bibinfo{person}{Ivan
  Fernandez}, \bibinfo{person}{Juan~Gómez Luna}, \bibinfo{person}{Nectarios
  Koziris}, \bibinfo{person}{Georgios Goumas}, {and} \bibinfo{person}{Onur
  Mutlu}.} \bibinfo{year}{2022}\natexlab{}.
\newblock \showarticletitle{SparseP: Towards Efficient Sparse Matrix Vector
  Multiplication on Real Processing-In-Memory Architectures}.
\newblock  \bibinfo{volume}{6}, \bibinfo{number}{1} (\bibinfo{year}{2022}).
\newblock
\href{https://doi.org/10.1145/3508041}{doi:\nolinkurl{10.1145/3508041}}


\bibitem[Gim et~al\mbox{.}(2024)]%
        {Gim2024promptcache}
\bibfield{author}{\bibinfo{person}{In Gim}, \bibinfo{person}{Guojun Chen},
  \bibinfo{person}{Seung-seob Lee}, \bibinfo{person}{Nikhil Sarda},
  \bibinfo{person}{Anurag Khandelwal}, {and} \bibinfo{person}{Lin Zhong}.}
  \bibinfo{year}{2024}\natexlab{}.
\newblock \showarticletitle{Prompt Cache: Modular Attention Reuse for
  Low-Latency Inference}. In \bibinfo{booktitle}{\emph{Proceedings of Machine
  Learning and Systems}} (2024),
  \bibfield{editor}{\bibinfo{person}{P.~Gibbons},
  \bibinfo{person}{G.~Pekhimenko}, {and} \bibinfo{person}{C.~De Sa}} (Eds.),
  Vol.~\bibinfo{volume}{6}. \bibinfo{pages}{325--338}.
\newblock
\urldef\tempurl%
\url{https://proceedings.mlsys.org/paper_files/paper/2024/file/a66caa1703fe34705a4368c3014c1966-Paper-Conference.pdf}
\showURL{%
\tempurl}


\bibitem[Gu et~al\mbox{.}(2025)]%
        {Gu2025ASPLOSCENT}
\bibfield{author}{\bibinfo{person}{Yufeng Gu}, \bibinfo{person}{Alireza
  Khadem}, \bibinfo{person}{Sumanth Umesh}, \bibinfo{person}{Ning Liang},
  \bibinfo{person}{Xavier Servot}, \bibinfo{person}{Onur Mutlu},
  \bibinfo{person}{Ravi Iyer}, {and} \bibinfo{person}{Reetuparna Das}.}
  \bibinfo{year}{2025}\natexlab{}.
\newblock \bibinfo{booktitle}{\emph{PIM Is All You Need: A CXL-Enabled GPU-Free
  System for Large Language Model Inference}}.
\newblock \bibinfo{publisher}{Association for Computing Machinery},
  \bibinfo{pages}{862--881}.
\newblock
\showISBNx{9798400710797}
\urldef\tempurl%
\url{https://doi.org/10.1145/3676641.3716267}
\showURL{%
\tempurl}


\bibitem[Han et~al\mbox{.}(2025)]%
        {Han2025NearMemoryLLM}
\bibfield{author}{\bibinfo{person}{Sanghyeok Han}, \bibinfo{person}{Byungkuk
  Yoon}, \bibinfo{person}{Gyeonghwan Park}, \bibinfo{person}{Choungki Song},
  \bibinfo{person}{Dongkyun Kim}, {and} \bibinfo{person}{Jae-Joon Kim}.}
  \bibinfo{year}{2025}\natexlab{}.
\newblock \showarticletitle{Near-Memory LLM Inference Processor based on 3D
  DRAM-to-logic Hybrid Bonding}. In \bibinfo{booktitle}{\emph{2025 62nd
  ACM/IEEE Design Automation Conference (DAC)}} (2025).
  \bibinfo{publisher}{IEEE}, \bibinfo{pages}{1--7}.
\newblock
\href{https://doi.org/10.1109/DAC63849.2025.11132870}{doi:\nolinkurl{10.1109/DAC63849.2025.11132870}}


\bibitem[He et~al\mbox{.}(2025b)]%
        {he2025lpspecleveraginglpddrpim}
\bibfield{author}{\bibinfo{person}{Siyuan He}, \bibinfo{person}{Zhantong Zhu},
  \bibinfo{person}{Yandong He}, {and} \bibinfo{person}{Tianyu Jia}.}
  \bibinfo{year}{2025}\natexlab{b}.
\newblock \bibinfo{title}{LP-Spec: Leveraging LPDDR PIM for Efficient LLM
  Mobile Speculative Inference with Architecture-Dataflow Co-Optimization}.
\newblock
\showeprint[arXiv]{2508.07227}~[cs.AR]
\urldef\tempurl%
\url{https://arxiv.org/abs/2508.07227}
\showURL{%
\tempurl}


\bibitem[He et~al\mbox{.}(2025a)]%
        {He2025PAPI}
\bibfield{author}{\bibinfo{person}{Yintao He}, \bibinfo{person}{Haiyu Mao},
  \bibinfo{person}{Christina Giannoula}, \bibinfo{person}{Mohammad
  Sadrosadati}, \bibinfo{person}{Juan Gómez-Luna}, \bibinfo{person}{Huawei
  Li}, \bibinfo{person}{Xiaowei Li}, \bibinfo{person}{Ying Wang}, {and}
  \bibinfo{person}{Onur Mutlu}.} \bibinfo{year}{2025}\natexlab{a}.
\newblock \showarticletitle{PAPI: Exploiting Dynamic Parallelism in Large
  Language Model Decoding with a Processing-In-Memory-Enabled Computing
  System}. In \bibinfo{booktitle}{\emph{Proceedings of the 30th ACM
  International Conference on Architectural Support for Programming Languages
  and Operating Systems, Volume 2}} (Rotterdam, Netherlands, 2025)
  \emph{(\bibinfo{series}{ASPLOS '25})}. \bibinfo{publisher}{Association for
  Computing Machinery}, \bibinfo{pages}{766--782}.
\newblock
\href{https://doi.org/10.1145/3676641.3716009}{doi:\nolinkurl{10.1145/3676641.3716009}}


\bibitem[Heo et~al\mbox{.}(2024)]%
        {heo2024neupim}
\bibfield{author}{\bibinfo{person}{Guseul Heo}, \bibinfo{person}{Sangyeop Lee},
  \bibinfo{person}{Jaehong Cho}, \bibinfo{person}{Hyunmin Choi},
  \bibinfo{person}{Sanghyeon Lee}, \bibinfo{person}{Hyungkyu Ham},
  \bibinfo{person}{Gwangsun Kim}, \bibinfo{person}{Divya Mahajan}, {and}
  \bibinfo{person}{Jongse Park}.} \bibinfo{year}{2024}\natexlab{}.
\newblock \showarticletitle{NeuPIMs: NPU-PIM Heterogeneous Acceleration for
  Batched LLM Inferencing}. In \bibinfo{booktitle}{\emph{Proceedings of the
  29th ACM International Conference on Architectural Support for Programming
  Languages and Operating Systems, Volume 3}} (La Jolla, CA, USA, 2024)
  \emph{(\bibinfo{series}{ASPLOS '24})}. \bibinfo{publisher}{Association for
  Computing Machinery}, \bibinfo{pages}{722--737}.
\newblock
\showISBNx{9798400703867}
\href{https://doi.org/10.1145/3620666.3651380}{doi:\nolinkurl{10.1145/3620666.3651380}}


\bibitem[Hong et~al\mbox{.}(2026a)]%
        {Hong2026LoCaLUT}
\bibfield{author}{\bibinfo{person}{Junguk Hong}, \bibinfo{person}{Changmin
  Shin}, \bibinfo{person}{Sukjin Kim}, \bibinfo{person}{Si~Ung Noh},
  \bibinfo{person}{Taehee Kwon}, \bibinfo{person}{Seongyeon Park},
  \bibinfo{person}{Hanjun Kim}, \bibinfo{person}{Youngsok Kim}, {and}
  \bibinfo{person}{Jinho Lee}.} \bibinfo{year}{2026}\natexlab{a}.
\newblock \showarticletitle{LoCaLUT: Harnessing Capacity-Computation Tradeoffs
  for LUT-Based Inference in DRAM-PIM}. In \bibinfo{booktitle}{\emph{2026 IEEE
  International Symposium on High Performance Computer Architecture (HPCA)}}
  (2026). \bibinfo{publisher}{IEEE}, \bibinfo{pages}{1--16}.
\newblock
\href{https://doi.org/10.1109/HPCA68181.2026.11408523}{doi:\nolinkurl{10.1109/HPCA68181.2026.11408523}}


\bibitem[Hong et~al\mbox{.}(2026b)]%
        {Hong2026REPA}
\bibfield{author}{\bibinfo{person}{Yang Hong}, \bibinfo{person}{Junlong Yang},
  \bibinfo{person}{Bo Peng}, {and} \bibinfo{person}{Jianguo Yao}.}
  \bibinfo{year}{2026}\natexlab{b}.
\newblock \showarticletitle{REPA: Reconfigurable PIM for the Joint Acceleration
  of KV Cache Offloading and Processing}. In
  \bibinfo{booktitle}{\emph{Proceedings of the 31st ACM International
  Conference on Architectural Support for Programming Languages and Operating
  Systems, Volume 2}} (Pittsburgh, PA, USA, 2026)
  \emph{(\bibinfo{series}{ASPLOS '26})}. \bibinfo{publisher}{Association for
  Computing Machinery}, \bibinfo{pages}{1622--1639}.
\newblock
\href{https://doi.org/10.1145/3779212.3790212}{doi:\nolinkurl{10.1145/3779212.3790212}}


\bibitem[Hooper et~al\mbox{.}(2024)]%
        {Hooper2024NeurIPSKVQuant}
\bibfield{author}{\bibinfo{person}{Coleman Hooper}, \bibinfo{person}{Sehoon
  Kim}, \bibinfo{person}{Hiva Mohammadzadeh}, \bibinfo{person}{Michael~W.
  Mahoney}, \bibinfo{person}{Yakun~Sophia Shao}, \bibinfo{person}{Kurt
  Keutzer}, {and} \bibinfo{person}{Amir Gholami}.}
  \bibinfo{year}{2024}\natexlab{}.
\newblock \showarticletitle{{KVQuant}: Towards 10 Million Context Length {LLM}
  Inference with {KV} Cache Quantization}. In
  \bibinfo{booktitle}{\emph{Advances in Neural Information Processing Systems
  (NeurIPS)}} (2024).
\newblock


\bibitem[Hu et~al\mbox{.}(2025b)]%
        {Hu2025MANT}
\bibfield{author}{\bibinfo{person}{Weiming Hu}, \bibinfo{person}{Haoyan Zhang},
  \bibinfo{person}{Cong Guo}, \bibinfo{person}{Yu Feng},
  \bibinfo{person}{Renyang Guan}, \bibinfo{person}{Zhendong Hua},
  \bibinfo{person}{Zihan Liu}, \bibinfo{person}{Yue Guan},
  \bibinfo{person}{Minyi Guo}, {and} \bibinfo{person}{Jingwen Leng}.}
  \bibinfo{year}{2025}\natexlab{b}.
\newblock \showarticletitle{{M-ANT}: Efficient Low-bit Group Quantization for
  {LLMs} via Mathematically Adaptive Numerical Type}.
\newblock  (\bibinfo{year}{2025}).
\newblock


\bibitem[Hu et~al\mbox{.}(2025a)]%
        {Hu2025PLAIN}
\bibfield{author}{\bibinfo{person}{Yiwei Hu}, \bibinfo{person}{Fangxin Liu},
  \bibinfo{person}{Zongwu Wang}, \bibinfo{person}{Yilong Zhao},
  \bibinfo{person}{Tao Yang}, \bibinfo{person}{Li Jiang}, {and}
  \bibinfo{person}{Haibing Guan}.} \bibinfo{year}{2025}\natexlab{a}.
\newblock \showarticletitle{PLAIN: Leveraging High Internal Bandwidth in PIM
  for Accelerating Large Language Model Inference via Mixed-Precision
  Quantization}. In \bibinfo{booktitle}{\emph{2025 IEEE/ACM International
  Conference On Computer Aided Design (ICCAD)}} (2025). \bibinfo{pages}{1--9}.
\newblock
\href{https://doi.org/10.1109/ICCAD66269.2025.11240991}{doi:\nolinkurl{10.1109/ICCAD66269.2025.11240991}}


\bibitem[Huang et~al\mbox{.}(2025a)]%
        {Huang2025HDMoE}
\bibfield{author}{\bibinfo{person}{Haochen Huang}, \bibinfo{person}{Shuzhang
  Zhong}, \bibinfo{person}{Zhe Zhang}, \bibinfo{person}{Shuangchen Li},
  \bibinfo{person}{Dimin Niu}, \bibinfo{person}{Hongzhong Zheng},
  \bibinfo{person}{Runsheng Wang}, {and} \bibinfo{person}{Meng Li}.}
  \bibinfo{year}{2025}\natexlab{a}.
\newblock \bibinfo{title}{HD-MoE: Hybrid and Dynamic Parallelism for
  Mixture-of-Expert LLMs with 3D Near-Memory Processing}.
\newblock
\showeprint[arXiv]{2509.09420}~[cs.PF]
\urldef\tempurl%
\url{https://arxiv.org/abs/2509.09420}
\showURL{%
\tempurl}


\bibitem[Huang et~al\mbox{.}(2026b)]%
        {Huang2026HybridSpec}
\bibfield{author}{\bibinfo{person}{Zongle Huang}, \bibinfo{person}{Wenbin Jia},
  \bibinfo{person}{Yaolei Li}, \bibinfo{person}{Xinyuan Lin},
  \bibinfo{person}{Shupei Fan}, \bibinfo{person}{Chen Tang},
  \bibinfo{person}{Shuwen Deng}, {and} \bibinfo{person}{Yongpan Liu}.}
  \bibinfo{year}{2026}\natexlab{b}.
\newblock \showarticletitle{HybridSpec: Exploiting Hybrid-bonding Memory to
  Accelerate LLM Serving through Heterogeneous Architecture and Speculative
  Decoding}. In \bibinfo{booktitle}{\emph{53rd Annual International Symposium
  on Computer Architecture (ISCA '26)}} (2026).
\newblock
\urldef\tempurl%
\url{https://www.iscaconf.org/isca2026/program/}
\showURL{%
\tempurl}
\newblock
\shownote{ISCA 2026 technical program}.


\bibitem[Imani et~al\mbox{.}(2019)]%
        {Imani2019FloatPIM}
\bibfield{author}{\bibinfo{person}{Mohsen Imani}, \bibinfo{person}{Saransh
  Gupta}, \bibinfo{person}{Yeseong Kim}, {and} \bibinfo{person}{Tajana
  Rosing}.} \bibinfo{year}{2019}\natexlab{}.
\newblock \showarticletitle{FloatPIM: In-Memory Acceleration of Deep Neural
  Network Training with High Precision}. In \bibinfo{booktitle}{\emph{2019
  ACM/IEEE 46th Annual International Symposium on Computer Architecture
  (ISCA)}} (2019). \bibinfo{pages}{802--815}.
\newblock


\bibitem[Jang et~al\mbox{.}(2025b)]%
        {Jang2025PIMRAG}
\bibfield{author}{\bibinfo{person}{Je-Woo Jang}, \bibinfo{person}{Junyong Oh},
  \bibinfo{person}{Youngbae Kong}, \bibinfo{person}{Jae-Youn Hong},
  \bibinfo{person}{Sung-Hyuk Cho}, \bibinfo{person}{Jeongyeol Lee},
  \bibinfo{person}{Hoeseok Yang}, {and} \bibinfo{person}{Joon-Sung Yang}.}
  \bibinfo{year}{2025}\natexlab{b}.
\newblock \showarticletitle{Accelerating Retrieval Augmented Language Model via
  PIM and PNM Integration}. In \bibinfo{booktitle}{\emph{Proceedings of the
  58th IEEE/ACM International Symposium on Microarchitecture}} (2025)
  \emph{(\bibinfo{series}{MICRO '25})}. \bibinfo{publisher}{Association for
  Computing Machinery}, \bibinfo{pages}{246--262}.
\newblock
\showISBNx{9798400715730}
\href{https://doi.org/10.1145/3725843.3756020}{doi:\nolinkurl{10.1145/3725843.3756020}}


\bibitem[Jang et~al\mbox{.}(2025a)]%
        {Jang2025PIMPAL}
\bibfield{author}{\bibinfo{person}{Yoonho Jang}, \bibinfo{person}{Hyeongjun
  Cho}, \bibinfo{person}{Yesin Ryu}, \bibinfo{person}{Jungrae Kim}, {and}
  \bibinfo{person}{Seokin Hong}.} \bibinfo{year}{2025}\natexlab{a}.
\newblock \showarticletitle{PIMPAL: Accelerating LLM Inference on Edge Devices
  via In-DRAM Arithmetic Lookup}. In \bibinfo{booktitle}{\emph{2025 62nd
  ACM/IEEE Design Automation Conference (DAC)}} (2025).
  \bibinfo{publisher}{IEEE}, \bibinfo{pages}{1--7}.
\newblock
\href{https://doi.org/10.1109/DAC63849.2025.11133391}{doi:\nolinkurl{10.1109/DAC63849.2025.11133391}}


\bibitem[JEDEC(2025)]%
        {JEDEC2025HBM4}
\bibfield{author}{\bibinfo{person}{JEDEC}.} \bibinfo{year}{2025}\natexlab{}.
\newblock \bibinfo{title}{High Bandwidth Memory ({HBM4}) {DRAM}}.
\newblock
\urldef\tempurl%
\url{https://www.jedec.org/standards-documents/docs/jesd270-4a}
\showURL{%
\tempurl}


\bibitem[Jeun et~al\mbox{.}(2025)]%
        {jeun2025lpddr5foldpim}
\bibfield{author}{\bibinfo{person}{Kyoungho Jeun}, \bibinfo{person}{Hyeonu
  Kim}, {and} \bibinfo{person}{Eojin Lee}.} \bibinfo{year}{2025}\natexlab{}.
\newblock \showarticletitle{Fold-PIM: A Cost-Efficient LPDDR5-Based PIM for
  On-Device SLMs}.
\newblock  \bibinfo{volume}{24}, \bibinfo{number}{1} (\bibinfo{year}{2025}),
  \bibinfo{pages}{185--188}.
\newblock
\href{https://doi.org/10.1109/LCA.2025.3566692}{doi:\nolinkurl{10.1109/LCA.2025.3566692}}


\bibitem[Ke et~al\mbox{.}(2020a)]%
        {Ke2020RecNMPNearMemoryProcessing}
\bibfield{author}{\bibinfo{person}{Liu Ke}, \bibinfo{person}{Udit Gupta},
  \bibinfo{person}{Benjamin~Youngjae Cho}, \bibinfo{person}{David Brooks},
  \bibinfo{person}{Vikas Chandra}, \bibinfo{person}{Utku Diril},
  \bibinfo{person}{Amin Firoozshahian}, \bibinfo{person}{Kim Hazelwood},
  \bibinfo{person}{Bill Jia}, \bibinfo{person}{Hsien-Hsin~S. Lee},
  \bibinfo{person}{Meng Li}, \bibinfo{person}{Bert Maher},
  \bibinfo{person}{Dheevatsa Mudigere}, \bibinfo{person}{Maxim Naumov},
  \bibinfo{person}{Martin Schatz}, \bibinfo{person}{Mikhail Smelyanskiy},
  \bibinfo{person}{Xiaodong Wang}, \bibinfo{person}{Brandon Reagen},
  \bibinfo{person}{Carole-Jean Wu}, \bibinfo{person}{Mark Hempstead}, {and}
  \bibinfo{person}{Xuan Zhang}.} \bibinfo{year}{2020}\natexlab{a}.
\newblock \showarticletitle{RecNMP: Accelerating Personalized Recommendation
  with Near-Memory Processing}. In \bibinfo{booktitle}{\emph{2020 ACM/IEEE 47th
  Annual International Symposium on Computer Architecture (ISCA)}} (2020).
  \bibinfo{pages}{790--803}.
\newblock
\href{https://doi.org/10.1109/ISCA45697.2020.00070}{doi:\nolinkurl{10.1109/ISCA45697.2020.00070}}


\bibitem[Ke et~al\mbox{.}(2022b)]%
        {Ke2022AxDIMM}
\bibfield{author}{\bibinfo{person}{Liu Ke}, \bibinfo{person}{Xuan Zhang},
  \bibinfo{person}{Jinin So}, \bibinfo{person}{Jong-Geon Lee},
  \bibinfo{person}{Shin-Haeng Kang}, \bibinfo{person}{Sukhan Lee},
  \bibinfo{person}{Songyi Han}, \bibinfo{person}{YeonGon Cho},
  \bibinfo{person}{Jin~Hyun Kim}, \bibinfo{person}{Yongsuk Kwon},
  \bibinfo{person}{KyungSoo Kim}, \bibinfo{person}{Jin Jung},
  \bibinfo{person}{Ilkwon Yun}, \bibinfo{person}{Sung~Joo Park},
  \bibinfo{person}{Hyunsun Park}, \bibinfo{person}{Joonho Song},
  \bibinfo{person}{Jeonghyeon Cho}, \bibinfo{person}{Kyomin Sohn},
  \bibinfo{person}{Nam~Sung Kim}, {and} \bibinfo{person}{Hsien-Hsin~S. Lee}.}
  \bibinfo{year}{2022}\natexlab{b}.
\newblock \showarticletitle{Near-Memory Processing in Action: Accelerating
  Personalized Recommendation With AxDIMM}.
\newblock  \bibinfo{volume}{42}, \bibinfo{number}{1} (\bibinfo{year}{2022}),
  \bibinfo{pages}{116--127}.
\newblock
\href{https://doi.org/10.1109/MM.2021.3097700}{doi:\nolinkurl{10.1109/MM.2021.3097700}}


\bibitem[Kim et~al\mbox{.}(2025f)]%
        {Kim2025PIMQuant}
\bibfield{author}{\bibinfo{person}{Byeori Kim}, \bibinfo{person}{Changhun Lee},
  \bibinfo{person}{Gwangsun Kim}, {and} \bibinfo{person}{Eunhyeok Park}.}
  \bibinfo{year}{2025}\natexlab{f}.
\newblock \showarticletitle{Cost-Effective Extension of DRAM-PIM for Group-Wise
  LLM Quantization}.
\newblock  \bibinfo{volume}{24}, \bibinfo{number}{1} (\bibinfo{year}{2025}),
  \bibinfo{pages}{53--56}.
\newblock
\href{https://doi.org/10.1109/LCA.2025.3532682}{doi:\nolinkurl{10.1109/LCA.2025.3532682}}


\bibitem[Kim et~al\mbox{.}(2016a)]%
        {Kim2016NeuroCube}
\bibfield{author}{\bibinfo{person}{Duckhwan Kim}, \bibinfo{person}{Jaeha Kung},
  \bibinfo{person}{Sek Chai}, \bibinfo{person}{Sudhakar Yalamanchili}, {and}
  \bibinfo{person}{Saibal Mukhopadhyay}.} \bibinfo{year}{2016}\natexlab{a}.
\newblock \showarticletitle{Neurocube: A Programmable Digital Neuromorphic
  Architecture with High-Density 3D Memory}. In \bibinfo{booktitle}{\emph{2016
  ACM/IEEE 43rd Annual International Symposium on Computer Architecture
  (ISCA)}} (2016). \bibinfo{pages}{380--392}.
\newblock
\href{https://doi.org/10.1109/ISCA.2016.41}{doi:\nolinkurl{10.1109/ISCA.2016.41}}


\bibitem[Kim et~al\mbox{.}(2025h)]%
        {Kim2025NMPPaK}
\bibfield{author}{\bibinfo{person}{Heewoo Kim}, \bibinfo{person}{Sanjay
  Sri~Vallabh Singapuram}, \bibinfo{person}{Haojie Ye}, \bibinfo{person}{Joseph
  Izraelevitz}, \bibinfo{person}{Trevor Mudge}, \bibinfo{person}{Ronald~G.
  Dreslinski}, {and} \bibinfo{person}{Nishil Talati}.}
  \bibinfo{year}{2025}\natexlab{h}.
\newblock \showarticletitle{NMP-PaK: Near-Memory Processing Acceleration of
  Scalable De Novo Genome Assembly}. In \bibinfo{booktitle}{\emph{Proceedings
  of the 52nd Annual International Symposium on Computer Architecture (ISCA
  '25)}} (New York, NY, USA, 2025) \emph{(\bibinfo{series}{ISCA '25})}.
  \bibinfo{publisher}{Association for Computing Machinery},
  \bibinfo{pages}{1834--1847}.
\newblock
\href{https://doi.org/10.1145/3695053.3731056}{doi:\nolinkurl{10.1145/3695053.3731056}}


\bibitem[Kim et~al\mbox{.}(2025i)]%
        {Kim2025Anaheim}
\bibfield{author}{\bibinfo{person}{Jongmin Kim}, \bibinfo{person}{Sungmin Yun},
  \bibinfo{person}{Hyesung Ji}, \bibinfo{person}{Wonseok Choi},
  \bibinfo{person}{Sangpyo Kim}, {and} \bibinfo{person}{Jung~Ho Ahn}.}
  \bibinfo{year}{2025}\natexlab{i}.
\newblock \showarticletitle{Anaheim: Architecture and Algorithms for Processing
  Fully Homomorphic Encryption in Memory}. In \bibinfo{booktitle}{\emph{2025
  IEEE International Symposium on High Performance Computer Architecture
  (HPCA)}} (2025). \bibinfo{publisher}{IEEE}, \bibinfo{pages}{1158--1173}.
\newblock
\href{https://doi.org/10.1109/HPCA61900.2025.00089}{doi:\nolinkurl{10.1109/HPCA61900.2025.00089}}


\bibitem[Kim et~al\mbox{.}(2021b)]%
        {KwonHBM2PIM}
\bibfield{author}{\bibinfo{person}{Jin~Hyun Kim}, \bibinfo{person}{Shin-haeng
  Kang}, \bibinfo{person}{Sukhan Lee}, \bibinfo{person}{Hyeonsu Kim},
  \bibinfo{person}{Woongjae Song}, \bibinfo{person}{Yuhwan Ro},
  \bibinfo{person}{Seungwon Lee}, \bibinfo{person}{David Wang},
  \bibinfo{person}{Hyunsung Shin}, \bibinfo{person}{Bengseng Phuah},
  \bibinfo{person}{Jihyun Choi}, \bibinfo{person}{Jinin So},
  \bibinfo{person}{YeonGon Cho}, \bibinfo{person}{JoonHo Song},
  \bibinfo{person}{Jangseok Choi}, \bibinfo{person}{Jeonghyeon Cho},
  \bibinfo{person}{Kyomin Sohn}, \bibinfo{person}{Youngsoo Sohn},
  \bibinfo{person}{Kwangil Park}, {and} \bibinfo{person}{Nam~Sung Kim}.}
  \bibinfo{year}{2021}\natexlab{b}.
\newblock \showarticletitle{Aquabolt-XL: Samsung HBM2-PIM with in-memory
  processing for ML accelerators and beyond}. In \bibinfo{booktitle}{\emph{2021
  IEEE Hot Chips 33 Symposium (HCS)}} (2021). \bibinfo{pages}{1--26}.
\newblock
\href{https://doi.org/10.1109/HCS52781.2021.9567191}{doi:\nolinkurl{10.1109/HCS52781.2021.9567191}}


\bibitem[Kim et~al\mbox{.}(2023c)]%
        {Kim2023SamsungPIMPNMTransformerAI}
\bibfield{author}{\bibinfo{person}{Jin~Hyun Kim}, \bibinfo{person}{Yuhwan Ro},
  \bibinfo{person}{Jinin So}, \bibinfo{person}{Sukhan Lee},
  \bibinfo{person}{Shin-haeng Kang}, \bibinfo{person}{YeonGon Cho},
  \bibinfo{person}{Hyeonsu Kim}, \bibinfo{person}{Byeongho Kim},
  \bibinfo{person}{Kyungsoo Kim}, \bibinfo{person}{Sangsoo Park},
  \bibinfo{person}{Jin-Seong Kim}, \bibinfo{person}{Sanghoon Cha},
  \bibinfo{person}{Won-Jo Lee}, \bibinfo{person}{Jin Jung},
  \bibinfo{person}{Jong-Geon Lee}, \bibinfo{person}{Jieun Lee},
  \bibinfo{person}{JoonHo Song}, \bibinfo{person}{Seungwon Lee},
  \bibinfo{person}{Jeonghyeon Cho}, \bibinfo{person}{Jaehoon Yu}, {and}
  \bibinfo{person}{Kyomin Sohn}.} \bibinfo{year}{2023}\natexlab{c}.
\newblock \showarticletitle{Samsung PIM/PNM for Transfmer Based AI : Energy
  Efficiency on PIM/PNM Cluster}. In \bibinfo{booktitle}{\emph{2023 IEEE Hot
  Chips 35 Symposium (HCS)}} (2023). \bibinfo{pages}{1--31}.
\newblock
\href{https://doi.org/10.1109/HCS59251.2023.10254711}{doi:\nolinkurl{10.1109/HCS59251.2023.10254711}}


\bibitem[Kim et~al\mbox{.}(2024d)]%
        {kim2024mondemixtureneardataexperts}
\bibfield{author}{\bibinfo{person}{Taehyun Kim}, \bibinfo{person}{Kwanseok
  Choi}, \bibinfo{person}{Youngmock Cho}, \bibinfo{person}{Jaehoon Cho},
  \bibinfo{person}{Hyuk-Jae Lee}, {and} \bibinfo{person}{Jaewoong Sim}.}
  \bibinfo{year}{2024}\natexlab{d}.
\newblock \bibinfo{title}{MoNDE: Mixture of Near-Data Experts for Large-Scale
  Sparse Models}.
\newblock
\showeprint[arXiv]{2405.18832}~[cs.LG]
\urldef\tempurl%
\url{https://arxiv.org/abs/2405.18832}
\showURL{%
\tempurl}


\bibitem[Kim et~al\mbox{.}(2025e)]%
        {Kim2025EOD}
\bibfield{author}{\bibinfo{person}{Taehwan Kim}, \bibinfo{person}{Yunki Han},
  \bibinfo{person}{Seohye Ha}, \bibinfo{person}{Jiwan Kim}, {and}
  \bibinfo{person}{Lee-Sup Kim}.} \bibinfo{year}{2025}\natexlab{e}.
\newblock \showarticletitle{EOD: Enabling Low Latency GNN Inference via
  Near-Memory Concatenate Aggregation}. In
  \bibinfo{booktitle}{\emph{Proceedings of the 52nd Annual International
  Symposium on Computer Architecture (ISCA '25)}} (New York, NY, USA, 2025)
  \emph{(\bibinfo{series}{ISCA '25})}. \bibinfo{publisher}{Association for
  Computing Machinery}, \bibinfo{pages}{1125--1139}.
\newblock
\href{https://doi.org/10.1145/3695053.3731083}{doi:\nolinkurl{10.1145/3695053.3731083}}


\bibitem[Kim et~al\mbox{.}(2025g)]%
        {Kim2025Pimba}
\bibfield{author}{\bibinfo{person}{Wonung Kim}, \bibinfo{person}{Yubin Lee},
  \bibinfo{person}{Yoonsung Kim}, \bibinfo{person}{Jinwoo Hwang},
  \bibinfo{person}{Seongryong Oh}, \bibinfo{person}{Jiyong Jung},
  \bibinfo{person}{Aziz Huseynov}, \bibinfo{person}{Woong~Gyu Park},
  \bibinfo{person}{Chang~Hyun Park}, \bibinfo{person}{Divya Mahajan}, {and}
  \bibinfo{person}{Jongse Park}.} \bibinfo{year}{2025}\natexlab{g}.
\newblock \showarticletitle{Pimba: A Processing-in-Memory Acceleration for
  Post-Transformer Large Language Model Serving}. In
  \bibinfo{booktitle}{\emph{Proceedings of the 58th IEEE/ACM International
  Symposium on Microarchitecture}} (Seoul, Republic of Korea, 2025)
  \emph{(\bibinfo{series}{MICRO '25})}. \bibinfo{publisher}{Association for
  Computing Machinery}, \bibinfo{pages}{292--307}.
\newblock
\showISBNx{9798400715730}
\href{https://doi.org/10.1145/3725843.3756121}{doi:\nolinkurl{10.1145/3725843.3756121}}


\bibitem[Kwon et~al\mbox{.}(2026e)]%
        {kwon2026pimphony}
\bibfield{author}{\bibinfo{person}{Hyucksung Kwon}, \bibinfo{person}{Kyungmo
  Koo}, \bibinfo{person}{Janghyeon Kim}, \bibinfo{person}{Woongkyu Lee},
  \bibinfo{person}{Minjae Lee}, \bibinfo{person}{Gyeonggeun Jung},
  \bibinfo{person}{Hyungdeok Lee}, \bibinfo{person}{Yousub Jung},
  \bibinfo{person}{Jaehan Park}, \bibinfo{person}{Yosub Song},
  \bibinfo{person}{Byeongsu Yang}, \bibinfo{person}{Haerang Choi},
  \bibinfo{person}{Guhyun Kim}, \bibinfo{person}{Jongsoon Won},
  \bibinfo{person}{Woojae Shin}, \bibinfo{person}{Changhyun Kim},
  \bibinfo{person}{Gyeongcheol Shin}, \bibinfo{person}{Yongkee Kwon},
  \bibinfo{person}{Ilkon Kim}, \bibinfo{person}{Euicheol Lim},
  \bibinfo{person}{John Kim}, {and} \bibinfo{person}{Jungwook Choi}.}
  \bibinfo{year}{2026}\natexlab{e}.
\newblock \showarticletitle{PIMphony: Overcoming Bandwidth and Capacity
  Inefficiency in PIM-Based Long-Context LLM Inference System}. In
  \bibinfo{booktitle}{\emph{2026 IEEE International Symposium on High
  Performance Computer Architecture (HPCA)}} (2026). \bibinfo{pages}{1--21}.
\newblock
\href{https://doi.org/10.1109/HPCA68181.2026.11408592}{doi:\nolinkurl{10.1109/HPCA68181.2026.11408592}}


\bibitem[Kwon et~al\mbox{.}(2023d)]%
        {Kwon2023vLLM}
\bibfield{author}{\bibinfo{person}{Woosuk Kwon}, \bibinfo{person}{Zhuohan Li},
  \bibinfo{person}{Siyuan Zhuang}, \bibinfo{person}{Ying Sheng},
  \bibinfo{person}{Lianmin Zheng}, \bibinfo{person}{Cody~Hao Yu},
  \bibinfo{person}{Joseph Gonzalez}, \bibinfo{person}{Hao Zhang}, {and}
  \bibinfo{person}{Ion Stoica}.} \bibinfo{year}{2023}\natexlab{d}.
\newblock \showarticletitle{Efficient Memory Management for Large Language
  Model Serving with PagedAttention}. In \bibinfo{booktitle}{\emph{Proceedings
  of the 29th Symposium on Operating Systems Principles}} (Koblenz, Germany,
  2023) \emph{(\bibinfo{series}{SOSP '23})}. \bibinfo{publisher}{Association
  for Computing Machinery}, \bibinfo{pages}{611--626}.
\newblock
\showISBNx{9798400702297}
\href{https://doi.org/10.1145/3600006.3613165}{doi:\nolinkurl{10.1145/3600006.3613165}}


\bibitem[Kwon et~al\mbox{.}(2019a)]%
        {Kwon2019TensorDIMM}
\bibfield{author}{\bibinfo{person}{Youngeun Kwon}, \bibinfo{person}{Yunjae
  Lee}, {and} \bibinfo{person}{Minsoo Rhu}.} \bibinfo{year}{2019}\natexlab{a}.
\newblock \showarticletitle{TensorDIMM: A Practical Near-Memory Processing
  Architecture for Embeddings and Tensor Operations in Deep Learning}. In
  \bibinfo{booktitle}{\emph{Proceedings of the 52nd Annual IEEE/ACM
  International Symposium on Microarchitecture}} (Columbus, OH, USA, 2019)
  \emph{(\bibinfo{series}{MICRO '52})}. \bibinfo{publisher}{Association for
  Computing Machinery}, \bibinfo{pages}{740--753}.
\newblock
\showISBNx{9781450369381}
\href{https://doi.org/10.1145/3352460.3358284}{doi:\nolinkurl{10.1145/3352460.3358284}}


\bibitem[Kwon et~al\mbox{.}(2022c)]%
        {Kwon2022GDDR6AiM}
\bibfield{author}{\bibinfo{person}{Yongkee Kwon}, \bibinfo{person}{Kornijcuk
  Vladimir}, \bibinfo{person}{Nahsung Kim}, \bibinfo{person}{Woojae Shin},
  \bibinfo{person}{Jongsoon Won}, \bibinfo{person}{Minkyu Lee},
  \bibinfo{person}{Hyunha Joo}, \bibinfo{person}{Haerang Choi},
  \bibinfo{person}{Guhyun Kim}, \bibinfo{person}{Byeongju An},
  \bibinfo{person}{Jeongbin Kim}, \bibinfo{person}{Jaewook Lee},
  \bibinfo{person}{Ilkon Kim}, \bibinfo{person}{Jaehan Park},
  \bibinfo{person}{Chanwook Park}, \bibinfo{person}{Yosub Song},
  \bibinfo{person}{Byeongsu Yang}, \bibinfo{person}{Hyungdeok Lee},
  \bibinfo{person}{Seho Kim}, \bibinfo{person}{Daehan Kwon},
  \bibinfo{person}{Seongju Lee}, \bibinfo{person}{Kyuyoung Kim},
  \bibinfo{person}{Sanghoon Oh}, \bibinfo{person}{Joonhong Park},
  \bibinfo{person}{Gimoon Hong}, \bibinfo{person}{Dongyoon Ka},
  \bibinfo{person}{Kyudong Hwang}, \bibinfo{person}{Jeongje Park},
  \bibinfo{person}{Kyeongpil Kang}, \bibinfo{person}{Jungyeon Kim},
  \bibinfo{person}{Junyeol Jeon}, \bibinfo{person}{Myeongjun Lee},
  \bibinfo{person}{Minyoung Shin}, \bibinfo{person}{Minhwan Shin},
  \bibinfo{person}{Jaekyung Cha}, \bibinfo{person}{Changson Jung},
  \bibinfo{person}{Kijoon Chang}, \bibinfo{person}{Chunseok Jeong},
  \bibinfo{person}{Euicheol Lim}, \bibinfo{person}{Il Park},
  \bibinfo{person}{Junhyun Chun}, {and} \bibinfo{person}{Sk Hynix}.}
  \bibinfo{year}{2022}\natexlab{c}.
\newblock \showarticletitle{System Architecture and Software Stack for
  GDDR6-AiM}. In \bibinfo{booktitle}{\emph{2022 IEEE Hot Chips 34 Symposium
  (HCS)}} (2022). \bibinfo{pages}{1--25}.
\newblock
\href{https://doi.org/10.1109/HCS55958.2022.9895629}{doi:\nolinkurl{10.1109/HCS55958.2022.9895629}}


\bibitem[Kwon et~al\mbox{.}(2021b)]%
        {Samsung2021ISSCCHBM2PIM}
\bibfield{author}{\bibinfo{person}{Young-Cheon Kwon}, \bibinfo{person}{Suk~Han
  Lee}, \bibinfo{person}{Jaehoon Lee}, \bibinfo{person}{Sang-Hyuk Kwon},
  \bibinfo{person}{Je~Min Ryu}, \bibinfo{person}{Jong-Pil Son},
  \bibinfo{person}{O Seongil}, \bibinfo{person}{Hak-Soo Yu},
  \bibinfo{person}{Haesuk Lee}, \bibinfo{person}{Soo~Young Kim},
  \bibinfo{person}{Youngmin Cho}, \bibinfo{person}{Jin~Guk Kim},
  \bibinfo{person}{Jongyoon Choi}, \bibinfo{person}{Hyun-Sung Shin},
  \bibinfo{person}{Jin Kim}, \bibinfo{person}{BengSeng Phuah},
  \bibinfo{person}{HyoungMin Kim}, \bibinfo{person}{Myeong~Jun Song},
  \bibinfo{person}{Ahn Choi}, \bibinfo{person}{Daeho Kim},
  \bibinfo{person}{SooYoung Kim}, \bibinfo{person}{Eun-Bong Kim},
  \bibinfo{person}{David Wang}, \bibinfo{person}{Shinhaeng Kang},
  \bibinfo{person}{Yuhwan Ro}, \bibinfo{person}{Seungwoo Seo},
  \bibinfo{person}{JoonHo Song}, \bibinfo{person}{Jaeyoun Youn},
  \bibinfo{person}{Kyomin Sohn}, {and} \bibinfo{person}{Nam~Sung Kim}.}
  \bibinfo{year}{2021}\natexlab{b}.
\newblock \showarticletitle{25.4 A 20nm 6GB Function-In-Memory DRAM, Based on
  HBM2 with a 1.2TFLOPS Programmable Computing Unit Using Bank-Level
  Parallelism, for Machine Learning Applications}. In
  \bibinfo{booktitle}{\emph{2021 IEEE International Solid-State Circuits
  Conference (ISSCC)}} (2021), Vol.~\bibinfo{volume}{64}.
  \bibinfo{pages}{350--352}.
\newblock
\href{https://doi.org/10.1109/ISSCC42613.2021.9365862}{doi:\nolinkurl{10.1109/ISSCC42613.2021.9365862}}


\bibitem[Lee et~al\mbox{.}(2025d)]%
        {lee2025pimutation}
\bibfield{author}{\bibinfo{person}{Dongin Lee}, \bibinfo{person}{Enhyeok Jang},
  \bibinfo{person}{Seungwoo Choi}, \bibinfo{person}{Junwoong An},
  \bibinfo{person}{Cheolhwan Kim}, {and} \bibinfo{person}{Won~Woo Ro}.}
  \bibinfo{year}{2025}\natexlab{d}.
\newblock \bibinfo{title}{PIMutation: Exploring the Potential of PIM
  Architecture for Quantum Circuit Simulation}.
\newblock
\showeprint[arXiv]{2503.00668}~[quant-ph]
\urldef\tempurl%
\url{https://arxiv.org/abs/2503.00668}
\showURL{%
\tempurl}


\bibitem[Lee et~al\mbox{.}(2025c)]%
        {Lee2025PAISE}
\bibfield{author}{\bibinfo{person}{Hyojung Lee}, \bibinfo{person}{Daehyeon
  Baek}, \bibinfo{person}{Jimyoung Son}, \bibinfo{person}{Jieun Choi},
  \bibinfo{person}{Kihyo Moon}, {and} \bibinfo{person}{Minsung Jang}.}
  \bibinfo{year}{2025}\natexlab{c}.
\newblock \showarticletitle{PAISE: PIM-Accelerated Inference Scheduling Engine
  for Transformer-based LLM}. In \bibinfo{booktitle}{\emph{2025 IEEE
  International Symposium on High Performance Computer Architecture (HPCA)}}
  (2025). \bibinfo{pages}{1707--1719}.
\newblock
\href{https://doi.org/10.1109/HPCA61900.2025.00126}{doi:\nolinkurl{10.1109/HPCA61900.2025.00126}}


\bibitem[Lee et~al\mbox{.}(2022b)]%
        {Lee2022ISSCCHBM3}
\bibfield{author}{\bibinfo{person}{S. Lee} {et~al\mbox{.}}}
  \bibinfo{year}{2022}\natexlab{b}.
\newblock \showarticletitle{A 192-Gb 12-High 896-{GB/s} {HBM3} {DRAM} with a
  {TSV} Auto-Calibration Scheme and Machine-Learning-Based Layout
  Optimization}. In \bibinfo{booktitle}{\emph{Proc. IEEE International
  Solid-State Circuits Conference (ISSCC)}} (2022). \bibinfo{pages}{176--178}.
\newblock
\href{https://doi.org/10.1109/ISSCC42614.2022.9731562}{doi:\nolinkurl{10.1109/ISSCC42614.2022.9731562}}


\bibitem[Lee et~al\mbox{.}(2021a)]%
        {Lee2021HBM2PIMSamsung}
\bibfield{author}{\bibinfo{person}{Sukhan Lee}, \bibinfo{person}{Shin-haeng
  Kang}, \bibinfo{person}{Jaehoon Lee}, \bibinfo{person}{Hyeonsu Kim},
  \bibinfo{person}{Eojin Lee}, \bibinfo{person}{Seungwoo Seo},
  \bibinfo{person}{Hosang Yoon}, \bibinfo{person}{Seungwon Lee},
  \bibinfo{person}{Kyounghwan Lim}, \bibinfo{person}{Hyunsung Shin},
  \bibinfo{person}{Jinhyun Kim}, \bibinfo{person}{O Seongil},
  \bibinfo{person}{Anand Iyer}, \bibinfo{person}{David Wang},
  \bibinfo{person}{Kyomin Sohn}, {and} \bibinfo{person}{Nam~Sung Kim}.}
  \bibinfo{year}{2021}\natexlab{a}.
\newblock \showarticletitle{Hardware Architecture and Software Stack for PIM
  Based on Commercial DRAM Technology : Industrial Product}. In
  \bibinfo{booktitle}{\emph{2021 ACM/IEEE 48th Annual International Symposium
  on Computer Architecture (ISCA)}} (2021). \bibinfo{pages}{43--56}.
\newblock
\href{https://doi.org/10.1109/ISCA52012.2021.00013}{doi:\nolinkurl{10.1109/ISCA52012.2021.00013}}


\bibitem[Li et~al\mbox{.}(2025d)]%
        {Li2025H2LLM}
\bibfield{author}{\bibinfo{person}{Cong Li}, \bibinfo{person}{Yihan Yin},
  \bibinfo{person}{Xintong Wu}, \bibinfo{person}{Jingchen Zhu},
  \bibinfo{person}{Zhutianya Gao}, \bibinfo{person}{Dimin Niu},
  \bibinfo{person}{Qiang Wu}, \bibinfo{person}{Xin Si}, \bibinfo{person}{Yuan
  Xie}, \bibinfo{person}{Chen Zhang}, {and} \bibinfo{person}{Guangyu Sun}.}
  \bibinfo{year}{2025}\natexlab{d}.
\newblock \showarticletitle{H$^{2}$-LLM: Hardware-Dataflow Co-Exploration for
  Heterogeneous Hybrid-Bonding-based Low-Batch LLM Inference}. In
  \bibinfo{booktitle}{\emph{Proceedings of the 52nd Annual International
  Symposium on Computer Architecture (ISCA '25)}} (New York, NY, USA, 2025)
  \emph{(\bibinfo{series}{ISCA '25})}. \bibinfo{publisher}{Association for
  Computing Machinery}, \bibinfo{pages}{194--210}.
\newblock
\href{https://doi.org/10.1145/3695053.3731008}{doi:\nolinkurl{10.1145/3695053.3731008}}


\bibitem[Li et~al\mbox{.}(2024a)]%
        {Li2024PIMDL}
\bibfield{author}{\bibinfo{person}{Cong Li}, \bibinfo{person}{Zhe Zhou},
  \bibinfo{person}{Yang Wang}, \bibinfo{person}{Fan Yang},
  \bibinfo{person}{Ting Cao}, \bibinfo{person}{Mao Yang}, \bibinfo{person}{Yun
  Liang}, {and} \bibinfo{person}{Guangyu Sun}.}
  \bibinfo{year}{2024}\natexlab{a}.
\newblock \showarticletitle{PIM-DL: Expanding the Applicability of Commodity
  DRAM-PIMs for Deep Learning via Algorithm-System Co-Optimization}. In
  \bibinfo{booktitle}{\emph{Proceedings of the 29th ACM International
  Conference on Architectural Support for Programming Languages and Operating
  Systems, Volume 2}} (La Jolla, CA, USA, 2024) \emph{(\bibinfo{series}{ASPLOS
  '24})}. \bibinfo{publisher}{Association for Computing Machinery},
  \bibinfo{pages}{879--896}.
\newblock
\showISBNx{9798400703850}
\href{https://doi.org/10.1145/3620665.3640376}{doi:\nolinkurl{10.1145/3620665.3640376}}


\bibitem[Li et~al\mbox{.}(2024b)]%
        {cong2024specpim}
\bibfield{author}{\bibinfo{person}{Cong Li}, \bibinfo{person}{Zhe Zhou},
  \bibinfo{person}{Size Zheng}, \bibinfo{person}{Jiaxi Zhang},
  \bibinfo{person}{Yun Liang}, {and} \bibinfo{person}{Guangyu Sun}.}
  \bibinfo{year}{2024}\natexlab{b}.
\newblock \showarticletitle{SpecPIM: Accelerating Speculative Inference on
  PIM-Enabled System via Architecture-Dataflow Co-Exploration}. In
  \bibinfo{booktitle}{\emph{Proceedings of the 29th ACM International
  Conference on Architectural Support for Programming Languages and Operating
  Systems, Volume 3}} (La Jolla, CA, USA, 2024) \emph{(\bibinfo{series}{ASPLOS
  '24})}. \bibinfo{publisher}{Association for Computing Machinery},
  \bibinfo{pages}{950--965}.
\newblock
\showISBNx{9798400703867}
\href{https://doi.org/10.1145/3620666.3651352}{doi:\nolinkurl{10.1145/3620666.3651352}}


\bibitem[Li et~al\mbox{.}(2026e)]%
        {Li2026HybridPIM}
\bibfield{author}{\bibinfo{person}{Hongyi Li}, \bibinfo{person}{Songchen Ma},
  \bibinfo{person}{Huanyu Qu}, \bibinfo{person}{Weihao Zhang},
  \bibinfo{person}{Jia Chen}, \bibinfo{person}{Junfeng Lin},
  \bibinfo{person}{Fengbin Tu}, {and} \bibinfo{person}{Rong Zhao}.}
  \bibinfo{year}{2026}\natexlab{e}.
\newblock \showarticletitle{Bridging Efficiency and Scalability in LLM System
  via 3D Hybrid PIM with 2D In-Transit Computation}. In
  \bibinfo{booktitle}{\emph{53rd Annual International Symposium on Computer
  Architecture (ISCA '26)}} (2026).
\newblock
\urldef\tempurl%
\url{https://www.iscaconf.org/isca2026/program/}
\showURL{%
\tempurl}
\newblock
\shownote{ISCA 2026 technical program}.


\bibitem[Li et~al\mbox{.}(2025c)]%
        {Li2025ANSMET}
\bibfield{author}{\bibinfo{person}{Yiwei Li}, \bibinfo{person}{Yuxin Jin},
  \bibinfo{person}{Boyu Tian}, \bibinfo{person}{Huanchen Zhang}, {and}
  \bibinfo{person}{Mingyu Gao}.} \bibinfo{year}{2025}\natexlab{c}.
\newblock \showarticletitle{ANSMET: Approximate Nearest Neighbor Search with
  Near-Memory Processing and Hybrid Early Termination}. In
  \bibinfo{booktitle}{\emph{Proceedings of the 52nd Annual International
  Symposium on Computer Architecture (ISCA '25)}} (New York, NY, USA, 2025)
  \emph{(\bibinfo{series}{ISCA '25})}. \bibinfo{publisher}{Association for
  Computing Machinery}, \bibinfo{pages}{1093--1107}.
\newblock
\href{https://doi.org/10.1145/3695053.3731013}{doi:\nolinkurl{10.1145/3695053.3731013}}


\bibitem[Lin et~al\mbox{.}(2025)]%
        {Lin2025Ironman}
\bibfield{author}{\bibinfo{person}{Chenqi Lin}, \bibinfo{person}{Kang Yang},
  \bibinfo{person}{Tianshi Xu}, \bibinfo{person}{Ling Liang},
  \bibinfo{person}{Yufei Wang}, \bibinfo{person}{Zhaohui Chen},
  \bibinfo{person}{Runsheng Wang}, \bibinfo{person}{Mingyu Gao}, {and}
  \bibinfo{person}{Meng Li}.} \bibinfo{year}{2025}\natexlab{}.
\newblock \showarticletitle{Ironman: Accelerating Oblivious Transfer Extension
  for Privacy-Preserving AI with Near-Memory Processing}. In
  \bibinfo{booktitle}{\emph{Proceedings of the 58th IEEE/ACM International
  Symposium on Microarchitecture}} (Seoul, Republic of Korea, 2025)
  \emph{(\bibinfo{series}{MICRO '25})}. \bibinfo{publisher}{Association for
  Computing Machinery}, \bibinfo{pages}{354--368}.
\newblock
\showISBNx{9798400715730}
\href{https://doi.org/10.1145/3725843.3756025}{doi:\nolinkurl{10.1145/3725843.3756025}}


\bibitem[Liu et~al\mbox{.}(2025e)]%
        {Liu2025SeIM}
\bibfield{author}{\bibinfo{person}{Chaoqiang Liu}, \bibinfo{person}{Dan Chen},
  \bibinfo{person}{Yu Huang}, \bibinfo{person}{Wenjing Xiao},
  \bibinfo{person}{Haifeng Liu}, \bibinfo{person}{Yi Zhang},
  \bibinfo{person}{Huize Li}, \bibinfo{person}{Xiaofei Liao}, {and}
  \bibinfo{person}{Hai Jin}.} \bibinfo{year}{2025}\natexlab{e}.
\newblock \showarticletitle{SeIM: In-Memory Acceleration for Approximate
  Nearest Neighbor Search}. In \bibinfo{booktitle}{\emph{2025 62nd ACM/IEEE
  Design Automation Conference (DAC)}} (2025). \bibinfo{publisher}{IEEE},
  \bibinfo{pages}{1--7}.
\newblock
\href{https://doi.org/10.1109/DAC63849.2025.11132620}{doi:\nolinkurl{10.1109/DAC63849.2025.11132620}}


\bibitem[Liu et~al\mbox{.}(2026k)]%
        {Liu2026MERIDIAN}
\bibfield{author}{\bibinfo{person}{Chaoqiang Liu}, \bibinfo{person}{Yu Huang},
  \bibinfo{person}{Haifeng Liu}, \bibinfo{person}{Yi Zhang},
  \bibinfo{person}{Qihang Qiu}, \bibinfo{person}{Xueqi Li},
  \bibinfo{person}{Long Zheng}, \bibinfo{person}{Xiaofei Liao},
  \bibinfo{person}{Hai Jin}, {and} \bibinfo{person}{Jingling Xue}.}
  \bibinfo{year}{2026}\natexlab{k}.
\newblock \showarticletitle{MERIDIAN: In-Memory Acceleration for RAG with
  Document Attention Decomposition}. In \bibinfo{booktitle}{\emph{53rd Annual
  International Symposium on Computer Architecture (ISCA '26)}} (2026).
\newblock
\urldef\tempurl%
\url{https://www.iscaconf.org/isca2026/program/}
\showURL{%
\tempurl}
\newblock
\shownote{ISCA 2026 technical program}.


\bibitem[Liu et~al\mbox{.}(2025h)]%
        {Liu2025HeterRAG}
\bibfield{author}{\bibinfo{person}{Chaoqiang Liu}, \bibinfo{person}{Haifeng
  Liu}, \bibinfo{person}{Dan Chen}, \bibinfo{person}{Yu Huang},
  \bibinfo{person}{Yi Zhang}, \bibinfo{person}{Wenjing Xiao},
  \bibinfo{person}{Xiaofei Liao}, {and} \bibinfo{person}{Hai Jin}.}
  \bibinfo{year}{2025}\natexlab{h}.
\newblock \showarticletitle{HeterRAG: Heterogeneous Processing-in-Memory
  Acceleration for Retrieval-augmented Generation}. In
  \bibinfo{booktitle}{\emph{Proceedings of the 52nd Annual International
  Symposium on Computer Architecture (ISCA '25)}} (New York, NY, USA, 2025)
  \emph{(\bibinfo{series}{ISCA '25})}. \bibinfo{publisher}{Association for
  Computing Machinery}, \bibinfo{pages}{884--898}.
\newblock
\href{https://doi.org/10.1145/3695053.3731089}{doi:\nolinkurl{10.1145/3695053.3731089}}


\bibitem[Liu et~al\mbox{.}(2023b)]%
        {Liu2023ISCAAccel}
\bibfield{author}{\bibinfo{person}{Haifeng Liu}, \bibinfo{person}{Long Zheng},
  \bibinfo{person}{Yu Huang}, \bibinfo{person}{Chaoqiang Liu},
  \bibinfo{person}{Xiangyu Ye}, \bibinfo{person}{Jingrui Yuan},
  \bibinfo{person}{Xiaofei Liao}, \bibinfo{person}{Hai Jin}, {and}
  \bibinfo{person}{Jingling Xue}.} \bibinfo{year}{2023}\natexlab{b}.
\newblock \showarticletitle{Accelerating Personalized Recommendation with
  Cross-level Near-Memory Processing}. In \bibinfo{booktitle}{\emph{Proceedings
  of the 50th Annual International Symposium on Computer Architecture}}
  (Orlando, FL, USA, 2023) \emph{(\bibinfo{series}{ISCA '23})}.
  \bibinfo{publisher}{Association for Computing Machinery}.
\newblock
\showISBNx{9798400700958}
\href{https://doi.org/10.1145/3579371.3589101}{doi:\nolinkurl{10.1145/3579371.3589101}}


\bibitem[Liu et~al\mbox{.}(2024d)]%
        {Liu2024NearCXL}
\bibfield{author}{\bibinfo{person}{Haifeng Liu}, \bibinfo{person}{Long Zheng},
  \bibinfo{person}{Yu Huang}, \bibinfo{person}{Jingyi Zhou},
  \bibinfo{person}{Chaoqiang Liu}, \bibinfo{person}{Runze Wang},
  \bibinfo{person}{Xiaofei Liao}, \bibinfo{person}{Hai Jin}, {and}
  \bibinfo{person}{Jingling Xue}.} \bibinfo{year}{2024}\natexlab{d}.
\newblock \showarticletitle{Enabling Efficient Large Recommendation Model
  Training with Near CXL Memory Processing}. In \bibinfo{booktitle}{\emph{2024
  ACM/IEEE 51st Annual International Symposium on Computer Architecture
  (ISCA)}} (2024). \bibinfo{pages}{382--395}.
\newblock
\href{https://doi.org/10.1109/ISCA59077.2024.00036}{doi:\nolinkurl{10.1109/ISCA59077.2024.00036}}


\bibitem[Liu et~al\mbox{.}(2021a)]%
        {Liu2021ENMCNearMemoryClassification}
\bibfield{author}{\bibinfo{person}{Liu Liu}, \bibinfo{person}{Jilan Lin},
  \bibinfo{person}{Zheng Qu}, \bibinfo{person}{Yufei Ding}, {and}
  \bibinfo{person}{Yuan Xie}.} \bibinfo{year}{2021}\natexlab{a}.
\newblock \showarticletitle{ENMC: Extreme Near-Memory Classification via
  Approximate Screening}. In \bibinfo{booktitle}{\emph{MICRO-54: 54th Annual
  IEEE/ACM International Symposium on Microarchitecture}} (Virtual Event,
  Greece, 2021) \emph{(\bibinfo{series}{MICRO '21})}.
  \bibinfo{publisher}{Association for Computing Machinery},
  \bibinfo{pages}{1309--1322}.
\newblock
\showISBNx{9781450385572}
\href{https://doi.org/10.1145/3466752.3480090}{doi:\nolinkurl{10.1145/3466752.3480090}}


\bibitem[Liu et~al\mbox{.}(2025i)]%
        {Liu2025NDPDIMM}
\bibfield{author}{\bibinfo{person}{Lian Liu}, \bibinfo{person}{Shixin Zhao},
  \bibinfo{person}{Bing Li}, \bibinfo{person}{Haimeng Ren},
  \bibinfo{person}{Zhaohui Xu}, \bibinfo{person}{Mengdi Wang},
  \bibinfo{person}{Xiaowei Li}, \bibinfo{person}{Yinhe Han}, {and}
  \bibinfo{person}{Ying Wang}.} \bibinfo{year}{2025}\natexlab{i}.
\newblock \showarticletitle{Make LLM Inference Affordable to Everyone:
  Augmenting GPU Memory with NDP-DIMM}. In \bibinfo{booktitle}{\emph{2025 IEEE
  International Symposium on High Performance Computer Architecture (HPCA)}}
  (2025). \bibinfo{publisher}{IEEE}, \bibinfo{pages}{1751--1765}.
\newblock
\href{https://doi.org/10.1109/HPCA61900.2025.00129}{doi:\nolinkurl{10.1109/HPCA61900.2025.00129}}


\bibitem[Liu et~al\mbox{.}(2026l)]%
        {Liu2026PAM}
\bibfield{author}{\bibinfo{person}{Lian Liu}, \bibinfo{person}{Shixin Zhao},
  \bibinfo{person}{Yutian Zhou}, \bibinfo{person}{Yintao He},
  \bibinfo{person}{Mengdi Wang}, \bibinfo{person}{Yinhe Han}, {and}
  \bibinfo{person}{Ying Wang}.} \bibinfo{year}{2026}\natexlab{l}.
\newblock \bibinfo{title}{PAM: Processing Across Memory Hierarchy for Efficient
  KV-centric LLM Serving System}.
\newblock
\showeprint[arXiv]{2602.11521}~[cs.AR]
\urldef\tempurl%
\url{https://arxiv.org/abs/2602.11521}
\showURL{%
\tempurl}


\bibitem[Liu et~al\mbox{.}(2026j)]%
        {Liu2026CHIME}
\bibfield{author}{\bibinfo{person}{Qingyuan Liu}, \bibinfo{person}{Liyan Chen},
  \bibinfo{person}{Haocheng Wang}, \bibinfo{person}{Yanning Yang},
  \bibinfo{person}{Dong Du}, \bibinfo{person}{Zhigang Mao},
  \bibinfo{person}{Naifeng Jing}, \bibinfo{person}{Yubin Xia}, {and}
  \bibinfo{person}{Haibo Chen}.} \bibinfo{year}{2026}\natexlab{j}.
\newblock \showarticletitle{CHIME: A Case for Efficient Long-Context
  Attention-FC Disaggregated Inference with DIMM-PIM}. In
  \bibinfo{booktitle}{\emph{53rd Annual International Symposium on Computer
  Architecture (ISCA '26)}} (2026).
\newblock
\urldef\tempurl%
\url{https://www.iscaconf.org/isca2026/program/}
\showURL{%
\tempurl}
\newblock
\shownote{ISCA 2026 technical program}.


\bibitem[Liu et~al\mbox{.}(2025f)]%
        {liu2025l3dimmpim}
\bibfield{author}{\bibinfo{person}{Qingyuan Liu}, \bibinfo{person}{Liyan Chen},
  \bibinfo{person}{Yanning Yang}, \bibinfo{person}{Haocheng Wang},
  \bibinfo{person}{Dong Du}, \bibinfo{person}{Zhigang Mao},
  \bibinfo{person}{Naifeng Jing}, \bibinfo{person}{Yubin Xia}, {and}
  \bibinfo{person}{Haibo Chen}.} \bibinfo{year}{2025}\natexlab{f}.
\newblock \bibinfo{title}{L3: DIMM-PIM Integrated Architecture and Coordination
  for Scalable Long-Context LLM Inference}.
\newblock
\showeprint[arXiv]{2504.17584}~[cs.AR]
\urldef\tempurl%
\url{https://arxiv.org/abs/2504.17584}
\showURL{%
\tempurl}


\bibitem[Liu et~al\mbox{.}(2025g)]%
        {Liu2025McPAL}
\bibfield{author}{\bibinfo{person}{Shiwei Liu}, \bibinfo{person}{Zhirui Huang},
  \bibinfo{person}{Jiangnan Yu}, \bibinfo{person}{Qi Liu}, {and}
  \bibinfo{person}{Chixiao Chen}.} \bibinfo{year}{2025}\natexlab{g}.
\newblock \showarticletitle{McPAL: Scaling Unstructured Sparse Inference with
  Multi-Chiplet HBM-PIM Architecture for LLMs}. In
  \bibinfo{booktitle}{\emph{2025 62nd ACM/IEEE Design Automation Conference
  (DAC)}} (2025). \bibinfo{publisher}{IEEE}, \bibinfo{pages}{1--7}.
\newblock
\href{https://doi.org/10.1109/DAC63849.2025.11132914}{doi:\nolinkurl{10.1109/DAC63849.2025.11132914}}


\bibitem[Liu et~al\mbox{.}(2024c)]%
        {Liu2024ICMLKIVI}
\bibfield{author}{\bibinfo{person}{Zirui Liu}, \bibinfo{person}{Jiayi Yuan},
  \bibinfo{person}{Hongye Jin}, \bibinfo{person}{Shaochen Zhong},
  \bibinfo{person}{Zhaozhuo Xu}, \bibinfo{person}{Vladimir Braverman},
  \bibinfo{person}{Beidi Chen}, {and} \bibinfo{person}{Xia Hu}.}
  \bibinfo{year}{2024}\natexlab{c}.
\newblock \showarticletitle{{KIVI}: A Tuning-Free Asymmetric 2bit Quantization
  for {KV} Cache}. In \bibinfo{booktitle}{\emph{Proceedings of the 41st
  International Conference on Machine Learning (ICML)}} (2024).
\newblock


\bibitem[Luo et~al\mbox{.}(2024)]%
        {Luo2024Ramulator2}
\bibfield{author}{\bibinfo{person}{Haocong Luo}, \bibinfo{person}{Yahya~Can
  Tuğrul}, \bibinfo{person}{F.~Nisa Bostancı}, \bibinfo{person}{Ataberk
  Olgun}, \bibinfo{person}{A.~Giray Yağlıkçı}, {and} \bibinfo{person}{Onur
  Mutlu}.} \bibinfo{year}{2024}\natexlab{}.
\newblock \showarticletitle{Ramulator 2.0: A Modern, Modular, and Extensible
  DRAM Simulator}.
\newblock  \bibinfo{volume}{23}, \bibinfo{number}{1} (\bibinfo{year}{2024}),
  \bibinfo{pages}{112--116}.
\newblock
\href{https://doi.org/10.1109/LCA.2023.3333759}{doi:\nolinkurl{10.1109/LCA.2023.3333759}}


\bibitem[Ma et~al\mbox{.}(2024)]%
        {Ma2024Moctopus}
\bibfield{author}{\bibinfo{person}{Ruoyan Ma}, \bibinfo{person}{Shengan Zheng},
  \bibinfo{person}{Guifeng Wang}, \bibinfo{person}{Jin Pu},
  \bibinfo{person}{Yifan Hua}, \bibinfo{person}{Wentao Wang}, {and}
  \bibinfo{person}{Linpeng Huang}.} \bibinfo{year}{2024}\natexlab{}.
\newblock \showarticletitle{Accelerating Regular Path Queries over Graph
  Database with Processing-in-Memory}. In \bibinfo{booktitle}{\emph{Proceedings
  of the 61st ACM/IEEE Design Automation Conference}} (San Francisco, CA, USA,
  2024) \emph{(\bibinfo{series}{DAC '24})}. \bibinfo{publisher}{Association for
  Computing Machinery}.
\newblock
\href{https://doi.org/10.1145/3649329.3656235}{doi:\nolinkurl{10.1145/3649329.3656235}}


\bibitem[Malekar et~al\mbox{.}(2025)]%
        {malekar2025pimllmhighthroughputhybridpim}
\bibfield{author}{\bibinfo{person}{Jinendra Malekar}, \bibinfo{person}{Peyton
  Chandarana}, \bibinfo{person}{Md~Hasibul Amin}, \bibinfo{person}{Mohammed~E.
  Elbtity}, {and} \bibinfo{person}{Ramtin Zand}.}
  \bibinfo{year}{2025}\natexlab{}.
\newblock \bibinfo{title}{PIM-LLM: A High-Throughput Hybrid PIM Architecture
  for 1-bit LLMs}.
\newblock
\showeprint[arXiv]{2504.01994}~[cs.AR]
\urldef\tempurl%
\url{https://arxiv.org/abs/2504.01994}
\showURL{%
\tempurl}


\bibitem[{Marvell Technology}(2024)]%
        {marvell2024customhbm}
\bibfield{author}{\bibinfo{person}{{Marvell Technology}}.}
  \bibinfo{year}{2024}\natexlab{}.
\newblock \bibinfo{title}{{Custom HBM: What Is It and Why It's the Future}}.
\newblock
  \bibinfo{howpublished}{\url{https://www.marvell.com/blogs/custom-hbm-what-is-it-and-why-its-the-future.html}}.
\newblock
\newblock
\shownote{Accessed: 2026-06-17}.


\bibitem[Matsushima et~al\mbox{.}(2026)]%
        {Matsushima2026AQPIM}
\bibfield{author}{\bibinfo{person}{Kosuke Matsushima},
  \bibinfo{person}{Yasuyuki Okoshi}, \bibinfo{person}{Masato Motomura}, {and}
  \bibinfo{person}{Daichi Fujiki}.} \bibinfo{year}{2026}\natexlab{}.
\newblock \showarticletitle{AQPIM: Breaking the PIM Capacity Wall for LLMs with
  In-Memory Activation Quantization}. In \bibinfo{booktitle}{\emph{2026 IEEE
  International Symposium on High Performance Computer Architecture (HPCA)}}
  (2026). \bibinfo{publisher}{IEEE}, \bibinfo{pages}{1--17}.
\newblock
\href{https://doi.org/10.1109/HPCA68181.2026.11408452}{doi:\nolinkurl{10.1109/HPCA68181.2026.11408452}}


\bibitem[Nair et~al\mbox{.}(2026)]%
        {Nair2026Raptor}
\bibfield{author}{\bibinfo{person}{Prashant~J. Nair}, \bibinfo{person}{Ramyad
  Hadidi}, \bibinfo{person}{Subramani Ganesh}, \bibinfo{person}{Sangamesh
  Kodge}, \bibinfo{person}{Shubhankit Rathore}, \bibinfo{person}{Neil
  Thanawala}, \bibinfo{person}{Nikitha Reddy}, \bibinfo{person}{Gyanesh
  Saharia}, \bibinfo{person}{Vinayak Patankar}, \bibinfo{person}{Arun Tiruvur},
  \bibinfo{person}{Nithesh Kurella}, {and} \bibinfo{person}{Sudeep Bhoja}.}
  \bibinfo{year}{2026}\natexlab{}.
\newblock \showarticletitle{Early Silicon of Raptor: The First 3D-DRAM
  Accelerator for Generative Inference}. In \bibinfo{booktitle}{\emph{53rd
  Annual International Symposium on Computer Architecture (ISCA '26)}} (2026).
\newblock
\urldef\tempurl%
\url{https://www.iscaconf.org/isca2026/program/}
\showURL{%
\tempurl}
\newblock
\shownote{ISCA 2026 technical program}.


\bibitem[Nam et~al\mbox{.}(2025)]%
        {Nam2025AffinityPIM}
\bibfield{author}{\bibinfo{person}{Kevin Nam}, \bibinfo{person}{Heonhui Jung},
  \bibinfo{person}{Hyunyoung Oh}, {and} \bibinfo{person}{Yunheung Paek}.}
  \bibinfo{year}{2025}\natexlab{}.
\newblock \showarticletitle{Affinity-based Optimizations for TFHE on
  Processing-in-DRAM}. In \bibinfo{booktitle}{\emph{Proceedings of the 30th ACM
  International Conference on Architectural Support for Programming Languages
  and Operating Systems, Volume 2}} (Rotterdam, Netherlands, 2025)
  \emph{(\bibinfo{series}{ASPLOS '25})}. \bibinfo{publisher}{Association for
  Computing Machinery}, \bibinfo{pages}{16--31}.
\newblock
\href{https://doi.org/10.1145/3676641.3716246}{doi:\nolinkurl{10.1145/3676641.3716246}}


\bibitem[Pan et~al\mbox{.}(2025)]%
        {Pan2025Stratum}
\bibfield{author}{\bibinfo{person}{Yue Pan}, \bibinfo{person}{Zihan Xia},
  \bibinfo{person}{Po-Kai Hsu}, \bibinfo{person}{Lanxiang Hu},
  \bibinfo{person}{Hyungyo Kim}, \bibinfo{person}{Janak Sharda},
  \bibinfo{person}{Minxuan Zhou}, \bibinfo{person}{Nam~Sung Kim},
  \bibinfo{person}{Shimeng Yu}, \bibinfo{person}{Tajana Rosing}, {and}
  \bibinfo{person}{Mingu Kang}.} \bibinfo{year}{2025}\natexlab{}.
\newblock \showarticletitle{Stratum: System-Hardware Co-Design with Tiered
  Monolithic 3D-Stackable DRAM for Efficient MoE Serving}. In
  \bibinfo{booktitle}{\emph{Proceedings of the 58th IEEE/ACM International
  Symposium on Microarchitecture}} (Seoul, Republic of Korea, 2025)
  \emph{(\bibinfo{series}{MICRO '25})}. \bibinfo{publisher}{Association for
  Computing Machinery}, \bibinfo{pages}{1--17}.
\newblock
\showISBNx{9798400715730}
\href{https://doi.org/10.1145/3725843.3756043}{doi:\nolinkurl{10.1145/3725843.3756043}}


\bibitem[Park et~al\mbox{.}(2024a)]%
        {Park2024ASPLOSAttAcc}
\bibfield{author}{\bibinfo{person}{Jaehyun Park}, \bibinfo{person}{Jaewan
  Choi}, \bibinfo{person}{Kwanhee Kyung}, \bibinfo{person}{Michael~Jaemin Kim},
  \bibinfo{person}{Yongsuk Kwon}, \bibinfo{person}{Nam~Sung Kim}, {and}
  \bibinfo{person}{Jung~Ho Ahn}.} \bibinfo{year}{2024}\natexlab{a}.
\newblock \showarticletitle{AttAcc! Unleashing the Power of PIM for Batched
  Transformer-based Generative Model Inference}. In
  \bibinfo{booktitle}{\emph{Proceedings of the 29th ACM International
  Conference on Architectural Support for Programming Languages and Operating
  Systems, Volume 2}} (La Jolla, CA, USA, 2024) \emph{(\bibinfo{series}{ASPLOS
  '24})}. \bibinfo{publisher}{Association for Computing Machinery},
  \bibinfo{pages}{103--119}.
\newblock
\showISBNx{9798400703850}
\href{https://doi.org/10.1145/3620665.3640422}{doi:\nolinkurl{10.1145/3620665.3640422}}


\bibitem[Park et~al\mbox{.}(2024b)]%
        {park2024lpddrcxl}
\bibfield{author}{\bibinfo{person}{Sang-Soo Park}, \bibinfo{person}{KyungSoo
  Kim}, \bibinfo{person}{Jinin So}, \bibinfo{person}{Jin Jung},
  \bibinfo{person}{Jonggeon Lee}, \bibinfo{person}{Kyoungwan Woo},
  \bibinfo{person}{Nayeon Kim}, \bibinfo{person}{Younghyun Lee},
  \bibinfo{person}{Hyungyo Kim}, \bibinfo{person}{Yongsuk Kwon},
  \bibinfo{person}{Jinhyun Kim}, \bibinfo{person}{Jieun Lee},
  \bibinfo{person}{YeonGon Cho}, \bibinfo{person}{Yongmin Tai},
  \bibinfo{person}{Jeonghyeon Cho}, \bibinfo{person}{Hoyoung Song},
  \bibinfo{person}{Jung~Ho Ahn}, {and} \bibinfo{person}{Nam~Sung Kim}.}
  \bibinfo{year}{2024}\natexlab{b}.
\newblock \showarticletitle{An LPDDR-based CXL-PNM Platform for TCO-efficient
  Inference of Transformer-based Large Language Models}. In
  \bibinfo{booktitle}{\emph{2024 IEEE International Symposium on
  High-Performance Computer Architecture (HPCA)}} (2024).
  \bibinfo{pages}{970--982}.
\newblock
\href{https://doi.org/10.1109/HPCA57654.2024.00078}{doi:\nolinkurl{10.1109/HPCA57654.2024.00078}}


\bibitem[Park et~al\mbox{.}(2025c)]%
        {Park2025FHENDI}
\bibfield{author}{\bibinfo{person}{Yongmo Park}, \bibinfo{person}{Aporva
  Amarnath}, \bibinfo{person}{Subhankar Pal}, \bibinfo{person}{Karthik
  Swaminathan}, \bibinfo{person}{Alper Buyuktosunoglu}, \bibinfo{person}{Hayim
  Shaul}, \bibinfo{person}{Ehud Aharoni}, \bibinfo{person}{Nir Drucker},
  \bibinfo{person}{Wei~D. Lu}, \bibinfo{person}{Omri Soceanu}, {and}
  \bibinfo{person}{Pradip Bose}.} \bibinfo{year}{2025}\natexlab{c}.
\newblock \showarticletitle{FHENDI: A Near-DRAM Accelerator for
  Compiler-Generated Fully Homomorphic Encryption Applications}. In
  \bibinfo{booktitle}{\emph{2025 IEEE International Symposium on High
  Performance Computer Architecture (HPCA)}} (2025). \bibinfo{publisher}{IEEE},
  \bibinfo{pages}{1127--1142}.
\newblock
\href{https://doi.org/10.1109/HPCA61900.2025.00087}{doi:\nolinkurl{10.1109/HPCA61900.2025.00087}}


\bibitem[Qin et~al\mbox{.}(2025)]%
        {Kimi2025MoonCake}
\bibfield{author}{\bibinfo{person}{Ruoyu Qin}, \bibinfo{person}{Zheming Li},
  \bibinfo{person}{Weiran He}, \bibinfo{person}{Jialei Cui},
  \bibinfo{person}{Feng Ren}, \bibinfo{person}{Mingxing Zhang},
  \bibinfo{person}{Yongwei Wu}, \bibinfo{person}{Weimin Zheng}, {and}
  \bibinfo{person}{Xinran Xu}.} \bibinfo{year}{2025}\natexlab{}.
\newblock \showarticletitle{MOONCAKE: trading more storage for less computation
  — a KVCache-centric architecture for serving LLM chatbot}. In
  \bibinfo{booktitle}{\emph{Proceedings of the 23rd USENIX Conference on File
  and Storage Technologies}} (Santa Clara, CA, USA, 2025)
  \emph{(\bibinfo{series}{FAST '25})}. \bibinfo{publisher}{USENIX Association}.
\newblock
\showISBNx{978-1-939133-45-8}


\bibitem[Quinn et~al\mbox{.}(2025a)]%
        {Quinn2025LongSight}
\bibfield{author}{\bibinfo{person}{Derrick Quinn}, \bibinfo{person}{E.~Ezgi
  Y{"u}cel}, \bibinfo{person}{Jinkwon Kim}, \bibinfo{person}{Jos{'e}~F.
  Mart{'i}nez}, {and} \bibinfo{person}{Mohammad Alian}.}
  \bibinfo{year}{2025}\natexlab{a}.
\newblock \showarticletitle{LongSight: Compute-Enabled Memory to Accelerate
  Large-Context LLMs via Sparse Attention}. In
  \bibinfo{booktitle}{\emph{Proceedings of the 58th IEEE/ACM International
  Symposium on Microarchitecture}} (Seoul, Republic of Korea, 2025)
  \emph{(\bibinfo{series}{MICRO '25})}. \bibinfo{publisher}{Association for
  Computing Machinery}, \bibinfo{pages}{34--48}.
\newblock
\showISBNx{9798400715730}
\href{https://doi.org/10.1145/3725843.3756062}{doi:\nolinkurl{10.1145/3725843.3756062}}


\bibitem[Quinn et~al\mbox{.}(2025b)]%
        {Quinn2025DReX}
\bibfield{author}{\bibinfo{person}{Derrick Quinn}, \bibinfo{person}{E.~Ezgi
  Yücel}, \bibinfo{person}{Martin Prammer}, \bibinfo{person}{Zhenxing Fan},
  \bibinfo{person}{Kevin Skadron}, \bibinfo{person}{Jignesh~M. Patel},
  \bibinfo{person}{José~F. Martínez}, {and} \bibinfo{person}{Mohammad
  Alian}.} \bibinfo{year}{2025}\natexlab{b}.
\newblock \showarticletitle{DReX: Accurate and Scalable Dense Retrieval
  Acceleration via Algorithmic-Hardware Codesign}. In
  \bibinfo{booktitle}{\emph{Proceedings of the 52nd Annual International
  Symposium on Computer Architecture (ISCA '25)}} (New York, NY, USA, 2025)
  \emph{(\bibinfo{series}{ISCA '25})}. \bibinfo{publisher}{Association for
  Computing Machinery}, \bibinfo{pages}{1108--1124}.
\newblock
\href{https://doi.org/10.1145/3695053.3731079}{doi:\nolinkurl{10.1145/3695053.3731079}}


\bibitem[{Rambus}(2026)]%
        {rambus2026hbm4e}
\bibfield{author}{\bibinfo{person}{{Rambus}}.} \bibinfo{year}{2026}\natexlab{}.
\newblock \bibinfo{title}{{Rambus Sets New Benchmark for AI Memory Performance
  with Industry-Leading HBM4E Controller IP}}.
\newblock
  \bibinfo{howpublished}{\url{https://www.rambus.com/rambus-sets-new-benchmark-for-ai-memory-performance-with-industry-leading-hbm4e-controller-ip/}}.
\newblock
\newblock
\shownote{Accessed: 2026-06-17}.


\bibitem[Saed et~al\mbox{.}(2025)]%
        {Saed2025RayN}
\bibfield{author}{\bibinfo{person}{Mohammadreza Saed},
  \bibinfo{person}{Prashant~J. Nair}, {and} \bibinfo{person}{Tor~M. Aamodt}.}
  \bibinfo{year}{2025}\natexlab{}.
\newblock \showarticletitle{RayN: Ray Tracing Acceleration with Near-memory
  Computing}. In \bibinfo{booktitle}{\emph{Proceedings of the 58th IEEE/ACM
  International Symposium on Microarchitecture}} (Seoul, Republic of Korea,
  2025) \emph{(\bibinfo{series}{MICRO '25})}. \bibinfo{publisher}{Association
  for Computing Machinery}, \bibinfo{pages}{277--291}.
\newblock
\showISBNx{9798400715730}
\href{https://doi.org/10.1145/3725843.3756067}{doi:\nolinkurl{10.1145/3725843.3756067}}


\bibitem[{Samsung Semiconductor}(2026)]%
        {samsung2026hbm}
\bibfield{author}{\bibinfo{person}{{Samsung Semiconductor}}.}
  \bibinfo{year}{2026}\natexlab{}.
\newblock \bibinfo{title}{{HBM | DRAM | Samsung Semiconductor Global}}.
\newblock
  \bibinfo{howpublished}{\url{https://semiconductor.samsung.com/dram/hbm/}}.
\newblock
\newblock
\shownote{Accessed: 2026-06-17}.


\bibitem[Seo et~al\mbox{.}(2024a)]%
        {IanusNPUPIM2024}
\bibfield{author}{\bibinfo{person}{Minseok Seo}, \bibinfo{person}{Xuan~Truong
  Nguyen}, \bibinfo{person}{Seok~Joong Hwang}, \bibinfo{person}{Yongkee Kwon},
  \bibinfo{person}{Guhyun Kim}, \bibinfo{person}{Chanwook Park},
  \bibinfo{person}{Ilkon Kim}, \bibinfo{person}{Jaehan Park},
  \bibinfo{person}{Jeongbin Kim}, \bibinfo{person}{Woojae Shin},
  \bibinfo{person}{Jongsoon Won}, \bibinfo{person}{Haerang Choi},
  \bibinfo{person}{Kyuyoung Kim}, \bibinfo{person}{Daehan Kwon},
  \bibinfo{person}{Chunseok Jeong}, \bibinfo{person}{Sangheon Lee},
  \bibinfo{person}{Yongseok Choi}, \bibinfo{person}{Wooseok Byun},
  \bibinfo{person}{Seungcheol Baek}, \bibinfo{person}{Hyuk-Jae Lee}, {and}
  \bibinfo{person}{John Kim}.} \bibinfo{year}{2024}\natexlab{a}.
\newblock \showarticletitle{IANUS: Integrated Accelerator based on NPU-PIM
  Unified Memory System}. In \bibinfo{booktitle}{\emph{Proceedings of the 29th
  ACM International Conference on Architectural Support for Programming
  Languages and Operating Systems, Volume 3}} (La Jolla, CA, USA, 2024)
  \emph{(\bibinfo{series}{ASPLOS '24})}. \bibinfo{publisher}{Association for
  Computing Machinery}, \bibinfo{pages}{545--560}.
\newblock
\showISBNx{9798400703867}
\href{https://doi.org/10.1145/3620666.3651324}{doi:\nolinkurl{10.1145/3620666.3651324}}


\bibitem[Seo et~al\mbox{.}(2025b)]%
        {Seo2025FACIL}
\bibfield{author}{\bibinfo{person}{Seong~Hoon Seo}, \bibinfo{person}{Junghoon
  Kim}, \bibinfo{person}{Donghyun Lee}, \bibinfo{person}{Seonah Yoo},
  \bibinfo{person}{Seokwon Moon}, \bibinfo{person}{Yeonhong Park}, {and}
  \bibinfo{person}{Jae~W. Lee}.} \bibinfo{year}{2025}\natexlab{b}.
\newblock \showarticletitle{FACIL: Flexible DRAM Address Mapping for SoC-PIM
  Cooperative On-device LLM Inference}. In \bibinfo{booktitle}{\emph{2025 IEEE
  International Symposium on High Performance Computer Architecture (HPCA)}}
  (2025). \bibinfo{publisher}{IEEE}, \bibinfo{pages}{1720--1733}.
\newblock
\href{https://doi.org/10.1109/HPCA61900.2025.00127}{doi:\nolinkurl{10.1109/HPCA61900.2025.00127}}


\bibitem[Shin et~al\mbox{.}(2025)]%
        {Shin2025Piccolo}
\bibfield{author}{\bibinfo{person}{Changmin Shin}, \bibinfo{person}{Jaeyong
  Song}, \bibinfo{person}{Hongsun Jang}, \bibinfo{person}{Dogeun Kim},
  \bibinfo{person}{Jun Sung}, \bibinfo{person}{Taehee Kwon},
  \bibinfo{person}{Jae~Hyung Ju}, \bibinfo{person}{Frank Liu},
  \bibinfo{person}{Yeonkyu Choi}, {and} \bibinfo{person}{Jinho Lee}.}
  \bibinfo{year}{2025}\natexlab{}.
\newblock \showarticletitle{Piccolo: Large-Scale Graph Processing with
  Fine-Grained In-Memory Scatter-Gather}. In \bibinfo{booktitle}{\emph{2025
  IEEE International Symposium on High Performance Computer Architecture
  (HPCA)}} (2025). \bibinfo{publisher}{IEEE}, \bibinfo{pages}{641--656}.
\newblock
\href{https://doi.org/10.1109/HPCA61900.2025.00055}{doi:\nolinkurl{10.1109/HPCA61900.2025.00055}}


\bibitem[Song et~al\mbox{.}(2018)]%
        {Zhuo2018GraphR}
\bibfield{author}{\bibinfo{person}{Linghao Song}, \bibinfo{person}{Youwei
  Zhuo}, \bibinfo{person}{Xuehai Qian}, \bibinfo{person}{Hai Li}, {and}
  \bibinfo{person}{Yiran Chen}.} \bibinfo{year}{2018}\natexlab{}.
\newblock \showarticletitle{GraphR: Accelerating Graph Processing Using ReRAM}.
  In \bibinfo{booktitle}{\emph{2018 IEEE International Symposium on High
  Performance Computer Architecture (HPCA)}} (2018). \bibinfo{pages}{531--543}.
\newblock
\href{https://doi.org/10.1109/HPCA.2018.00052}{doi:\nolinkurl{10.1109/HPCA.2018.00052}}


\bibitem[Sun et~al\mbox{.}(2025)]%
        {Sun2025LincolnR5}
\bibfield{author}{\bibinfo{person}{Weiyi Sun}, \bibinfo{person}{Mingyu Gao},
  \bibinfo{person}{Zhaoshi Li}, \bibinfo{person}{Aoyang Zhang},
  \bibinfo{person}{Iris~Ying Chou}, \bibinfo{person}{Jianfeng Zhu},
  \bibinfo{person}{Shaojun Wei}, {and} \bibinfo{person}{Leibo Liu}.}
  \bibinfo{year}{2025}\natexlab{}.
\newblock \showarticletitle{Lincoln: Real-Time 50~100B LLM Inference on
  Consumer Devices with LPDDR-Interfaced, Compute-Enabled Flash Memory}.
\newblock  (\bibinfo{year}{2025}), \bibinfo{pages}{1734--1750}.
\newblock
\urldef\tempurl%
\url{https://api.semanticscholar.org/CorpusID:277635661}
\showURL{%
\tempurl}


\bibitem[{UCIe Consortium}(2026)]%
        {ucie2026specifications}
\bibfield{author}{\bibinfo{person}{{UCIe Consortium}}.}
  \bibinfo{year}{2026}\natexlab{}.
\newblock \bibinfo{title}{{UCIe Specifications}}.
\newblock
  \bibinfo{howpublished}{\url{https://www.uciexpress.org/specifications}}.
\newblock
\newblock
\shownote{Accessed: 2026-06-17}.


\bibitem[Wang et~al\mbox{.}(2025a)]%
        {Wang2025KVCacheWild}
\bibfield{author}{\bibinfo{person}{Jiahao Wang}, \bibinfo{person}{Jinbo Han},
  \bibinfo{person}{Xingda Wei}, \bibinfo{person}{Sijie Shen},
  \bibinfo{person}{Dingyan Zhang}, \bibinfo{person}{Chenguang Fang},
  \bibinfo{person}{Rong Chen}, \bibinfo{person}{Wenyuan Yu}, {and}
  \bibinfo{person}{Haibo Chen}.} \bibinfo{year}{2025}\natexlab{a}.
\newblock \showarticletitle{KVCache Cache in the Wild: Characterizing and
  Optimizing KVCache Cache at a Large Cloud Provider}. In
  \bibinfo{booktitle}{\emph{Proceedings of the 2025 USENIX Annual Technical
  Conference}} (2025). \bibinfo{publisher}{USENIX Association},
  \bibinfo{pages}{465--480}.
\newblock
\urldef\tempurl%
\url{https://www.usenix.org/conference/atc25/presentation/wang-jiahao}
\showURL{%
\tempurl}


\bibitem[Wang et~al\mbox{.}(2026b)]%
        {Wang2026AdaptiveDraft}
\bibfield{author}{\bibinfo{person}{Runze Wang}, \bibinfo{person}{Qinggang
  Wang}, \bibinfo{person}{Haifeng Liu}, \bibinfo{person}{Long Zheng},
  \bibinfo{person}{Xiaofei Liao}, \bibinfo{person}{Hai Jin}, {and}
  \bibinfo{person}{Jingling Xue}.} \bibinfo{year}{2026}\natexlab{b}.
\newblock \showarticletitle{Adaptive Draft Sequence Length: Enhancing
  Speculative Decoding Throughput on PIM-Enabled Systems}. In
  \bibinfo{booktitle}{\emph{2026 IEEE International Symposium on High
  Performance Computer Architecture (HPCA)}} (2026). \bibinfo{publisher}{IEEE},
  \bibinfo{pages}{1--15}.
\newblock
\href{https://doi.org/10.1109/HPCA68181.2026.11408598}{doi:\nolinkurl{10.1109/HPCA68181.2026.11408598}}


\bibitem[Xiong et~al\mbox{.}(2022)]%
        {Xiong2022SecNDP}
\bibfield{author}{\bibinfo{person}{Wenjie Xiong}, \bibinfo{person}{Liu Ke},
  \bibinfo{person}{Dimitrije Jankov}, \bibinfo{person}{Michael Kounavis},
  \bibinfo{person}{Xiaochen Wang}, \bibinfo{person}{Eric Northup},
  \bibinfo{person}{Jie~Amy Yang}, \bibinfo{person}{Bilge Acun},
  \bibinfo{person}{Carole~Jean Wu}, \bibinfo{person}{Ping Tak~Peter Tang},
  \bibinfo{person}{G.~Edward Suh}, \bibinfo{person}{Xuan Zhang}, {and}
  \bibinfo{person}{Hsien Hsin~S. Lee}.} \bibinfo{year}{2022}\natexlab{}.
\newblock \showarticletitle{SecNDP: Secure Near-Data Processing with Untrusted
  Memory}. In \bibinfo{booktitle}{\emph{Proceedings - 2022 IEEE International
  Symposium on High-Performance Computer Architecture, HPCA 2022}} (Virtual,
  Online, Korea, Republic of, 2022) \emph{(\bibinfo{series}{Proceedings -
  International Symposium on High-Performance Computer Architecture},
  Vol.~\bibinfo{volume}{2022-April})}. \bibinfo{publisher}{IEEE Computer
  Society}, \bibinfo{pages}{244--258}.
\newblock
\showISBNx{9781665420273}
\href{https://doi.org/10.1109/HPCA53966.2022.00025}{doi:\nolinkurl{10.1109/HPCA53966.2022.00025}}


\bibitem[Yenimol et~al\mbox{.}(2026)]%
        {Yenimol2026CoGraf}
\bibfield{author}{\bibinfo{person}{Mehmetali~Semi Yenimol},
  \bibinfo{person}{Anirban Nag}, \bibinfo{person}{Chang~Hyun Park}, {and}
  \bibinfo{person}{David Black-Schaffer}.} \bibinfo{year}{2026}\natexlab{}.
\newblock \showarticletitle{CoGraf: Fully Accelerating Graph Applications with
  Fine-Grained PIM}. In \bibinfo{booktitle}{\emph{Proceedings of the 31st ACM
  International Conference on Architectural Support for Programming Languages
  and Operating Systems, Volume 2}} (Pittsburgh, PA, USA, 2026)
  \emph{(\bibinfo{series}{ASPLOS '26})}. \bibinfo{publisher}{Association for
  Computing Machinery}, \bibinfo{pages}{409--425}.
\newblock
\href{https://doi.org/10.1145/3779212.3790142}{doi:\nolinkurl{10.1145/3779212.3790142}}


\bibitem[Yun et~al\mbox{.}(2024)]%
        {Yun2024Duplex}
\bibfield{author}{\bibinfo{person}{Sungmin Yun}, \bibinfo{person}{Kwanhee
  Kyung}, \bibinfo{person}{Juhwan Cho}, \bibinfo{person}{Jaewan Choi},
  \bibinfo{person}{Jongmin Kim}, \bibinfo{person}{Byeongho Kim},
  \bibinfo{person}{Sukhan Lee}, \bibinfo{person}{Kyomin Sohn}, {and}
  \bibinfo{person}{Jung~Ho Ahn}.} \bibinfo{year}{2024}\natexlab{}.
\newblock \bibinfo{title}{Duplex: A Device for Large Language Models with
  Mixture of Experts, Grouped Query Attention, and Continuous Batching}.
\newblock
\showeprint[arXiv]{2409.01141}~[cs.AR]
\urldef\tempurl%
\url{https://arxiv.org/abs/2409.01141}
\showURL{%
\tempurl}


\bibitem[Zhang et~al\mbox{.}(2018a)]%
        {Zhuo2018GraphP}
\bibfield{author}{\bibinfo{person}{Mingxing Zhang}, \bibinfo{person}{Youwei
  Zhuo}, \bibinfo{person}{Chao Wang}, \bibinfo{person}{Mingyu Gao},
  \bibinfo{person}{Yongwei Wu}, \bibinfo{person}{Kang Chen},
  \bibinfo{person}{Christos Kozyrakis}, {and} \bibinfo{person}{Xuehai Qian}.}
  \bibinfo{year}{2018}\natexlab{a}.
\newblock \showarticletitle{GraphP: Reducing Communication for PIM-Based Graph
  Processing with Efficient Data Partition}. In \bibinfo{booktitle}{\emph{2018
  IEEE International Symposium on High Performance Computer Architecture
  (HPCA)}} (2018). \bibinfo{pages}{544--557}.
\newblock
\href{https://doi.org/10.1109/HPCA.2018.00053}{doi:\nolinkurl{10.1109/HPCA.2018.00053}}


\bibitem[Zhang et~al\mbox{.}(2025c)]%
        {Zhang2025FENIX}
\bibfield{author}{\bibinfo{person}{Tengyu Zhang}, \bibinfo{person}{Chenqi Lin},
  \bibinfo{person}{Jiangrui Yu}, \bibinfo{person}{Yi Chen},
  \bibinfo{person}{Shuwen Deng}, {and} \bibinfo{person}{Meng Li}.}
  \bibinfo{year}{2025}\natexlab{c}.
\newblock \showarticletitle{(Invited) FENIX: Flexible and Efficient Hybrid
  HE/MPC Acceleration with Near-Memory Processing}. In
  \bibinfo{booktitle}{\emph{2025 IEEE/ACM International Conference On Computer
  Aided Design (ICCAD)}} (2025). \bibinfo{pages}{1--9}.
\newblock
\href{https://doi.org/10.1109/ICCAD66269.2025.11240701}{doi:\nolinkurl{10.1109/ICCAD66269.2025.11240701}}


\bibitem[Zhang et~al\mbox{.}(2025b)]%
        {Zhang2025OSDIDiffKV}
\bibfield{author}{\bibinfo{person}{Yanqi Zhang}, \bibinfo{person}{Yuwei Hu},
  \bibinfo{person}{Runyuan Zhao}, \bibinfo{person}{John~C.S. Lui}, {and}
  \bibinfo{person}{Haibo Chen}.} \bibinfo{year}{2025}\natexlab{b}.
\newblock \showarticletitle{{DiffKV}: Differentiated Memory Management for
  Large Language Models with Parallel {KV} Compaction}. In
  \bibinfo{booktitle}{\emph{Proceedings of the 19th USENIX Symposium on
  Operating Systems Design and Implementation (OSDI)}} (2025).
  \bibinfo{publisher}{USENIX Association}.
\newblock


\bibitem[Zhao et~al\mbox{.}(2025)]%
        {Zhao2025PUSHtap}
\bibfield{author}{\bibinfo{person}{Yilong Zhao}, \bibinfo{person}{Mingyu Gao},
  \bibinfo{person}{Huanchen Zhang}, \bibinfo{person}{Fangxin Liu},
  \bibinfo{person}{Gongye Chen}, \bibinfo{person}{He Xian},
  \bibinfo{person}{Haibing Guan}, {and} \bibinfo{person}{Li Jiang}.}
  \bibinfo{year}{2025}\natexlab{}.
\newblock \showarticletitle{PUSHtap: PIM-based In-Memory HTAP with Unified Data
  Storage Format}. In \bibinfo{booktitle}{\emph{Proceedings of the 30th ACM
  International Conference on Architectural Support for Programming Languages
  and Operating Systems, Volume 3}} (Rotterdam, Netherlands, 2025)
  \emph{(\bibinfo{series}{ASPLOS '25})}. \bibinfo{publisher}{Association for
  Computing Machinery}, \bibinfo{pages}{179--194}.
\newblock
\href{https://doi.org/10.1145/3676642.3736120}{doi:\nolinkurl{10.1145/3676642.3736120}}


\bibitem[Zhou et~al\mbox{.}(2021a)]%
        {Zhou2021MAT}
\bibfield{author}{\bibinfo{person}{Minxuan Zhou}, \bibinfo{person}{Yunhui Guo},
  \bibinfo{person}{Weihong Xu}, \bibinfo{person}{Bin Li},
  \bibinfo{person}{Kevin~W. Eliceiri}, {and} \bibinfo{person}{Tajana Rosing}.}
  \bibinfo{year}{2021}\natexlab{a}.
\newblock \showarticletitle{MAT: Processing In-Memory Acceleration for
  Long-Sequence Attention}. In \bibinfo{booktitle}{\emph{2021 58th ACM/IEEE
  Design Automation Conference (DAC)}} (2021). \bibinfo{pages}{25--30}.
\newblock
\href{https://doi.org/10.1109/DAC18074.2021.9586212}{doi:\nolinkurl{10.1109/DAC18074.2021.9586212}}


\bibitem[Zhou et~al\mbox{.}(2022c)]%
        {Zhou2022TransPIM}
\bibfield{author}{\bibinfo{person}{Minxuan Zhou}, \bibinfo{person}{Weihong Xu},
  \bibinfo{person}{Jaeyoung Kang}, {and} \bibinfo{person}{Tajana Rosing}.}
  \bibinfo{year}{2022}\natexlab{c}.
\newblock \showarticletitle{TransPIM: A Memory-based Acceleration via
  Software-Hardware Co-Design for Transformer}. In
  \bibinfo{booktitle}{\emph{2022 IEEE International Symposium on
  High-Performance Computer Architecture (HPCA)}} (2022).
  \bibinfo{pages}{1071--1085}.
\newblock
\href{https://doi.org/10.1109/HPCA53966.2022.00082}{doi:\nolinkurl{10.1109/HPCA53966.2022.00082}}


\bibitem[Zhou et~al\mbox{.}(2022b)]%
        {zhou2022gnnearacceleratingfullbatchtraining}
\bibfield{author}{\bibinfo{person}{Zhe Zhou}, \bibinfo{person}{Cong Li},
  \bibinfo{person}{Xuechao Wei}, \bibinfo{person}{Xiaoyang Wang}, {and}
  \bibinfo{person}{Guangyu Sun}.} \bibinfo{year}{2022}\natexlab{b}.
\newblock \bibinfo{title}{GNNear: Accelerating Full-Batch Training of Graph
  Neural Networks with Near-Memory Processing}.
\newblock
\showeprint[arXiv]{2111.00680}~[cs.LG]
\urldef\tempurl%
\url{https://arxiv.org/abs/2111.00680}
\showURL{%
\tempurl}


\bibitem[Zhuo et~al\mbox{.}(2021b)]%
        {Zhuo2021DistGraphPIM}
\bibfield{author}{\bibinfo{person}{Youwei Zhuo}, \bibinfo{person}{Jingji Chen},
  \bibinfo{person}{Gengyu Rao}, \bibinfo{person}{Qinyi Luo},
  \bibinfo{person}{Yanzhi Wang}, \bibinfo{person}{Hailong Yang},
  \bibinfo{person}{Depei Qian}, {and} \bibinfo{person}{Xuehai Qian}.}
  \bibinfo{year}{2021}\natexlab{b}.
\newblock \showarticletitle{Distributed Graph Processing System and
  Processing-in-memory Architecture with Precise Loop-carried Dependency
  Guarantee}.
\newblock  \bibinfo{volume}{37}, \bibinfo{number}{1–4}
  (\bibinfo{year}{2021}).
\newblock
\showISSN{0734-2071}
\href{https://doi.org/10.1145/3453681}{doi:\nolinkurl{10.1145/3453681}}


\bibitem[Zhuo et~al\mbox{.}(2019a)]%
        {Zhuo2019GraphQ}
\bibfield{author}{\bibinfo{person}{Youwei Zhuo}, \bibinfo{person}{Chao Wang},
  \bibinfo{person}{Mingxing Zhang}, \bibinfo{person}{Rui Wang},
  \bibinfo{person}{Dimin Niu}, \bibinfo{person}{Yanzhi Wang}, {and}
  \bibinfo{person}{Xuehai Qian}.} \bibinfo{year}{2019}\natexlab{a}.
\newblock \showarticletitle{GraphQ: Scalable PIM-Based Graph Processing}. In
  \bibinfo{booktitle}{\emph{Proceedings of the 52nd Annual IEEE/ACM
  International Symposium on Microarchitecture}} (Columbus, OH, USA, 2019)
  \emph{(\bibinfo{series}{MICRO '52})}. \bibinfo{publisher}{Association for
  Computing Machinery}, \bibinfo{pages}{712--725}.
\newblock
\showISBNx{9781450369381}
\href{https://doi.org/10.1145/3352460.3358256}{doi:\nolinkurl{10.1145/3352460.3358256}}


\bibitem[Zou et~al\mbox{.}(2026)]%
        {Zou2026NasZip}
\bibfield{author}{\bibinfo{person}{Cheng Zou}, \bibinfo{person}{Shuo Yang},
  \bibinfo{person}{Chen Nie}, \bibinfo{person}{Yu Zou}, \bibinfo{person}{Yu
  He}, \bibinfo{person}{Chao Jiang}, \bibinfo{person}{Limin Xiao},
  \bibinfo{person}{Weifeng Zhang}, {and} \bibinfo{person}{Zhezhi He}.}
  \bibinfo{year}{2026}\natexlab{}.
\newblock \showarticletitle{NasZip: Software and Hardware Co-design to
  Accelerate Approximate Nearest Neighbor Search with DIMM-based Near-Data
  Processing}. In \bibinfo{booktitle}{\emph{53rd Annual International Symposium
  on Computer Architecture (ISCA '26)}} (2026).
\newblock
\urldef\tempurl%
\url{https://www.iscaconf.org/isca2026/program/}
\showURL{%
\tempurl}
\newblock
\shownote{ISCA 2026 technical program}.


\end{thebibliography}

\end{document}